\documentclass[12pt]{article}

\usepackage{amssymb,amsmath,amsfonts,eurosym,ulem,graphicx,color,setspace,sectsty,comment,footmisc,pdflscape,array,hyperref,makecell,longtable,pdflscape,booktabs,subcaption,siunitx,appendix,placeins,rotating, pgfplots, indentfirst,xstring, bm}
\usepackage[labelfont=bf]{caption}
\usepackage[longtable]{multirow}
\usepackage[blocks]{authblk}
\usepackage{lmodern}
\usepackage{setspace}
\usepackage{float}

\usepackage{tikz}
\usetikzlibrary{positioning, shapes.geometric, calc, arrows.meta, decorations.pathmorphing, fit, backgrounds}
\usepackage[
    backend=biber,
    style=authoryear,
    natbib=true,
    defernumbers=true
]{biblatex}
\newcolumntype{L}[1]{>{\raggedright\let\newline\\arraybackslash\hspace{0pt}}m{#1}}
\newcolumntype{C}[1]{>{\centering\let\newline\\arraybackslash\hspace{0pt}}m{#1}}
\newcolumntype{R}[1]{>{\raggedleft\let\newline\\arraybackslash\hspace{0pt}}m{#1}}
\newcommand{\st}[1]{\makebox[0pt][l]{\textsuperscript{#1}}}

\def\doublespacing{\setstretch{2}}
\usepackage[
  letterpaper,
  margin=1in
]{geometry}

\newcommand{\Acknowledgement} {
We would like to thank Hyeng Keun Koo, Kwangmin Jung, Dojoon Park (discussant), JinGi Ha, Jeonggyu Huh, Kyoung-Kuk Kim (discussant), Thummim Cho, Mark Loewenstein, Hong Liu, and seminar participants at the 2024 Spring Joint Conference of Korean Operations Research and Management Science Society and Korean Institute of Industrial Engineers, 2025 Asia-Pacific Association of Finance International Conference, 2025 Korean Finance Association Fall Conference, 2025 4th Workshop on Financial Mathematics and Engineering (Pusan National University), 2025 Korea Derivatives Association Fall Conference, 2025 Annual Conference on Asia-Pacific Financial Markets (CAFM), for helpful discussions and insightful comments. This work was supported by the Ministry of Education of the Republic of Korea and the National Research Foundation of Korea [NRF-2022S1A3A2A02089950]. Kim (E-mail: \texttt{changeun120@postech.ac.kr}) is at Department of Industrial and Management Engineering, POSTECH; Jeong (E-mail: \texttt{younwoo48@postech.ac.kr}) is at Graduate School of Artificial Intelligence, POSTECH; Jang (E-mail: \texttt{bonggyujang@postech.ac.kr}) is at Department of Industrial and Management Engineering, POSTECH, Korea University Business School. Correspondence concerning this article should be addressed to Bong-Gyu Jang, Department of Industrial and Management Engineering, POSTECH, Pohang 37673, Republic of Korea. E-mail: \texttt{bonggyujang@postech.ac.kr}.
}

\begin{document}

\begin{refsection}

\title{Interpretable Deep Learning for Stock Returns:\\ A Consensus-Bottleneck Asset Pricing Model
\thanks{\Acknowledgement}
}
\author{Changeun Kim \qquad Younwoo Jeong \qquad Bong-Gyu Jang}
\newcommand{\firstdraftdate}{March 5, 2025}
\date{}

\begin{titlepage}
\maketitle

\begin{center}
    \bigskip\bigskip\bigskip\bigskip
    {\small This draft: April 24, 2026 \\ First draft: \firstdraftdate}
\end{center}
\vfill
\setcounter{page}{0}
\thispagestyle{empty}
\end{titlepage}

\begin{titlepage}
    \begin{center}
        \LARGE
        Interpretable Deep Learning for Stock Returns: \\ A Consensus-Bottleneck Asset Pricing Model
        \vspace{1em}
    \end{center}

\begin{abstract}

\noindent We introduce the \textit{Consensus-Bottleneck Asset Pricing Model} (CB-APM), which embeds aggregate analyst consensus as a structural bottleneck, treating professional beliefs as a sufficient statistic for the market's high-dimensional information set. Unlike post-hoc explainability approaches, CB-APM achieves interpretability-by-design: the bottleneck constraint functions as an endogenous regularizer that simultaneously improves out-of-sample predictive accuracy and anchors inference to economically interpretable drivers. Portfolios sorted on CB-APM forecasts exhibit a strong monotonic return gradient, robust across macroeconomic regimes. Pricing diagnostics further reveal that the learned consensus encodes priced variation not spanned by canonical factor models, identifying belief-driven risk heterogeneity that standard linear frameworks systematically miss.

\vspace{0in}
\vspace{0in}
\end{abstract}

\thispagestyle{empty}
\end{titlepage}

\pagebreak \newpage

\section{Introduction} \label{sec:introduction}

Empirical asset pricing has long relied on statistical modeling to explain stock returns, often within the framework of factor-based models such as those proposed by \citeauthor{fama1993common} (\citeyear{fama1993common}, \citeyear{fama2015five}) and \textcite{carhart1997persistence}. These models aim to enhance explanatory power by identifying systematic risk factors that drive returns. However, despite decades of research, the ability of traditional models to predict future stock returns remains constrained, particularly in out-of-sample settings \parencite{ang2007stock, campbell2008predicting, cochrane2008dog}. The proliferation of new factors, often referred to as the ``factor zoo'' \parencite{cochrane2011presidential}, has further complicated the landscape, raising concerns about robustness, data mining, and the true economic relevance of many proposed predictors.

Recent work addresses these challenges using machine learning to extract signals from the factor zoo. Drawing from a number of studies on stock return predictors, \footnote{See \textcite{welch2008comprehensive}, \textcite{green2013supraview}, \textcite{hou2015digesting}, \textcite{harvey2016and}, \textcite{he2017intermediary}, \textcite{green2017characteristics},  \textcite{gu2020empirical}, \textcite{feng2020taming}, \textcite{freyberger2020dissecting},\textcite{bybee2023narrative} and \textcite{jensen2023there}.} seminal work of \textcite{gu2020empirical} proposes a ``return prediction model" that integrates traditional asset pricing empirical frameworks and theories with the rapidly evolving field of machine learning. By utilizing a variety of machine learning algorithms including neural networks, and leveraging a high-dimensional set of predictive factors, their results significantly contribute to the literature by showing the effectiveness of nonlinear and complex modeling on empirical asset pricing. Several subsequent studies utilize the conceptual formulation of this study across diverse financial markets and assets such as bonds \parencite{bianchi2021bond}, cryptocurrencies \parencite{jaquart2021short, fang2024ascertaining} and foreign stock exchanges \parencite{leippold2022machine}. Theoretical studies have also emerged to justify the use of machine learning in empirical asset pricing. For instance, \textcite{kelly2024virtue} illustrates how model complexity can be instrumental in achieving superior performance in cross-sectional return prediction, demonstrated through a simple example of penalized linear regression. While return prediction models benefit from machine learning approaches due to their empirical flexibility, deep learning has also proven successful in approximating ``asset pricing factor models" \citep[see, e.g.,][]{feng2018deep, gu2021autoencoder, chen2024deep}. 

\subsection{Beyond black boxes: the virtue of interpretability}

Despite the strong evidence that deep learning approaches illustrate evident potential in capturing the complex topology of predictor structures, a critical limitation remains: Can the results from these models be considered trustworthy? \textcite{rudin2022interpretable} highlights such critical issue with machine learning black box models, 
\begin{quote}
   \textit{Black box models often predict the right answer for the wrong reason (the ``Clever Hans" phenomenon), leading to excellent performance in training but poor performance in practice.}
\end{quote}
Recent studies in machine learning asset pricing frequently employ models that are not interpretable,\footnote{While the finance literature often uses the terms ``explainability'' and ``interpretability'' interchangeably, we maintain a strict conceptual distinction to ensure the reliability of our economic inferences. To keep the focus on the economic mechanisms of belief aggregation, we provide a detailed taxonomy of AI transparency in Internet Appendix~\ref{sec:interpret}. This section clarifies why our framework prioritizes \textit{interpretability-by-design} over the \textit{post-hoc} explainability (XAI) common in recent black-box models. Such rigor is essential to verify that the CB-APM is capturing priced fundamental signals rather than suffering from the ``Clever Hans'' phenomenon, where models achieve high accuracy through spurious, non-economic correlations.} which raises concerns about relying on complex machine learning algorithms in empirical asset pricing without a clear understanding of why and how these models arrive at their conclusions. Furthermore, these papers often attempt to interpret the prediction results based on the learned models and derive economic implications. However, \textcite{rudin2019stop} argues that such analyses are solely based on post-hoc explanations that should be considered as fitting narratives to the outcomes. Despite growing interest in interpretable machine learning and trustworthy Artificial Intelligence (AI), a notable gap persists in applying and validating these approaches within asset pricing, beyond traditional regression or decision-tree models. In particular, existing machine-learning frameworks rarely achieve both strong predictive performance and economic interpretability. To address this gap, we propose the Consensus-Bottleneck Asset Pricing Model (CB-APM), a framework that employs a partially interpretable neural architecture to predict future stock returns while preserving clear economic structure. In particular, the model explicitly represents the intermediate layer of market expectations---how information is transformed into beliefs---thereby providing a natural foundation for anchoring the architecture in analyst consensus.

\subsection{The analyst as an information intermediary}

The selection of analyst consensus as the anchor for this ``economic structure'' is predicated on its role as the primary information aggregator in capital markets. Our approach builds upon two established pillars of financial economics: the rational expectations hypothesis and empirically documented relationships between analyst consensus information and asset prices. Rational expectations, proposed by \textcite{muth1961rational}, posit that market participants form forecasts using all available historical information. Sell-side analysts serve as the primary ``cognitive filters'' in this process; they do not merely process data but act as vital information intermediaries who synthesize high-dimensional ``hard'' signals with qualitative ``soft'' information.

Empirical evidence from sell-side analysts supports this mechanism, linking information processing to observed forecasting behavior. \textcite{lovell1986tests} shows economic agents systematically incorporate public information into earnings forecasts, while \textcite{lim2001rationality} establishes predictable patterns in analysts' forecast revisions consistent with Bayesian updating. Crucially, \textcite{jegadeesh2004analyzing} identify specific style factors, including momentum, growth prospects, and trading volume, that systematically influence analysts' stock recommendations, suggesting a quantifiable link between firm characteristics and consensus formation.  \textcite{diether2002differences} show that forecast dispersion  is priced, while \textcite{sorescu2006cross} document pronounced price reactions to revisions. \textcite{barber2001can} further demonstrates the economic significance of consensus recommendations, showing that strategies based on the most and least favorable recommendations yield significant abnormal gross returns.

However, the efficacy of relying on these aggregated measures is nuanced, as their predictive value is critically moderated by the underlying heterogeneity of beliefs and inherent institutional biases. For instance, \textcite{palley2025effect} demonstrate that the informativeness of consensus depends on dispersion, while \citet{van2023man} find that conditional expectations are, on average, upwardly biased. Despite behavioral complexities, analyst consensus remains a critical mediator for future returns. \footnote{Recent evidence suggests a synergy between human and machine intelligence; \citet{cao2024man} show that combining AI's computational power with the human capacity to synthesize soft institutional information yields the most accurate forecasts. \citet{van2023man} demonstrate that when machine learning is used to successfully isolate forecast biases, these signals are predictive of both returns and corporate financing decisions.} Rather than blindly replicating these biases, CB-APM is designed to treat consensus as a sufficient statistic for the market's information set. By treating consensus as the bottleneck, the model acknowledges that while individual analysts may err, the collective consensus represents the final aggregation of information before it is incorporated into market prices. This allows the model to filter out high-dimensional patterns that market experts, incentivized to find alpha, have already deemed irrelevant for asset pricing.

The CB-APM framework operationalizes these insights through a concept-bottleneck architecture inspired by \textcite{koh2020concept}, prioritizing interpretability-by-design. This architecture serves as a structural filter that disciplines the ``factor zoo'', ensuring the model only utilizes characteristics that are salient enough to influence the expectations of market participants. By anchoring the latent states to observable analyst consensus, we effectively prevent the model from exploiting spurious correlations that lack a documented foundation in human belief formation. Building on the necessity to separate signal from noise, CB-APM is designed to recover the priced component of these expectations by explicitly filtering out the behavioral biases inherent in their aggregation. Its nonlinear ``consensus formation'' stage synthesizes firm characteristics and macroeconomic states into consensus-like latent expectations, reflecting the documented process through which analysts aggregate information. A subsequent linear ``pricing'' stage translates these learned expectations into expected returns, preserving interpretability through transparent economic loadings. By routing all predictive content through these latent expectations, the framework imposes an inherent information constraint that limits reliance on spurious high-dimensional patterns and anchors inference to economically interpretable drivers.

\subsection{Key contributions and research framework}

Our contributions are threefold. First, we introduce a concept-bottleneck framework that synthesizes the high-dimensional predictor set into interpretable, consensus-style expectations, providing a structured economic link between characteristics, analysts' beliefs, and expected returns. Second, we demonstrate that this architecture delivers economically large improvements in long-horizon return prediction across expanding-window evaluations. Third, we show that the learned consensus representations encode priced information that is only partially spanned by traditional factor models, offering new empirical insight into how belief heterogeneity and information aggregation shape risk premia. These contributions advance recent efforts to integrate interpretable machine learning with the core principles of empirical asset pricing.

To empirically validate the effectiveness of CB-APM, we assess its predictive performance and economic implications using a comprehensive dataset spanning from January 1994 to December 2023, consisting of 605{,}722 firm-month observations across 4{,}683 U.S. companies. The dataset integrates 114 firm-level predictors, 123 macroeconomic indicators, and 9 analysts' consensus variables including EPS forecast revisions and forecast dispersions. 

Our empirical analysis demonstrates that CB-APM delivers substantial improvements in both predictive performance and economic interpretability. First, in the cross-section of consensus and stock returns, incorporating consensus learning markedly enhances long-horizon return forecasts: CB-APM attains an out-of-sample $R^2$ of 10.46\% for annual returns, representing a significant improvement over a standard deep learning benchmark ($R^2=7.63\%$), while simultaneously achieving an average $R^2$ of 24.21\% in approximating analyst consensus variables. Second, portfolio-level analyses establish the model's economic relevance. Portfolios formed on out-of-sample CB-APM predictions display strongly monotonic payoff structures, with high-minus-low spreads approaching 2.3\% per month for regularized specifications ($\lambda \ge 0.3$). The double sorts on model-implied returns and analysts' earnings forecasts further reveal that the model internalizes both the informational and behavioral components embedded in analyst expectations. In particular, the expected-return spreads are largest in states characterized by analyst pessimism—low analysts' earnings forecasts levels—where expectation errors and mispricing are most pronounced, and they progressively shrink as analyst optimism increases. This state-dependent attenuation indicates that the CB-APM distills the priced component of forecasted earnings while appropriately adjusting for optimism-driven noise in analysts' beliefs. Third, long-short portfolios derived from the model’s forecasts achieve economically significant and stable out-of-sample performance, with mean monthly log returns rising from 1.53\% at $\lambda=0$ to 2.20\% at $\lambda=0.3$ and the annualized Sharpe ratio improving from 1.10 to 1.44. 

Beyond predictive performance, we evaluate the asset pricing implications of the CB-APM signals using Gibbons--Ross--Shanken (GRS) tests. Traditional factor models increasingly fail to price portfolios formed on CB-APM’s predicted returns as the consensus-bottleneck tightens, indicating that the model uncovers structured forms of nonlinear or interaction-based return heterogeneity that lie outside the linear span of standard factors. Portfolios sorted on individual consensus dimensions exhibit modest pricing errors, suggesting that belief-based signals reflect compressible yet economically meaningful combinations of characteristics. Taken together, these findings show that CB-APM extracts interpretable consensus representations that contain priced information only partially captured by existing factor models, positioning the framework as a complementary approach that links analysts' heterogeneous beliefs to expected returns in a transparent and theoretically coherent manner.

Collectively, these results establish CB-APM as a novel and effective framework that integrates interpretable deep learning with foundational principles of financial economics. Unlike prior machine learning approaches that prioritize accuracy at the expense of transparency, CB-APM demonstrates that interpretable architectures can preserve theoretical grounding while achieving strong empirical performance. By jointly modeling analysts' expectations and stock returns, our framework provides a principled means of disentangling forward-looking information embedded in firm characteristics and macroeconomic variables, yielding insights into how such information is aggregated and priced. This dual capacity, enhancing return predictability while maintaining an economically interpretable structure, constitutes the central contribution of our paper and advances the emerging literature on interpretable machine learning in finance.

The remainder of the paper is organized as follows. Section~\ref{sec:model} outlines the model, estimation procedure, and architecture. Section~\ref{sec:empirical} presents empirical results on predictive performance, portfolio-based economic implications, and the pricing content of the approximated consensuses. Section~\ref{sec:conclusion} concludes. The Internet Appendix provides additional robustness analyses and supplementary results.

\section{A Structural Representation of Belief Aggregation and Pricing} \label{sec:model}

\subsection{Mapping the factor zoo to structured market beliefs}

Similar to the \textcite{gu2020empirical}, the asset return prediction error model utilized in our work is formulated for the $h$-horizon forecasting problem as follows:

\begin{equation}
\label{eq:pred}
R_{i,t+h}
=
\mathbb{E}_t\!\left[ R_{i,t+h}\right]
+
\varepsilon_{i,t+h},
\end{equation}

where $R_{i,t+h}$ is the $h$-month return of asset $i$ excess of the risk-free rate at time $t+h$, and $\varepsilon_{i,t+h}$ is an error term. In this context, $h$ is used to assign the forecasting horizon, enabling the consideration of multi-horizon predictions, allowing CB-APM to model long-term dependencies. The expected excess return in equation \eqref{eq:pred} is defined as the expectation conditional on information sets,
$$
\mathbb{E}_t\!\left[ R_{i,t+h} \right]
=
\mathbb{E}\!\left[ R_{i,t+h} \mid \mathcal{I}^f_{i,t},\, \mathcal{I}^m_t \right].
$$
Here, $\mathcal{I}^f_{i,t}$ and $\mathcal{I}^m_t$ are the sets of firm-specific characteristics and macroeconomic predictors at time $t$, respectively. \footnote{A detailed description of the predictors comprising the information sets is provided in Section \ref{sec:data}, and a complete list of variables is available in Internet Appendix \ref{Appendix:data}. The macroeconomic information set $\mathcal{I}^m_t$ is represented empirically by a latent vector extracted through an autoencoder trained on macroeconomic variables, as described in Internet Appendix~\ref{Appendix:autoencoder}.} It is important to note that consensus information is deliberately excluded from $\mathcal{I}^f_{i,t}$. 

This framework is further developed by defining the functional form of the conditional expectation as a composite function,
\begin{equation}
\label{eq:cond}
\mathbb{E}\!\left[ R_{i,t+h} \mid \mathcal{I}^f_{i,t},\, \mathcal{I}^m_t \right]
=
g\!\left(
f\!\left( \mathcal{I}^f_{i,t},\, \mathcal{I}^m_t ; \phi \right)
; \theta
\right),
\end{equation}
where the function $f(\cdot)$ and $g(\cdot)$ are smooth functions parameterized by learnable parameters $\theta$ and $\phi$. The function $f(\cdot)$ is specifically designed to model the conditional expectation of analyst consensus. Then, the function $g(\cdot)$ models the expected return only using the features of approximated consensus from the previous step, creating a ``concept-bottleneck" within the prediction model. This empirical design is predicated on the understanding that both researchers in empirical asset pricing and financial analysts share the objective of assessing a firm's value and discerning the factors that influence these valuations. While analysts often have access to broader datasets, including some predictive signals that may not be publicly available or included in this article, the asset pricing panel can provide a high-quality, observable subset of information about firm fundamentals and macroeconomic conditions that is useful for approximating analyst consensus, as shown in the empirical results later on.  

In mathematical form, $f(\cdot)$ approximates the analyst consensus variables, denoted as $C_{i,t}$. 
$$
C_{i,t}
=
f\!\left( \mathcal{I}^f_{i,t},\, \mathcal{I}^m_t \,;\, \phi \right).
$$
Let the approximated value of $C_{i,t}$ and the parameter $\phi$ be $\hat{C}_{i,t}$ and $\hat{\phi}$, respectively, then,
$$
\hat{C}_{i,t}
=
f\!\left( \mathcal{I}^f_{i,t},\, \mathcal{I}^m_t \,;\, \hat{\phi} \right).
$$
Finally, the expected excess return is defined with function $g(\cdot)$ and the approximated $\hat{C}_{i,t}$ as below,
\begin{equation}
\label{eq:final_model}
\mathbb{E}_t\!\left[ R_{i,t+h} \right]
=
g\!\left( \hat{C}_{i,t} \,;\, \theta \right).
\end{equation}

As discussed in \textcite{daniel1997evidence}, the main limitation of the return prediction error model is the absence of economic constraints. For instance, the fundamental theorem of asset pricing constrains the arbitrage opportunity, which implies that the difference between the price of  an identical asset is improbable. This condition is referred to as ``the law of one price'' in asset pricing theory. In the cases without such condition, two different assets can have identical price despite disparate fundamental values. 

However, CB-APM diverges from the approach of return prediction modeling for several reasons. Firstly, it offers greater flexibility, accommodating diverse scenarios involving analyst estimates and future returns. Unlike factor models, which do not differentiate prices of identical risk factors, CB-APM acknowledges that similar analyst opinions across firms may yield distinct future returns. While we assume rational decision-making by analysts, as discussed in subsequent sections, it is prudent not to constrain such scenarios initially. Secondly, CB-APM facilitates a range of optimization approaches in approximating the asset pricing model. Unlike factor models, where the estimation process is mostly the extension of Fama--MacBeth regression \parencite{fama1973risk} restricting the integration of the entire expected return modeling process, CB-APM allows for a more holistic training process, avoiding multiple optimization procedures. Overall, given that the consensus-bottleneck represents a novel approach in asset pricing research, we aimed to maintain the underlying framework as simple and flexible as possible.

Although it is designed as intended, given that neural networks are well-recognized as ``universal approximators", the model can generate outcomes that need not align with economic theory. To overcome the limitation of the proposed prediction model, we apply stabilized optimization approaches proposed in the machine learning literature, such as regularization and scheduling. Such techniques are expected to function as ``universal constraints", improving empirical performance while preserving theoretical discipline. See Internet Appendix \ref{Appendix:hyper} for detailed discussions and experimental settings. 

\subsection{Joint estimation and endogenous bottleneck}

In this section, we provide the loss function of the model that simultaneously estimates the parameters of function $f(\cdot)$ and $g(\cdot)$ from equation \eqref{eq:cond} in a single optimization step. 

Given $\lambda>0$, the model's loss function is structured as a joint optimization task, represented by a weighted sum of two distinct loss functions: 
\begin{equation}
\label{eq:loss}
\mathcal{L}
=
\mathcal{L}_R
+
\lambda\, \langle\textbf{1}, \mathcal{L}_C\rangle,
\end{equation}
where the ``return loss" $\mathcal{L}_R$ is formulated as,
\begin{equation}
\label{eq:L_R}
\mathcal{L}_R(\phi,\theta)
=
\frac{1}{N T}
\sum_{i=1}^{N}
\sum_{t=1}^{T}
\Bigl(
R_{i,t+h}
-
g\!\left(
f\!\left( \mathcal{I}^f_{i,t},\, \mathcal{I}^m_t \,;\, \phi \right)
\,;\, \theta
\right)
\Bigr)^{2},
\end{equation}
and the ``consensus loss" $\mathcal{L}_C$ is formulated as,
\begin{equation}
\label{eq:L_C}
\mathcal{L}_C(\phi)
=
\frac{1}{N T}
\sum_{i=1}^{N}
\sum_{t=1}^{T}
\Bigl(
C_{i,t}
-
f\!\left( \mathcal{I}^f_{i,t},\, \mathcal{I}^m_t \,;\, \phi \right)
\Bigr)^{2}.
\end{equation}
$\mathcal{L}_R$ and $\mathcal{L}_C$ are cross-sectional mean squared errors (MSE) from a standard pooled OLS estimator, where $\lambda$ is a hyperparameter that assigns weight to the consensus loss, providing additional flexibility into the empirical design of the model. 

We also estimate a benchmark model taking $\lambda=0$ from equation \eqref{eq:loss}, which ignores learning analyst opinions by removing the consensus loss term, making the model identical to the na\"ive return prediction model. Although $f(\cdot)$ and $g(\cdot)$ are the model defined with a separate set of learnable parameters $\theta$ and $\phi$, they can be considered as a single neural network when $\lambda=0$ since the optimization procedures for each networks are not independent.

The strategy of jointly learning the consensus and excess return offers several advantages. Firstly, it tends to yield higher performance metrics due to the synergistic learning of interconnected variables. An alternative method might involve independent optimization, where the $f(\cdot)$ and $g(\cdot)$ are trained independently in two separate steps. However, this segmented approach often fails to capture the potential inter-dependencies between the consensus estimates and the resulting excess returns. Furthermore, since the information set $\mathcal{I}^f_{i,t}$ and $\mathcal{I}^m_t$ are not included in equation \eqref{eq:L_R}, the training of $g(\cdot)$ entirely depends on the quality of the extracted signals in approximated consensus, which makes training with the loss function $\mathcal{L}_R$ extremely challenging. 

Secondly, it provides deeper insights and more intuitive understanding of the underlying financial dynamics. Independently learning $f(\cdot)$ using the equation \eqref{eq:L_C} is not a novel concept and aligns with the existing literature supporting the evidence of the rational expectations hypothesis. As discussed in previous sections, the set of predictor signals used in this study is regarded to contain a significant amount of information sufficient to make ``rational" expectations,\footnote{Appendix~\ref{Appendix:ablation-consensus} formally evaluates this claim by examining the consensus-only specification corresponding to $\lambda \to \infty$ in Equation~\eqref{eq:loss}, showing that the model learns analysts' consensus variables remarkably well (out-of-sample $R^2 = 30.30\%$) even without any return-prediction objective. This validates the architectural design of the consensus-bottleneck and provides empirical support for the rational expectations interpretation underlying the model.} which simplifies the problem of approximating the opinions of individuals, compared to predicting future returns of assets. However, since analysts perform their analysis as their job, they must think and act beyond being merely ``rational"; they must be ``professional". Therefore, we posit that professional and successful analysts strive to make their estimates predict not only ``macroeconomic" consequences but also ``firm-specific" outcomes. More specifically, proficient analysts will make decisions that better predict the future returns of a firm's stocks. 

\subsection{Neural network architecture}

In this section, we outline the key structure of the model architecture. The overall framework of CB-APM is described in Figure \ref{fig:model_architecture}. The model consists of two main components; the consensus module and the prediction module. Each of these modules corresponds to the function $f(\cdot)$ and $g(\cdot)$ in equation \eqref{eq:cond}. A complete description of the implementation and computational details of the CB-APM is provided in Internet Appendix~\ref{Appendix:nn}.

\begin{center}
[Insert Figure~\ref{fig:model_architecture} here]
\end{center}

In the proposed model, the consensus module is designed as an arbitrary feedforward network, while the prediction module is restricted to a simple linear regression that receives consensus variables as inputs and yields the expected excess return. This design choice is critical for enhancing interpretability in particular. When both modules are complex feedforward networks with multiple hidden layers, the advantage of using a consensus-based approach diminishes since it creates two separate black-box models from a single black-box model. 

The loss functions, as defined in equations \eqref{eq:L_R} and \eqref{eq:L_C}, are computed using the outputs from the respective modules. Once we get the return loss from the return module, the final loss function is calculated via weighted sum of these two loss functions as described in equation \eqref{eq:loss}. The backpropagation in the CB-APM is conducted in a single step, utilizing the composite loss function in equation \eqref{eq:loss}, which simultaneously adjusts the weights in both the consensus and prediction modules.

\section{The Economic Gains of Interpretable Prediction} \label{sec:empirical}

\subsection{Data} \label{sec:data}
In this section, we briefly describe the dataset used in the empirical analysis. The dataset comes from four distinct sources, which are all publicly available at the moment. Firstly, we obtain open-source asset pricing panel data from \textcite{ChenZimmermann2021}, available to download on their website (\url{https://www.openassetpricing.com/}).\footnote{Data from \textcite{ChenZimmermann2021} undergoes several preprocessing steps including lagging, data sampling, data imputation, and rank normalization, as detailed in Internet Appendix \ref{Appendix:preprocess}.} It comprises 114 firm-level predictors consisting of diverse financial metrics such as accounting figures, 13F filings, trading activities, and derivatives data. 

\textcite{ChenZimmermann2021} also provides nine analyst consensus variables including EPS forecast revision (\textit{AnalystRevision}), Change in recommendation (\textit{ChangeInRecommendation}), Change in Forecast and Accrual (\textit{ChForecastAccrual}), Long-vs-short EPS forecasts (\textit{EarningsForecastDisparity}), Analyst earnings per share (\textit{FEPS}), EPS Forecast Dispersion (\textit{ForecastDispersion}), Earnings forecast revisions (\textit{REV6}), Analyst Value (\textit{AnalystValue}), and Analyst Optimism (\textit{AOP}).

Secondly, stock prices and firm size data are obtained from CRSP for firms listed on the NYSE, Amex, and Nasdaq. This dataset is synchronized with the firm list from the panel data provided by \textcite{ChenZimmermann2021}. 

Lastly, the macroeconomic variables are obtained from FRED-MD database \parencite{mccracken2016fred} and \textcite{welch2008comprehensive}. FRED-MD consists of 115 monthly predictors that include macroeconomic indicators reflecting the U.S. labor markets, consumption rates, monetary policies, etc. An additional set of 8 macroeconomic variables is constructed from the database maintained by \textcite{welch2008comprehensive} on Goyal's website (\url{https://sites.google.com/view/agoyal145}), following \textcite{gu2020empirical}. T-bill rate is also obtained from this dataset, which is used for calculating risk premiums. To incorporate macroeconomic information, we encode the aggregate state using a nonlinear embedding constructed via an autoencoder. This approach allows us to summarize high-dimensional macroeconomic variables into a compact latent representation. \footnote{Detailed construction is provided in Internet Appendix \ref{Appendix:autoencoder}.}

The final merged dataset consists of samples spanning from January 1994 to December 2023, with 605{,}722 firm-month observations from 4{,}683 U.S. companies. Detailed descriptions of the dataset components and their respective sources are provided in Internet Appendix \ref{Appendix:data}.

\subsection{Interpretability-accuracy amplification effect} \label{sec:cross-section}

This section presents empirical results on the cross-sectional prediction of stock returns and consensus variables. We evaluate predictive performance under varying forecast horizons $h$ from equation \eqref{eq:final_model} to assess the effectiveness of the consensus-bottleneck in asset pricing. \footnote{Out-of-sample $R^2$ is used as the primary evaluation metric and is defined as:
\[
R^{2}_{\text{return}}
=
1
-
\frac{
\sum_{i=1}^{N}
\sum_{t=1}^{T}
\bigl(
R_{i,t+h}
-
\hat{R}_{i,t+h}
\bigr)^{2}
}{
\sum_{i=1}^{N}
\sum_{t=1}^{T}
R_{i,t+h}^{2}
},
\]
for return prediction and,
\[
R^{2}_{\text{consensus}}
=
1
-
\frac{
\sum_{i=1}^{N}
\sum_{t=1}^{T}
\bigl(
C_{i,t}
-
\hat{C}_{i,t}
\bigr)^{2}
}{
\sum_{i=1}^{N}
\sum_{t=1}^{T}
C_{i,t}^{2}
},
\]
for consensus approximation, where $N$ and $T$ denote the number of firms and time periods, respectively.}

We evaluate model performance using an expanding-window scheme in which the training sample grows sequentially over time while validation and test sets remain fixed. Full implementation details are provided in Internet Appendix~\ref{Appendix:expanding}. This experimental setup allows us to assess the robustness of CB-APM under evolving market conditions, while ensuring that all predictions are strictly out-of-sample and free from look-ahead bias.

While much of the asset pricing literature emphasizes short-horizon return forecasts, sell-side analysts typically issue multi-quarter to annual forecasts. Consensus measures therefore reflect longer-term expectations about fundamentals and risk premia rather than short-term price fluctuations. Evaluating the consensus-bottleneck over horizons that align with analysts' forecast horizons is more economically relevant than using noisy short-term intervals. Accordingly, we focus on annual return prediction, consistent with prior studies on long-horizon predictability \parencite{gu2020empirical, leippold2022machine}.\footnote{The results for other forecasting horizons are provided in Internet Appendix \ref{Appendix:horizon}}

\begin{center}
[Insert Table~\ref{tab:annual} here]
\end{center}

Table \ref{tab:annual} reports the monthly out-of-sample $R^2$ values (in percentage) for both annual stock return prediction ($R_{t+12}$) and the approximation of analysts' consensus variables ($C_t$) across different values of the regularization parameter $\lambda$. The benchmark case ($\lambda=0$), which excludes consensus learning, yields an annual return $R^2$ of 7.63\%, serving as a baseline for evaluating the incremental benefits of integrating consensus prediction into the CB-APM framework.

Introducing consensus learning via $\lambda>0$ leads to a pronounced improvement in return predictability. The out-of-sample $R^2$ for annual returns rises steadily, peaking at 10.46\% when $\lambda=0.3$, a 37\% increase relative to the benchmark. While larger $\lambda$ values beyond $0.3$ result in a gradual decline in $R^2$, it is noteworthy that even at $\lambda=1.0$, the return forecasting accuracy remains above the benchmark case (9.37\% versus 7.63\%), demonstrating that the integration of consensus information provides robust predictive gains across all tested settings.

The consensus approximation results provide further insight into this regularization effect. Among the nine consensus variables, \textit{Analyst Earnings per Share} dominates, achieving an $R^2$ of 71.43\% at $\lambda=1.0$, followed by strong performance in \textit{EPS Forecast Dispersion} and \textit{Analyst Optimism}. These results corroborate empirical findings that earnings estimates and their associated dispersion contain salient information about future returns \parencite{diether2002differences, jegadeesh2004analyzing}. By contrast, \textit{Change in Recommendation} exhibits persistently negative $R^2$, consistent with prior evidence of limited incremental predictive content in recommendation changes once earnings revisions are accounted for.

The consensus average $R^2$ increases monotonically from 7.33\% at $\lambda=0.1$ to 24.21\% at $\lambda=1.0$, indicating that the model becomes progressively better at reconstructing analyst consensus as $\lambda$ grows. However, the modest decline in return $R^2$ beyond $\lambda=0.3$ reflects the trade-off inherent in joint optimization; while higher $\lambda$ emphasizes consensus approximation, return forecasting benefits most when consensus serves as an auxiliary concept rather than the dominant objective.

\begin{figure}[htbp]
    \centering
    \includegraphics[width=\columnwidth]{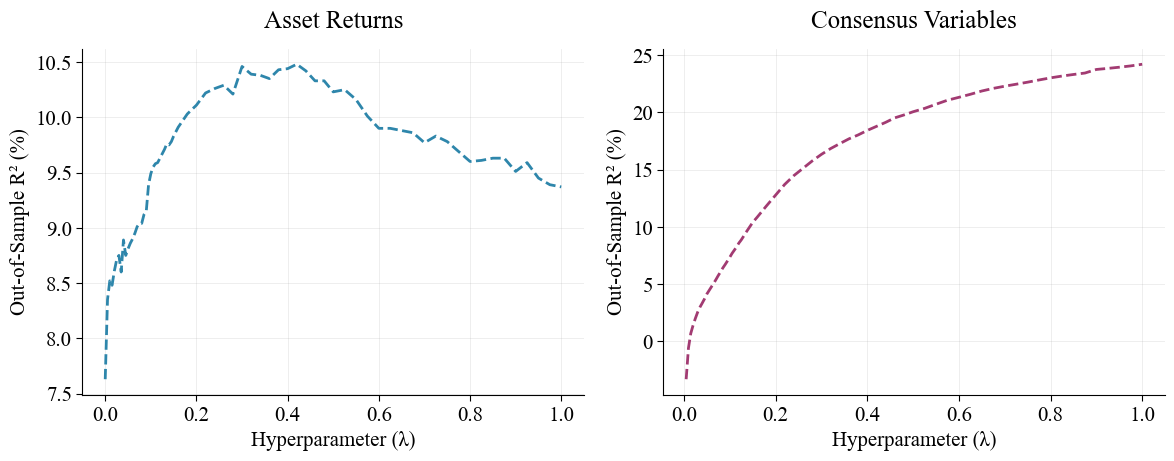}
    \caption{Out-of-sample $R^2$ of return predictions and consensus approximations. \\
    This figure presents monthly $R^2$ of annual stock return estimation (left) and average $R^2$ of analysts' consensus variable approximation (right) across the entire evaluation sets for different $\lambda$ settings. Return predictability improves sharply when consensus learning is introduced, peaking around $\lambda=0.3$-$0.4$, and remains above the benchmark even at $\lambda=1.0$. Consensus approximation accuracy increases monotonically with $\lambda$.}

    \label{fig:r2_12month}
\end{figure}

\begin{center}
[Insert Figure~\ref{fig:r2_12month} here]
\end{center}

Figure \ref{fig:r2_12month} complements Table \ref{tab:annual} by visualizing these trends. The left panel shows how return predictability improves sharply with the introduction of consensus learning, peaks around $\lambda=0.3$-$0.4$, and then tapers slightly while remaining above the benchmark even at $\lambda=1.0$. The right panel demonstrates the monotonic improvement in consensus approximation with increasing $\lambda$, eventually plateauing near 24\%. Together, these panels illustrate the trade-off, where moderate $\lambda$ balances return prediction and consensus learning most effectively, while larger $\lambda$ values shift focus toward consensus reconstruction.

Collectively, these results validate the core design of CB-APM that by incorporating consensus learning as a concept-bottleneck enhances return prediction while retaining interpretability. The model's ability to achieve robust gains across different market environments underscores both its practical relevance under realistic, expanding-window evaluation and its theoretical grounding in analyst-driven information aggregation.

While the out-of-sample $R^2$ metrics directly capture forecasting accuracy, they do not reveal how the joint loss function in equation \eqref{eq:loss} balances return prediction and consensus approximation during training. To address this, Internet Appendix~\ref{Appendix:optim} provides additional evidence on the optimization dynamics of CB-APM by reporting the in-sample MSE, which demonstrates that, at short horizons, increasing $\lambda$ introduces the expected trade-off between predictive accuracy and consensus reconstruction, whereas at longer horizons the two objectives reinforce each other, yielding what we term an ``interpretability-accuracy amplification effect''.

We further examine the role of macroeconomic conditioning in the CB-APM. An ablation analysis (Internet Appendix~\ref{Appendix:ablation-autoencoder}) shows that removing the macro embedding leads to a substantial deterioration in predictive performance, particularly under higher $\lambda$ values. This finding indicates that macroeconomic information is not merely auxiliary but a central component that stabilizes learning and improves out-of-sample generalization. Additional evidence (Internet Appendix~\ref{Appendix:autoencoder-representation}) confirms that the learned macroeconomic embeddings capture economically meaningful state dynamics, rather than statistical artifacts, thereby providing a structured conditioning variable for cross-sectional return prediction.

\subsection{Portfolio-based pricing validation}

We further examine the economic implications of the CB-APM through portfolio-level tests. While the preceding sections evaluated the model's predictive and explanatory power using out-of-sample $R^2$ metrics, these statistical measures alone do not reveal whether the predicted returns merely reflect transitory noise. Portfolio-based analyses provide a more direct and economically interpretable assessment of model performance by linking cross-sectional predictions to realized investment payoffs. In particular, if the CB-APM successfully extracts a priced component of expected returns from the consensus structure, portfolios formed on its predictions should yield monotonic and persistent return differentials across quantiles.

Our portfolio analysis proceeds in three steps. First, we perform single-sort tests that rank stocks by CB-APM-predicted annual returns to evaluate the model's raw cross-sectional discriminating power. These tests quantify whether higher model-implied expected returns translate into higher realized payoffs and whether the strength of this relationship varies with the degree of consensus regularization. Second, we conduct double-sort analyses that jointly sort stocks by both predicted returns and consensus variables to examine how the model's inferred expectations interact with, and potentially refine, traditional analyst forecasts. Finally, we form long-short portfolios based on out-of-sample CB-APM predictions to evaluate their risk-adjusted performance relative to benchmark strategies and to assess the model's practical value from an asset-management perspective.

These portfolio-level analyses allow us to connect the statistical accuracy of the CB-APM to its economic relevance. By translating predictive signals into realized return differentials, we can determine whether the consensus-bottleneck representation captures genuinely priced information, consistent with rational risk compensation, or reflects transitory deviations unrelated to systematic risk premia. The following subsections detail the construction of these portfolio tests and discuss their empirical results.

\subsubsection{Cross-sectional ordering and double-sort evidence}

For each month in the out-of-sample evaluation period, the CB-APM produces annual return forecasts for all stocks. Based on these out-of-sample predictions, stocks are ranked by their expected returns and assigned to ten value-weighted decile portfolios, ranging from the lowest (decile~1) to the highest (decile~10) predicted-return group. Portfolio constituents and weights are updated monthly as new forecasts become available, ensuring that portfolio formation relies exclusively on information observable at the prediction date. The realized monthly returns of each decile are then computed over the subsequent month, thereby evaluating the model's ex-ante forecasts in a strictly out-of-sample setting.

\begin{center}
[Insert Table~\ref{tab:single-sort} here]
\end{center}

The single-sort portfolio results in Table~\ref{tab:single-sort} reinforce the predictive validity of the CB-APM framework in the cross-section of returns. Average realized returns increase monotonically from the lowest to the highest predicted-return decile, with the bottom portfolios consistently yielding negative returns and the top portfolios earning approximately 1.3\% per month. The resulting high-minus-low (H--L) spreads range from 1.64\% for the naïve neural network ($\lambda=0$) to around 2.3\% for regularized CB-APM specifications ($\lambda \ge 0.3$). This progressive widening of the return differential highlights the model's ability to produce more economically meaningful and stable return rankings as the degree of consensus regularization increases. Beyond the level effects, the distribution of decile returns also becomes smoother and more monotonic as $\lambda$ rises, suggesting that the bottleneck constraint mitigates noise in the model-implied expected returns.

The patterns in portfolio payoffs align closely with the out-of-sample performance metrics reported in Table~\ref{tab:annual}. While the predictive $R^2$ for stock returns peaks around 10\% and remains relatively stable across higher $\lambda$ values, the $R^2$ for consensus variable approximation improves dramatically—from roughly 7\% at $\lambda=0.1$ to over 24\% at $\lambda=1.0$. This joint evidence implies that better recovery of analysts' consensus structure translates into more reliable expected-return forecasts. In other words, the improvement in cross-sectional pricing performance, as captured by the H--L spread, parallels the enhanced interpretability and generalization observed in the consensus approximation task. Together, the results indicate that the consensus-bottleneck regularization enables the model to balance flexibility and economic discipline, yielding forecasts that are both interpretable and empirically potent in explaining the cross-section of returns.

To further examine the pricing content embedded in CB-APM forecasts, we conduct a double-sorting exercise based on the model-implied expected returns and the analyst-based measure \textit{FEPS}. At each month in the out-of-sample period, all stocks are first assigned to quintiles using their CB-APM-approximated \textit{FEPS} levels. Within each \textit{FEPS} group, stocks are then independently sorted into quintiles by their CB-APM-predicted annual returns. This procedure yields 5×5 portfolios rebalanced monthly, ensuring that both the sorting signal and subsequent return evaluation rely strictly on information available at the prediction date. For each panel, the bottom and rightmost rows report high-minus-low (H--L) spreads along the predicted-return and consensus dimensions, measuring the incremental ordering power of CB-APM forecasts conditional on \textit{FEPS}.

\begin{center}
[Insert Table~\ref{tab:double-sort} here]
\end{center}

Table~\ref{tab:double-sort} shows that the CB-APM generates economically meaningful spreads across both sorting dimensions, highlighting an interaction between model-implied expected returns and analysts' earnings expectations. The \textit{FEPS} variable, the most recent I/B/E/S consensus forecast of next-fiscal-year earnings per share, is widely used as a standardized proxy for expected profitability. Prior evidence from \citet{Cen2006ForecastedEP} demonstrates that \textit{FEPS} predicts future returns even after controlling for common risk factors, with the premium concentrated among small and neglected firms and persisting without reversal. These patterns suggest that \textit{FEPS} embeds both valuable information about firm fundamentals and systematic expectation errors.

The double-sort design provides a natural setting to assess how the CB-APM processes this dual nature of analyst expectations. By construction, the model's consensus-bottleneck is designed to extract the priced component of forecasted earnings while mitigating noise arising from optimism-driven biases. This mechanism is consistent with recent evidence such as \citet{palley2025effect}, who document that consensus signals become unreliable when analyst dispersion is high, a condition strongly associated with stale or incentive-driven optimism. The state-dependent attenuation visible in Table~\ref{tab:double-sort}, where CB-APM's expected-return differentiation is largest in low-\textit{FEPS} states and diminishes as optimism rises, is precisely the pattern one would expect if behavioral components contaminate raw analyst forecasts while the model selectively filters them.

Across all regularization levels $\lambda$, mean realized returns increase monotonically from the lower-left (low \textit{FEPS}, low predicted return) to the upper-right (high \textit{FEPS}, high predicted return), confirming strong joint ordering power. Within each \textit{FEPS} quintile, the predicted-return portfolios exhibit clear monotonicity, with H--L spreads ranging from roughly 0.9\% to 2.5\% per month. These spreads peak at intermediate regularization strengths ($\lambda=0.3$--$0.6$), consistent with the interpretation that moderate consensus constraints balance flexibility with economic discipline, whereas very small $\lambda$ introduces noise and very large $\lambda$ ($>0.8$) leads to over-regularization.

More revealing is the cross-sectional pattern along the \textit{FEPS} dimension. The H--L spreads for \textit{FEPS} are positive among stocks with low model-predicted returns but turn negative among those with high predicted returns. This inversion indicates that firms with high analyst-forecasted earnings outperform in segments where the model sees limited return potential but underperform where the model projects high returns. Simultaneously, the magnitude of the expected-return H--L spread declines systematically from low to high \textit{FEPS} quintiles. Taken together, these findings imply that the CB-APM’s return signal is most potent precisely where analyst optimism is weakest, reinforcing the idea that the model distinguishes fundamental information from optimism-induced distortions.

These results extend the regularities documented by \citet{Cen2006ForecastedEP}. Although \textit{FEPS} generally predicts higher future returns, the largest expectation errors occur where forecasts are pessimistic, allowing the CB-APM to retain their predictive content while tempering the behavioral component. The observed reversals in the double-sort tables thus reflect not contradictions but adjustments: the CB-APM internalizes the asymmetric way markets react to forecasted earnings, preserving the informative component of \textit{FEPS} while reweighting it in states where optimism clouds the signal.

Overall, the evidence indicates that CB-APM forecasts complement rather than replicate the information in \textit{FEPS}. The consensus-bottleneck extracts the priced, risk-aligned component of analysts’ expectations while filtering optimism-related noise. The resulting reversal and attenuation patterns provide direct support for the interpretation that the CB-APM transforms raw forecasted earnings into a state-dependent pricing signal that refines, rather than contradicts, the analysts' consensus view.

\subsubsection{Risk-adjusted payoffs of long-short strategy}

We construct the long-short portfolio as follows. The first step involves generating monthly predicted annual returns for each stock within the universe from CB-APM. These predicted returns are then ranked from highest to lowest and sorted into deciles based on their values. Subsequently, a long portfolio is formed by purchasing the top 10\% of stocks with the highest predicted returns, while concurrently establishing a short portfolio by selling the bottom 10\% of stocks with the lowest predicted returns. Weighting of the stocks within each portfolio is executed based on the size of the firm, ensuring that larger firms are assigned higher weights. Then the long-short portfolio is rebalanced every month to uphold the desired exposure and maintain alignment with the initial strategy.

The long-short construction directly operationalizes the cross-sectional ordering evidence reported in Tables~\ref{tab:single-sort}. The monotonic increase in realized returns across predicted-return deciles translates naturally into economically significant long-short spreads.

To evaluate the risk-adjusted performance of the CB-APM portfolio, we compute seven portfolio metrics: monthly mean log return, standard deviation, cumulative log return, annualized Sharpe ratio, maximum one-month loss, maximum drawdown,\footnote{Maximum drawdown (Max~DD) is defined as the largest cumulative loss from a historical peak in portfolio wealth:
\[
\text{Max DD}
=\max_{t\in T}
\left(
1-
\frac{W_t}{
\max_{\tau\le t} W_\tau}
\right),
\qquad
W_t=\prod_{\tau=1}^{t}(1+R_\tau),
\]
where \(W_t\) denotes cumulative portfolio wealth at time \(t\).
This measure captures the worst peak-to-trough decline experienced over the sample period.} and turnover rate.\footnote{Portfolio turnover is calculated as,
\begin{equation}
\label{eq:turnover}
\text{Turnover}
=
\frac{1}{T_r}
\sum_{t \in T_r}
\left(
\sum_{i=1}^N
\left|
w_{i,t+1}
-
\frac{
w_{i,t}\left(1 + R_{i,t}\right)
}{
1 + \sum_{j=1}^N w_{j,t} R_{j,t}
}
\right|
\right),
\end{equation}
where $w_{i,t}$ denotes the portfolio weight of asset $i$ at time $t$,  
$R_{i,t}$ is its arithmetic monthly return,  
and $T_r \subset T$ denotes the set of rebalancing dates.} Monthly mean and cumulative returns quantify the overall profitability of the model, while the Sharpe ratio measures risk-adjusted performance by relating expected excess returns to return volatility. Maximum one-month loss and maximum drawdown capture downside risk by quantifying the worst historical losses, both in single periods and cumulatively. Finally, portfolio turnover measures the degree of portfolio rebalancing activity, which is directly linked to transaction costs and practical implementability.

Portfolio positions are formed using CB-APM's out-of-sample return forecasts, allowing the portfolio tests to evaluate genuine real-time predictability over a long-horizon target.

\begin{center}
[Insert Table~\ref{tab:portfolio-lambda} here]
\end{center}

The portfolio performance results in Table~\ref{tab:portfolio-lambda} mirror the statistical improvements in predictive and explanatory performance documented in Table~\ref{tab:annual}. As the hyperparameter~$\lambda$ increases to moderate values around~0.3--0.4, both out-of-sample return $R^2$ and consensus-approximation accuracy rise sharply, and this improvement translates directly into superior realized portfolio returns. Mean monthly log returns climb from~1.53\% at $\lambda=0$ to~2.20\% at $\lambda=0.3$, while the annualized Sharpe ratio concurrently increases from~1.10 to~1.44. This near one-to-one correspondence between predictive power and portfolio profitability substantiates the economic value of the consensus-bottleneck: the same mechanism that refines predictive signal extraction in-sample also enhances risk-adjusted returns out-of-sample.

Beyond moderate~$\lambda$ values, both predictive and portfolio metrics exhibit mild flattening, as excessive weighting on consensus reconstruction (\(\lambda>0.4\)) marginally reduces return~$R^2$ and diminishes economic gains. This pattern implies a practical upper bound to interpretability regularization, beyond which the model overemphasizes consensus consistency at the expense of direct return optimization. Nonetheless, even at high~$\lambda$ values, performance remains consistently above the benchmark, confirming that consensus learning contributes persistently to economically meaningful predictability rather than statistical overfitting.

Risk profiles exhibit a moderate but economically intuitive trade-off between profitability and downside exposure. As $\lambda$ increases to~0.3--0.4, maximum one-month losses rise slightly relative to the naïve network ($\lambda=0$), while remaining of similar magnitude at $\lambda=0.3$, which yields the highest Sharpe ratio. Maximum drawdowns, by contrast, are consistently lower than those of the S\&P~500 benchmark—staying below~21\% versus the market's~25\%—indicating that CB-APM's consensus-regularized predictions generate smoother long-term wealth trajectories. The modest increase in short-horizon losses is more than compensated by the substantial improvement in mean return and Sharpe ratio, implying enhanced efficiency on a risk-adjusted basis. Overall, the co-movement of predictive~$R^2$, Sharpe ratios, and drawdown behavior captures an economically meaningful balance between return amplification and risk containment, reflecting the emergence of stable, consensus-aligned risk premia rather than transient noise-fitting effects.

Portfolio turnover remains high, approximately~60\% per month, which is consistent with the characteristics of complex nonlinear architectures.\footnote{A formal transaction‐cost analysis based on the turnover definition in Equation~\eqref{eq:turnover} is provided in Internet Appendix~\ref{Appendix:TC}. The results show that the main economic conclusions are robust to proportional trading costs.} This observation aligns with the findings of \citet{gu2020empirical}, suggesting that neural-network-based return predictors typically produce higher turnover than linear or tree-based models due to their greater sensitivity to small shifts in cross-sectional signals. While \citet{kelly2024virtue} argue that out-of-sample predictive~$R^2$ and Sharpe ratios of characteristics-sorted portfolios may not always constitute decisive evidence of pricing relevance, the convergence of both statistical and economic measures in CB-APM suggests that its latent consensus components capture systematically priced information that conventional deep learning frameworks fail to isolate. Together, these results affirm that CB-APM's consensus-bottleneck not only improves explanatory power but also yields tangible, risk-adjusted portfolio benefits, linking interpretability and profitability within a unified empirical asset pricing framework.

\begin{figure}[htbp]
    \centering
    \includegraphics[width=\columnwidth]{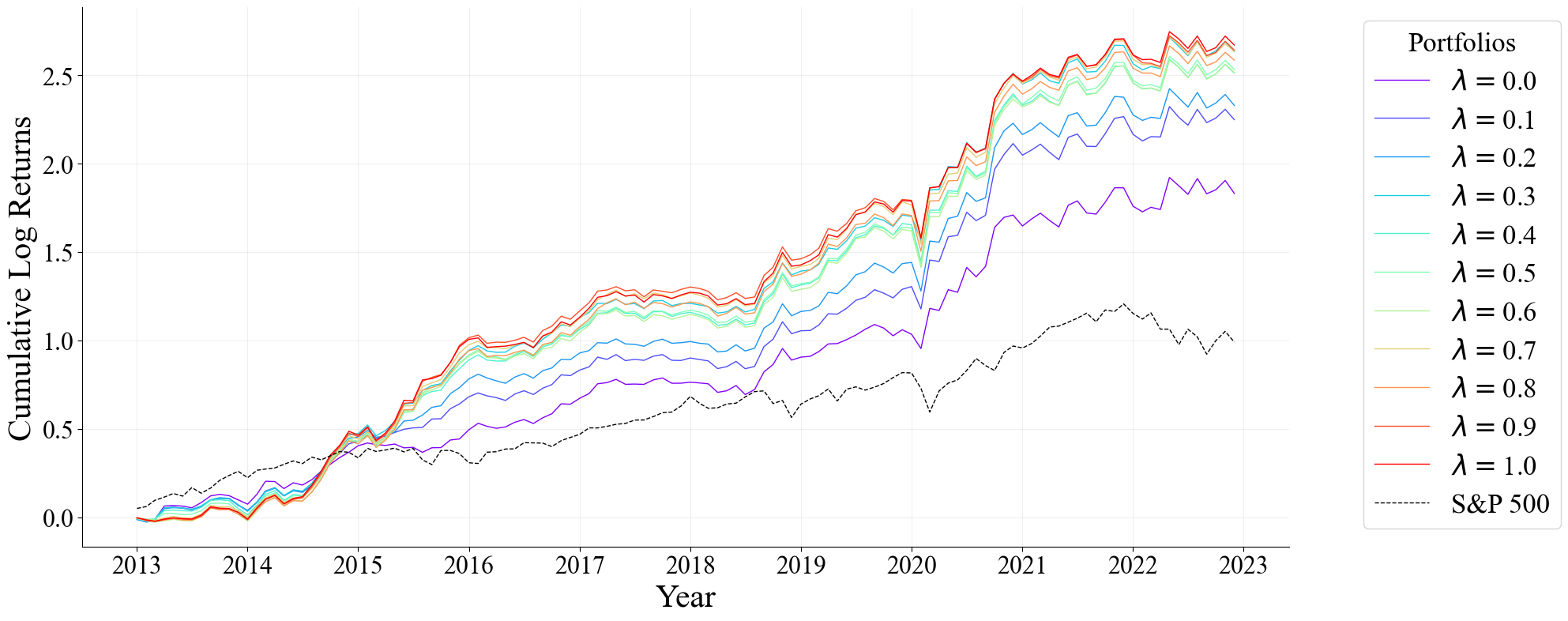}
    \caption{Out-of-sample cumulative returns of long-short decile portfolios. \\
    The figure plots cumulative log returns of value-weighted long-short decile portfolios formed from annual return forecasts, rebalanced monthly using out-of-sample predictions. Each line corresponds to a different hyperparameter $\lambda$, with the S\&P 500 index buy-and-hold strategy (dashed) as a benchmark. The na\"ive neural network ($\lambda=0$) outperforms the S\&P 500 benchmark, while CB-APM models with $\lambda>0$ deliver substantially higher performance than the na\"ive specification, underscoring the added value of consensus learning.}

    \label{fig:portfolio-12month}
\end{figure}

\begin{center}
[Insert Figure~\ref{fig:portfolio-12month} here]
\end{center}

Figure~\ref{fig:portfolio-12month} visualizes the cumulative out-of-sample performance of CB-APM long-short portfolios across different regularization strengths~$\lambda$. All neural-network portfolios substantially outperform the S\&P~500 buy-and-hold benchmark (black dashed line), demonstrating that the model’s predictive signals translate into economically meaningful excess returns. The na\"ive network ($\lambda=0$, purple line) already yields notable outperformance relative to the market, yet introducing the consensus-bottleneck regularization (\(\lambda>0\)) substantially elevates cumulative returns. Portfolio performance improves sharply up to~$\lambda\approx0.3$, after which cumulative returns remain at a comparably high level with minor oscillations across subsequent~$\lambda$ values. The best-performing specification at~$\lambda=1.0$ represents a continuation of this high-return plateau rather than a strict monotonic gain, highlighting the robustness of CB-APM's economic performance across a wide range of regularization intensities. This stability suggests that consensus regularization consistently enhances the model's predictive and economic relevance without overfitting to a narrow hyperparameter regime.

The figure further highlights the temporal robustness of CB-APM's performance. Even during adverse market conditions, notably the 2020 downturn, consensus-regularized portfolios experience smaller and more rapidly recovered drawdowns relative to both the market and the unregularized model, reflecting smoother wealth accumulation and improved resilience to macro shocks. The consistent separation between the consensus-based portfolios and the S\&P~500 benchmark indicates that the learned consensus representations capture priced information that is both persistent and broadly exploitable.

\subsection{Factor spanning and structural pricing diagnostics} \label{sec:pricing-error}

The preceding portfolio-based analyses demonstrate that the CB-APM's predicted signals translate into systematic cross-sectional return differentials. We now examine whether these signals possess asset pricing content when evaluated through the lens of linear factor models.\footnote{As a complementary diagnostic, we assess whether the CB-APM--implied consensus captures economically meaningful variation in expected returns through pooled panel regressions. The results, reported in Internet Appendix~\ref{Appendix:linear_regression}, indicate that the filtered consensus provides incremental explanatory content beyond raw analyst signals.} Specifically, we assess whether traditional factor models\footnote{ As reference models, we estimate the CAPM, the Fama--French three-factor model (FF3), the Carhart four-factor model, the Fama--French five-factor model (FF5), and the Fama--French six-factor model (FF6). All models are estimated using monthly excess returns, and all results are reported in-sample to maintain comparability with the standard evaluation framework in the factor-pricing literature.} can price the return patterns implied by the CB-APM's predictions and consensus-based characteristics. To do so, we employ the multivariate GRS test \parencite{gibbons1989test}, which jointly evaluates whether the intercepts ($\boldsymbol{\alpha}$) in time-series regressions are statistically different from zero.

A central element of the empirical design is the construction of tradable portfolios that proxy for the CB-APM signals. Specifically, we form value-weighted long--short portfolios by sorting firms into deciles based on predicted returns or consensus measures and taking the return spread between the highest and lowest deciles. These portfolios provide a tractable representation of the model's signals within a traditional factor-pricing framework and enable direct evaluation using GRS tests.

Importantly, the CB-APM portfolios used in these tests are constructed from the same model configuration employed in the empirical return-forecasting exercise. That is, we apply the trained CB-APM, optimized to forecast annual excess returns, to generate firm-level predicted returns and consensus representations, which are then used to form sorted portfolios and their corresponding long--short returns. This design ensures coherence across empirical sections: the factor-pricing analysis evaluates the economic content of the very signals that the CB-APM learns to use for long-horizon prediction.

We conduct two complementary GRS exercises. First, we form decile portfolios based on CB-APM predicted returns and test whether standard factor models can explain their realized returns. This analysis assesses whether the return patterns generated by the model are incremental to the span of existing factors. Second, we construct decile portfolios sorted on each individual consensus dimension and examine whether traditional models can price these portfolios. This exercise isolates which consensus channels are most and least aligned with traditional factor structures.

Each specification is evaluated using the GRS $F$-statistic, its associated $p$-value, and mean absolute and root-mean-squared pricing errors. All results are computed in-sample, consistent with empirical asset pricing conventions in which factor-pricing tests focus on explaining cross-sectional return patterns rather than forecasting performance. Together, these exercises provide a focused assessment of whether the consensus representations learned by the CB-APM contain distinct factor-pricing information or whether their explanatory power is largely captured by established benchmark models. As an additional robustness check, we examine the pricing performance of CB-APM-implied portfolios on standard benchmark test assets; results are reported in Internet Appendix~\ref{Appendix:benchmark-portfolio}.

Tables~\ref{tab:grs-predicted}--\ref{tab:grs-consensus} present in-sample GRS tests evaluating the pricing performance of the CB-APM relative to conventional factor models. Across all tests, the GRS $F$-statistic assesses the joint null hypothesis that all pricing errors ($\boldsymbol{\alpha}$) are zero, such that lower $F$-statistics and higher $p$-values indicate superior mean--variance efficiency. The accompanying mean absolute and root-mean-squared alphas summarize the magnitude of mispricing across the corresponding test assets.

\begin{center}
[Insert Table~\ref{tab:grs-predicted} here]
\end{center}

Table~\ref{tab:grs-predicted} examines whether traditional factor models can jointly price decile portfolios formed on the CB-APM's predicted return scores. When the consensus-bottleneck is weak (small $\lambda$), conventional factor models achieve moderate GRS statistics and economically small pricing errors, suggesting that a substantial portion of the model’s predictive content overlaps with standard style factors. As $\lambda$ increases, however, the GRS statistics rise sharply and the joint null of zero pricing errors is rejected uniformly. This monotonic deterioration indicates that stronger reliance on the consensus-bottleneck induces expected-return patterns that increasingly depart from the linear span of market, size, value, momentum, and profitability/investment factors. Conceptually, this aligns with evidence that machine-learning models often extract nonlinear or interaction-based transformations of firm characteristics that extend beyond linear factor structures \parencite{freyberger2020dissecting, gu2020empirical}. In particular, the CB-APM with a tight consensus constraint appears to generate forecasts that incorporate structured forms of return heterogeneity that are difficult to reconcile with the standard factor space.

\begin{center}
[Insert Table~\ref{tab:grs-consensus} here]
\end{center}

Complementing the predicted return results in Table~\ref{tab:grs-predicted}, Table~\ref{tab:grs-consensus} evaluates portfolios formed on the individual consensus signals at $\lambda = 1.0$. Several dimensions—most prominently \textit{Analyst Value}, \textit{Analyst Optimism}, and \textit{Analyst Earnings per Share}—produce relatively low GRS statistics and economically small pricing errors, suggesting strong alignment between these inferred consensus measures and established factor structures. Forecast-based and dispersion-based dimensions (such as \textit{EPS forecast dispersion} and related revisions) exhibit somewhat larger pricing errors, but even here the magnitudes remain concentrated in the range of a few basis points per month. These patterns reinforce the idea that much of the predictive information contained in analyst-derived consensus measures can be represented through low-dimensional combinations of characteristics, often with sparse or localized influence \parencite{chinco2019sparse}, while still accommodating nonlinear interactions and heterogeneous partitions \parencite{bryzgalova2025forest}. The CB-APM's consensus variables therefore fit naturally within the broader empirical finding that return-relevant structure can be extracted by compressing high-dimensional characteristics into well-organized representations.

The set of GRS tests reveals how the CB-APM relates to the traditional factor space. First, the ability of traditional factor models to price CB-APM–generated portfolios deteriorates as the consensus-bottleneck becomes more stringent, implying that the model's predictive signals progressively move outside the span of standard linear characteristics. This behavior is consistent with the broader view that, while the priced dimension of the SDF is relatively low, flexible methods can uncover structured forms of return heterogeneity that improve portfolio efficiency \parencite{cong2025growing} without reproducing the canonical factors directly. Second, portfolios sorted on individual consensus dimensions exhibit moderate but non-negligible pricing errors, suggesting that the learned signals contain meaningful information about expected returns but do not themselves constitute a new standalone factor system. 

This finding crucially aligns with the emerging methodological consensus that traditional characteristic–based sorting procedures fundamentally fail to capture the full mean–variance efficient (MVE) frontier due to their neglect of nonlinearity and asymmetric characteristic interactions. Recent goal-oriented machine learning approaches—most notably the Panel Tree (P-Tree) framework \parencite{cong2025growing} and the Asset Pricing Tree (AP-Tree) framework \parencite{bryzgalova2025forest}—demonstrate that test assets constructed by explicitly optimizing for SDF spanning or mean--variance efficiency are substantially harder to price with conventional factor models, often yielding extremely high GRS statistics. In this sense, the behavior observed in Table~\ref{tab:grs-predicted} echoes the insight that once test assets begin to reflect structured, state-dependent return heterogeneity, linear factor structures fail sharply.

The CB-APM achieves a conceptually parallel outcome, but through an economically structured consensus-bottleneck rather than recursive partitioning rules. By restricting predictive content to pass through interpretable consensus dimensions, the model induces return patterns that resemble the “goal-oriented” test assets emphasized in the tree-based literature—namely, portfolios that expose deficiencies in the linear factor span precisely because they encode higher-order interactions and conditional pricing structure. This makes the resulting portfolios harder to price not as a flaw, but as evidence that the model recovers meaningful variation in expected returns that traditional factor models systematically miss.

Overall, the evidence positions the CB-APM as a complementary asset pricing framework: it enhances cross-sectional return prediction by compressing analysts' heterogeneous beliefs into interpretable consensus signals that partially overlap with—but do not collapse onto—the priced dimensions emphasized in modern work on characteristic-based factor representations \citep[e.g.,][]{cochrane2011presidential}. At the same time, the results indicate that the CB-APM does not merely denoise or reweight analyst inputs. Instead, it isolates structured and economically relevant components of analyst-derived information that are priced in the cross-section. The model therefore reveals that analyst-based information contains priced elements that conventional factor models only partially span, and that the consensus-bottleneck organizes these elements into an interpretable and economically coherent representation. This places the CB-APM within a growing line of research showing that belief-based or characteristic-based signals can be synthesized into low-dimensional, economically meaningful components without reproducing the canonical factor structure directly.

\section{Conclusion} \label{sec:conclusion}

This study introduces the CB-APM, a novel framework that integrates interpretable deep learning with empirical asset pricing. By embedding a concept-bottleneck architecture into a neural network, CB-APM not only achieves competitive predictive accuracy in cross-sectional stock return forecasts but also provides transparent insights into the role of analysts' consensus in shaping risk premiums. Our empirical results demonstrate that interpretability and performance are not inherently conflicting. CB-APM outperforms conventional deep learning benchmarks in long-horizon forecasts while preserving a clear, economically grounded structure. By linking machine learning's predictive capabilities with the theoretical underpinnings of financial economics, and by demonstrating that interpretable deep learning can yield both statistical and economic validity, this work offers a blueprint for building models that are both high-performing and aligned with established asset pricing principles.

The success of CB-APM highlights three key implications for empirical finance. First, interpretable neural architectures can reconcile the flexibility of machine learning with economic reasoning, enabling researchers to assess whether models capture meaningful risk factors rather than spurious correlations. Second, embedding interpretability directly within model design fosters transparency and trust, addressing the skepticism that often surrounds ``black-box" methods in high-stakes financial applications. Third, by explicitly modeling analysts' consensus as a latent mediator between firm characteristics and returns, CB-APM sheds new light on how information aggregation mechanisms influence asset prices, aligning closely with rational expectations theory and empirical evidence on analyst behavior.

Future research can extend this framework in several promising directions. Incorporating additional economically meaningful bottlenecks, such as investor sentiment or narrative-driven pricing components \parencite{bybee2023narrative}, could further disentangle the sources of risk premiums and strengthen the theoretical interpretability of model outputs. Addressing practical constraints, such as data latency in analyst consensus measures or improving computational efficiency for large-scale implementation, would enhance CB-APM's applicability in real-world investment contexts. More broadly, as the ``factor zoo" continues to grow, interpretable frameworks like CB-APM will be instrumental in bridging data-driven discovery with economic theory, offering a structured approach to understanding how high-dimensional predictors translate into priced information. By demonstrating that interpretable AI can achieve both predictive accuracy and theoretical coherence, this study lays the groundwork for a new generation of financially grounded machine learning models, advancing the study of asset pricing in both academic research and practical decision-making.

{
\singlespacing
\printbibliography[title={References}]
}

\clearpage

\begin{sidewaystable}[p]
        \centering
        \small
        \renewcommand{\arraystretch}{1.3}
        \setlength{\tabcolsep}{6pt}
        \newcolumntype{Y}{>{\centering\arraybackslash}X}
        
        \caption{Out-of-sample $R^2$ for stock return and consensus approximations. \\
        This table reports monthly $R^2$ (\%) of annual stock return estimation and analysts' consensus variable approximation for different $\lambda$ settings. The out-of-sample $R^2$ is computed by concatenating realized values across all testing periods and comparing them with model predictions, thereby capturing performance over the full evaluation horizon. The Overall Results panel reports (i) the average $R^2$ across all consensus variables and (ii) the $R^2$ for annual stock return predictions based on the specified $\lambda$ regularization parameter.}
        \label{tab:annual}

        \begin{tabular}{
            S[table-format=1.3] 
            S[table-format=-2.2] 
            S[table-format=-2.2] 
            S[table-format=-2.2] 
            S[table-format=-2.2] 
            S[table-format=2.2]  
            S[table-format=2.2]  
            S[table-format=2.2]  
            S[table-format=2.2]  
            S[table-format=2.2]  
            S[table-format=2.2]  
            S[table-format=1.2]  
        }
            \toprule
            & \multicolumn{9}{c}{\small Consensus Variables} 
            & \multicolumn{2}{c}{\small Overall Results} \\
            \cmidrule(lr){2-10} \cmidrule(lr){11-12}
            {$\lambda$} & {\small \makecell{EPS \\ Forecast \\ Revision}} & {\small \makecell{Change \\ in \\ Rec-\\ommend-\\ation}} & {\small \makecell{Change \\in \\ Forecast \\ \& \\ Accrual}} & {\small \makecell{Long \\ vs \\ short \\ EPS \\ Forecasts}} & {\small \makecell{Analyst \\ Earnings \\ per \\ Share}} & {\small \makecell{EPS \\ Forecast \\ Dispersion}} & {\small \makecell{ Earnings \\ Forecast \\ Revisions}} & {\small \makecell{Analyst \\ Value}} & {\small \makecell{Analyst \\ Optimisim}} & {\small \makecell{Consensus \\ Average}} & {\small \makecell{Stock \\ Returns}} \\
            \midrule
            0 & {-} & {-} & {-} & {-} & {-} & {-} & {-} & {-} & {-} & {-} & 7.63 \\ 
            0.1 & 1.40 & -0.24 & 1.52 & 3.04 & 22.60 & 11.70 & 5.14 & 9.72 & 11.11 & 7.33 & 9.49 \\
            0.2 & 2.45 & -0.24 & 2.34 & 4.71 & 40.13 & 21.19 & 7.65 & 17.06 & 19.63 & 12.77 & 10.11 \\
            0.3 & 3.21 & -0.25 & 2.92 & 5.91 & 50.96 & 27.25 & 9.81 & 22.22 & 24.97 & 16.33 & 10.46 \\
            0.4 & 3.67 & -0.24 & 3.35 & 6.72 & 56.73 & 30.52 & 11.45 & 25.42 & 28.21 & 18.43 & 10.44 \\
            0.5 & 4.01 & -0.23 & 3.67 & 7.36 & 61.11 & 33.03 & 12.69 & 28.16 & 30.73 & 20.06 & 10.23 \\
            0.6   & 4.3 & -0.19 & 3.9 & 7.92 & 64.31 & 34.97 & 13.71 & 30.27 & 32.7 & 21.32 & 9.9 \\
            0.7   & 4.53 & -0.18 & 4.14 & 8.33 & 66.82 & 36.4 & 14.48 & 31.99         & 34.13 & 22.29  & 9.77 \\
            0.8   & 4.71 & -0.16 & 4.31 & 8.66 & 68.55 & 37.4 & 15.08 & 33.34         & 35.32 & 23.02  & 9.6     \\
            0.9   & 4.85 & -0.17 & 4.5 & 8.89 & 70.39 & 38.44 & 15.73 & 34.65         & 36.53 & 23.76 & 9.51    \\
            1.0   & 4.97 & -0.16 & 4.62 & 9.12 & 71.43 & 39.06 & 16.18 & 35.45         & 37.24 & 24.21  & 9.37 \\

            \bottomrule
        \end{tabular}
    \begin{flushleft}
    {\textit{Note:} Results are reported for a sampled subset of $\lambda$ settings due to redundancy.}
    \end{flushleft}
\end{sidewaystable}

\begin{table}[htbp]
    \centering
    \small
    \renewcommand{\arraystretch}{1.1}
    \setlength{\tabcolsep}{6pt}

    \caption{Realized monthly returns of out-of-sample single-sorted portfolios across $\lambda$. \\
    Each panel reports mean monthly realized returns (in percentage points) for monthly rebalanced decile portfolios, formed by sorting stocks on CB-APM-predicted annual returns. The bottom row (H--L) represents the spread between the highest- and lowest-decile portfolios.}
    \label{tab:single-sort}

    \vspace{0.2cm}
    
\begin{tabular*}{\textwidth}{@{\extracolsep{\fill}} l S[table-format=-1.4] S[table-format=-1.4] S[table-format=-1.4] S[table-format=-1.4] S[table-format=-1.4] S[table-format=-1.4] }
\toprule[0.6pt]
{\small $\lambda$} & \small 0.0 & \small 0.1 & \small 0.2 & \small 0.3 & \small 0.4 & \small 0.5 \\
\midrule
Low & -0.36 & -0.70 & -0.78 & -0.96 & -0.88 & -0.92 \\
2 & -0.27 & -0.24 & -0.26 & -0.24 & -0.25 & -0.23 \\
3 & 0.06 & -0.03 & -0.11 & -0.03 & -0.16 & -0.16 \\
4 & 0.14 & 0.13 & 0.22 & 0.20 & 0.30 & 0.31 \\
5 & 0.20 & 0.39 & 0.41 & 0.33 & 0.35 & 0.36 \\
6 & 0.30 & 0.36 & 0.51 & 0.52 & 0.52 & 0.36 \\
7 & 0.47 & 0.52 & 0.48 & 0.55 & 0.41 & 0.49 \\
8 & 0.64 & 0.72 & 0.61 & 0.62 & 0.79 & 0.79 \\
9 & 0.77 & 0.78 & 0.88 & 0.91 & 0.85 & 0.93 \\
High & 1.28 & 1.31 & 1.27 & 1.34 & 1.32 & 1.30 \\
\midrule[0.1pt]
H--L & 1.64 & 2.00 & 2.06 & 2.30 & 2.20 & 2.21 \\
\bottomrule[0.6pt]
\end{tabular*}

    \vspace{0.2cm}
    
\begin{tabular*}{\textwidth}{@{\extracolsep{\fill}} l S[table-format=-1.4] S[table-format=-1.4] S[table-format=-1.4] S[table-format=-1.4] S[table-format=-1.4] }
\toprule[0.6pt]
{\small $\lambda$} & \small 0.6 & \small 0.7 & \small 0.8 & \small 0.9 & \small 1.0 \\
\midrule
Low & -0.91 & -0.93 & -0.92 & -0.96 & -0.94 \\
2 & -0.29 & -0.29 & -0.33 & -0.27 & -0.31 \\
3 & -0.03 & 0.05 & 0.03 & -0.06 & -0.07 \\
4 & 0.29 & 0.22 & 0.30 & 0.25 & 0.27 \\
5 & 0.41 & 0.41 & 0.35 & 0.39 & 0.40 \\
6 & 0.32 & 0.27 & 0.29 & 0.37 & 0.34 \\
7 & 0.50 & 0.49 & 0.44 & 0.44 & 0.51 \\
8 & 0.72 & 0.85 & 0.85 & 0.84 & 0.79 \\
9 & 0.93 & 0.79 & 0.87 & 0.89 & 0.85 \\
High & 1.29 & 1.38 & 1.35 & 1.35 & 1.40 \\
\midrule[0.1pt]
H--L & 2.20 & 2.31 & 2.27 & 2.31 & 2.34 \\
\bottomrule[0.6pt]
\end{tabular*}

\end{table}

\clearpage

\begin{table}[!htbp]
\centering
\scriptsize
\setlength{\tabcolsep}{4.5pt}
\renewcommand{\arraystretch}{1.0}
\caption{Realized monthly returns of out-of-sample double-sorted portfolios across $\lambda$.\\
Each panel reports mean monthly realized returns (in percentage points) for monthly rebalanced 5$\times$5 portfolios 
sorted by the approximated \textit{Analyst earning per share} ($\mathbb{E}[\textit{FEPS}]$, rows) and predicted annual returns ($\mathbb{E}[R]$, columns), independently.
H--L denotes the high-minus-low spread across the corresponding dimension.}
\label{tab:double-sort}

\newcommand{\DSPanel}[2]{%
\begin{minipage}[t]{0.48\textwidth}
\centering
\textbf{Panel:} $\bm{\lambda=#1}$\par\vspace{1pt}
\begin{tabular}{@{}l@{\hspace{8pt}}*{6}{S[table-format=-1.2]}@{}}
\toprule
& \multicolumn{5}{c}{$\mathbb{E}_t[R_{i,t+h}]$} & \\
\cmidrule(lr){2-6}
{$\mathbb{E}_t[\textit{FEPS}_{i,t}]$} & {Low} & {2} & {3} & {4} & {High} & {H--L} \\
\midrule
#2
\bottomrule
\end{tabular}
\end{minipage}}

\DSPanel{0.1}{%
Low  & -0.65 & -0.43 &  0.04 &  0.68 &  1.47 &  2.12 \\
2    & -0.31 &  0.03 &  0.38 &  0.30 &  0.63 &  0.94 \\
3    & -0.27 &  0.39 &  0.35 &  0.40 &  0.69 &  0.96 \\
4    & -0.11 &  0.17 &  0.45 &  0.33 &  0.86 &  0.98 \\
High &  0.28 &  0.45 &  0.46 &  0.70 &  0.81 &  0.53 \\
\midrule
H--L &  0.93 &  0.88 &  0.41 &  0.02 & -0.66 & -1.59 \\
}\hfill
\DSPanel{0.2}{%
Low  & -0.73 & -0.41 & -0.04 &  0.67 &  1.54 &  2.27 \\
2    & -0.34 &  0.01 &  0.38 &  0.54 &  0.62 &  0.95 \\
3    & -0.27 &  0.27 &  0.47 &  0.45 &  0.72 &  0.99 \\
4    & -0.14 &  0.27 &  0.24 &  0.45 &  0.85 &  0.99 \\
High &  0.19 &  0.44 &  0.56 &  0.62 &  0.73 &  0.54 \\
\midrule
H--L &  0.92 &  0.85 &  0.60 & -0.05 & -0.81 & -1.73 \\
}

\vspace{4pt}

\DSPanel{0.3}{%
Low  & -0.90 & -0.24 & -0.04 &  0.61 &  1.66 &  2.55 \\
2    & -0.41 &  0.05 &  0.30 &  0.64 &  0.62 &  1.03 \\
3    & -0.40 &  0.25 &  0.35 &  0.50 &  0.78 &  1.18 \\
4    & -0.19 &  0.18 &  0.32 &  0.68 &  0.83 &  1.03 \\
High &  0.11 &  0.39 &  0.60 &  0.65 &  0.75 &  0.64 \\
\midrule
H--L &  1.01 &  0.63 &  0.64 &  0.04 & -0.90 & -1.91 \\
}\hfill
\DSPanel{0.4}{%
Low  & -0.83 & -0.40 & -0.05 &  0.67 &  1.61 &  2.44 \\
2    & -0.40 &  0.03 &  0.30 &  0.57 &  0.71 &  1.11 \\
3    & -0.31 &  0.30 &  0.35 &  0.44 &  0.81 &  1.11 \\
4    & -0.19 &  0.18 &  0.38 &  0.57 &  0.89 &  1.08 \\
High &  0.14 &  0.36 &  0.51 &  0.72 &  0.73 &  0.59 \\
\midrule
H--L &  0.97 &  0.76 &  0.57 &  0.05 & -0.88 & -1.86 \\
}

\vspace{4pt}

\DSPanel{0.5}{%
Low  & -0.77 & -0.29 & -0.01 &  0.34 &  1.68 &  2.45 \\
2    & -0.59 &  0.03 &  0.37 &  0.58 &  0.80 &  1.39 \\
3    & -0.26 &  0.24 &  0.42 &  0.52 &  0.86 &  1.12 \\
4    & -0.17 &  0.17 &  0.40 &  0.62 &  0.83 &  1.00 \\
High &  0.02 &  0.34 &  0.49 &  0.68 &  0.77 &  0.75 \\
\midrule
H--L &  0.79 &  0.63 &  0.49 &  0.34 & -0.91 & -1.70 \\
}\hfill
\DSPanel{0.6}{%
Low  & -0.75 & -0.31 & -0.03 &  0.54 &  1.56 &  2.31 \\
2    & -0.55 &  0.02 &  0.38 &  0.55 &  0.89 &  1.44 \\
3    & -0.29 &  0.36 &  0.30 &  0.51 &  0.90 &  1.19 \\
4    & -0.25 &  0.20 &  0.39 &  0.60 &  0.81 &  1.06 \\
High &  0.04 &  0.42 &  0.48 &  0.62 &  0.73 &  0.69 \\
\midrule
H--L &  0.79 &  0.73 &  0.51 &  0.08 & -0.83 & -1.62 \\
}

\vspace{4pt}

\DSPanel{0.7}{%
Low  & -0.88 & -0.27 &  0.12 &  0.58 &  1.51 &  2.39 \\
2    & -0.56 &  0.12 &  0.31 &  0.65 &  0.79 &  1.35 \\
3    & -0.23 &  0.26 &  0.27 &  0.50 &  0.94 &  1.17 \\
4    & -0.27 &  0.27 &  0.40 &  0.53 &  0.80 &  1.07 \\
High & -0.04 &  0.43 &  0.46 &  0.70 &  0.72 &  0.75 \\
\midrule
H--L &  0.84 &  0.69 &  0.33 &  0.11 & -0.79 & -1.63 \\
}\hfill
\DSPanel{0.8}{%
Low  & -0.81 & -0.27 &  0.09 &  0.56 &  1.57 &  2.38 \\
2    & -0.53 &  0.06 &  0.29 &  0.69 &  0.85 &  1.38 \\
3    & -0.23 &  0.20 &  0.33 &  0.51 &  0.92 &  1.15 \\
4    & -0.26 &  0.22 &  0.42 &  0.50 &  0.80 &  1.06 \\
High & -0.05 &  0.45 &  0.46 &  0.61 &  0.75 &  0.80 \\
\midrule
H--L &  0.76 &  0.72 &  0.37 &  0.04 & -0.81 & -1.58 \\
}

\vspace{4pt}

\DSPanel{0.9}{%
Low  & -0.93 & -0.08 & -0.10 &  0.64 &  1.62 &  2.55 \\
2    & -0.62 &  0.16 &  0.32 &  0.55 &  0.92 &  1.54 \\
3    & -0.26 &  0.27 &  0.32 &  0.54 &  0.93 &  1.19 \\
4    & -0.30 &  0.16 &  0.49 &  0.57 &  0.76 &  1.05 \\
High & -0.10 &  0.39 &  0.51 &  0.60 &  0.73 &  0.83 \\
\midrule
H--L &  0.82 &  0.47 &  0.61 & -0.05 & -0.90 & -1.72 \\
}\hfill
\DSPanel{1.0}{%
Low  & -0.94 & -0.16 & -0.01 &  0.66 &  1.63 &  2.57 \\
2    & -0.64 &  0.23 &  0.31 &  0.57 &  0.90 &  1.53 \\
3    & -0.21 &  0.24 &  0.40 &  0.53 &  0.91 &  1.11 \\
4    & -0.42 &  0.21 &  0.41 &  0.67 &  0.71 &  1.13 \\
High & -0.06 &  0.41 &  0.47 &  0.57 &  0.73 &  0.79 \\
\midrule
H--L &  0.89 &  0.57 &  0.47 & -0.09 & -0.90 & -1.79 \\
}

\end{table}

\clearpage

\begin{table}[htbp]
\centering
\small
\renewcommand{\arraystretch}{1.3}
\setlength{\tabcolsep}{6pt}

\caption{Out-of-sample portfolio performance of CB-APM long-short portfolios.\\
This table reports performance metrics for value-weighted CB-APM long-short portfolios under different hyperparameter $\lambda$. 
Mean ($\bar{r}$) and standard deviation ($\sigma(r)$) are computed from monthly log returns, and cumulative log return ($\sum_t r_t$) is aggregated over the full sample period. 
The Sharpe ratio ($\bar{R}/\sigma(R)$) is annualized using the standard $\sqrt{12}$ scaling, assuming a zero risk-free rate. 
Maximum one-month loss ($-\min(R)$) and maximum drawdown (Max~DD) are expressed in percentage terms, while Turnover denotes the average monthly portfolio turnover. 
The S\&P~500 index serves as a benchmark.}
\label{tab:portfolio-lambda}

\begin{tabular*}{\textwidth}{@{\extracolsep{\fill}} 
    S[table-format=1.1]
    S[table-format=-1.4]
    S[table-format=-1.4]
    S[table-format=-1.4]
    S[table-format=-1.4]
    S[table-format=-2.4]
    S[table-format=-2.4]
    S[table-format=-3.4]
}
\toprule
{\small $\lambda$} & 
{\small $\bar{r}$} & 
{\small $\sigma(r)$} & 
{\small $\sum_t r_t$} & 
{\small $\bar{R}/\sigma(R)$} & 
{\small $-\min(R)$} & 
{\small Max DD} & 
{\small Turnover} \\
\midrule
0 & 0.0153 & 0.0528 & 1.8318 & 1.0997 & 9.8654 & 12.7337 & 58.2867 \\
0.1 & 0.0187 & 0.0573 & 2.2488 & 1.2697 & 11.8965 & 12.8505 & 58.1035 \\
0.2 & 0.0194 & 0.0600 & 2.3292 & 1.2630 & 14.9458 & 14.9458 & 58.8336 \\
0.3 & 0.0220 & 0.0605 & 2.6347 & 1.4375 & 12.7858 & 13.4161 & 60.9016 \\
0.4 & 0.0209 & 0.0632 & 2.5125 & 1.3051 & 18.6285 & 19.2519 & 61.0962 \\
0.5 & 0.0211 & 0.0632 & 2.5325 & 1.3169 & 18.2476 & 20.1880 & 60.6515 \\
0.6 & 0.0210 & 0.0636 & 2.5156 & 1.2992 & 18.6622 & 20.1824 & 60.9962 \\
0.7 & 0.0219 & 0.0643 & 2.6332 & 1.3535 & 19.8222 & 21.2800 & 60.2769 \\
0.8 & 0.0215 & 0.0631 & 2.5858 & 1.3496 & 18.3169 & 19.0805 & 60.6148 \\
0.9 & 0.0220 & 0.0636 & 2.6423 & 1.3727 & 18.8771 & 19.7140 & 61.2379 \\
1.0 & 0.0223 & 0.0642 & 2.6709 & 1.3766 & 18.9305 & 19.1723 & 60.7656 \\
\midrule
\multicolumn{1}{l}{\small S\&P~500} & 0.0083 & 0.0428 & 0.9903 & 0.7028 & 12.5119 & 24.7695 & {--} \\
\bottomrule
\end{tabular*}
\begin{flushleft}
\textit{Note:}
$r_t$ and $R_t$ denote log and arithmetic returns, respectively, where $r_t=\ln(1+R_t)$ and $R_t=e^{r_t}-1$.
Metrics based on $r$ (e.g., $\bar{r}$, $\sigma(r)$, $\sum_t r_t$) are computed in log-return space for time additivity, 
whereas those based on $R$ (e.g., $\bar{R}/\sigma(R)$, $-\min(R)$, and Max~DD) use arithmetic returns to ensure interpretability in percentage terms.
Turnover is defined as the average absolute change in portfolio weights between rebalancing dates. All portfolios are value-weighted to reflect firm-size heterogeneity.
\end{flushleft}
\end{table}

\begin{table}[htbp]
    \centering
    \small
    \renewcommand{\arraystretch}{1.1}
    \setlength{\tabcolsep}{5.5pt}

    \caption{GRS tests for traditional factor models on CB-APM decile portfolios. \\
    This table reports in-sample GRS tests applied to decile portfolios constructed from CB-APM predicted return scores. 
    Each $\lambda$ corresponds to a distinct CB-APM specification that generates the test assets. 
    For each $\lambda$, we report the GRS $F$-statistic, its $p$-value, and mean absolute and root-mean-squared pricing errors (monthly and annualized). 
    All models are estimated using monthly excess returns.}
    \label{tab:grs-predicted}

\begin{tabular*}{\textwidth}{@{\extracolsep{\fill}}
    l
    S[table-format=1.1]
    S[table-format=1.6]
    S[table-format=1.3]
    S[table-format=1.4]
    S[table-format=1.4]
    S[table-format=1.4]
}
    \toprule
    {\small Factor Model} & {\small $\lambda$} & {\small GRS $F$} & {\small $p$-value} &
    {\small Mean$|{\alpha}|$ (M)} & {\small RMS$\alpha$ (M)} & {\small RMS$\alpha$ (A)} \\
    \midrule

    CAPM      & 0.0 & 1.9886 & 0.0466 & 0.0052 & 0.0056 & 0.0701 \\
    FF3       & 0.0 & 1.8914 & 0.0603 & 0.0054 & 0.0058 & 0.0722 \\
    Carhart4  & 0.0 & 1.9192 & 0.0565 & 0.0061 & 0.0065 & 0.0815 \\
    FF5       & 0.0 & 1.8657 & 0.0649 & 0.0052 & 0.0056 & 0.0694 \\
    FF6       & 0.0 & 1.8376 & 0.0700 & 0.0058 & 0.0061 & 0.0765 \\
    \midrule[0.1pt]

    CAPM      & 0.2 & 4.2663 & 0.0001 & 0.0059 & 0.0066 & 0.0812 \\
    FF3       & 0.2 & 4.1061 & 0.0002 & 0.0062 & 0.0068 & 0.0840 \\
    Carhart4  & 0.2 & 3.9454 & 0.0003 & 0.0069 & 0.0074 & 0.0923 \\
    FF5       & 0.2 & 4.1928 & 0.0001 & 0.0060 & 0.0066 & 0.0815 \\
    FF6       & 0.2 & 3.9469 & 0.0003 & 0.0065 & 0.0071 & 0.0878 \\
    \midrule[0.1pt]

    CAPM      & 0.4 & 4.6838 & 0.0000 & 0.0063 & 0.0071 & 0.0873 \\
    FF3       & 0.4 & 4.4962 & 0.0001 & 0.0065 & 0.0073 & 0.0904 \\
    Carhart4  & 0.4 & 4.3423 & 0.0001 & 0.0072 & 0.0079 & 0.0986 \\
    FF5       & 0.4 & 4.5035 & 0.0001 & 0.0064 & 0.0071 & 0.0880 \\
    FF6       & 0.4 & 4.2818 & 0.0001 & 0.0069 & 0.0076 & 0.0942 \\
    \midrule[0.1pt]

    CAPM      & 0.6 & 3.7890 & 0.0004 & 0.0065 & 0.0070 & 0.0866 \\
    FF3       & 0.6 & 3.6982 & 0.0005 & 0.0067 & 0.0073 & 0.0898 \\
    Carhart4  & 0.6 & 3.5643 & 0.0007 & 0.0073 & 0.0079 & 0.0983 \\
    FF5       & 0.6 & 3.9175 & 0.0003 & 0.0066 & 0.0071 & 0.0873 \\
    FF6       & 0.6 & 3.7039 & 0.0005 & 0.0070 & 0.0076 & 0.0939 \\
    \midrule[0.1pt]

    CAPM      & 0.8 & 4.0003 & 0.0002 & 0.0066 & 0.0072 & 0.0882 \\
    FF3       & 0.8 & 4.1132 & 0.0002 & 0.0068 & 0.0075 & 0.0920 \\
    Carhart4  & 0.8 & 4.1205 & 0.0002 & 0.0074 & 0.0081 & 0.1006 \\
    FF5       & 0.8 & 4.3215 & 0.0001 & 0.0067 & 0.0073 & 0.0896 \\
    FF6       & 0.8 & 4.2176 & 0.0001 & 0.0071 & 0.0078 & 0.0963 \\
    \midrule[0.1pt]

    CAPM      & 1.0 & 3.9004 & 0.0003 & 0.0067 & 0.0073 & 0.0902 \\
    FF3       & 1.0 & 3.6067 & 0.0006 & 0.0069 & 0.0076 & 0.0936 \\
    Carhart4  & 1.0 & 3.7242 & 0.0005 & 0.0075 & 0.0082 & 0.1022 \\
    FF5       & 1.0 & 3.8619 & 0.0003 & 0.0067 & 0.0074 & 0.0913 \\
    FF6       & 1.0 & 3.8339 & 0.0004 & 0.0072 & 0.0079 & 0.0980 \\
    \bottomrule
\end{tabular*}

\begin{flushleft}
\textit{Note:} 
Each $\lambda$ denotes a distinct CB-APM configuration used to generate decile-sorted test portfolios. 
The GRS $F$-statistic tests the joint null hypothesis that all pricing errors ($\boldsymbol{\alpha}$) are zero. 
Mean$|{\alpha}|$ and RMS$\alpha$ are reported in monthly (M) and annualized (A) terms. 
$p$-values are rounded to two decimal places; values below $0.005$ appear as $0.00$.
\end{flushleft}
\end{table}

\clearpage

\begin{small}
\renewcommand{\arraystretch}{0.6}
\setlength{\tabcolsep}{5.5pt}
\setlength{\LTleft}{0pt}
\setlength{\LTright}{0pt}

\begin{longtable}{@{\extracolsep{\fill}} p{1.8cm} *{6}{S[table-format=-2.4, table-column-width=1.9cm]}}

\caption{GRS tests for traditional factor models applied to portfolios sorted on CB-APM approximated consensus signals.\\
This table reports GRS test statistics for value-weighted decile portfolios formed on each dimension of the CB-APM's approximated consensus at $\lambda = 1.0$. 
For each consensus dimension, portfolio returns are regressed on the CAPM, Fama--French three-factor model, Carhart four-factor model, Fama--French five-factor model, and Fama--French six-factor model.
Reported are the GRS $F$-statistic, corresponding $p$-value, and mean absolute and root-mean-squared pricing errors, shown in both monthly and annualized units.}
\label{tab:grs-consensus}\\

\toprule
\textbf{Model} & {GRS $F$} & {$p$-value} &
{Mean$|{\alpha}|$ (M)} & {Mean$|{\alpha}|$ (A)} &
{RMS$\alpha$ (M)} & {RMS$\alpha$ (A)} \\
\midrule
\endfirsthead

\multicolumn{7}{l}{\footnotesize\textbf{Table \thetable:} GRS tests for traditional factor models on CB-APM consensus-sorted portfolios. (cont'd)}\\
\midrule
\textbf{Model} & {GRS $F$} & {$p$-value} &
{Mean$|{\alpha}|$ (M)} & {Mean$|{\alpha}|$ (A)} &
{RMS$\alpha$ (M)} & {RMS$\alpha$ (A)} \\
\midrule
\endhead

\midrule
\multicolumn{7}{r}{\textit{(cont'd on next page)}}\\
\midrule
\endfoot
\endlastfoot

\multicolumn{7}{l}{\textit{EPS forecast revision}} \\[3pt]
CAPM      & 1.3805 & 0.21 & 0.0046 & 0.0564 & 0.0047 & 0.0584 \\
FF3       & 1.3699 & 0.21 & 0.0050 & 0.0617 & 0.0051 & 0.0636 \\
Carhart4  & 1.4237 & 0.19 & 0.0058 & 0.0714 & 0.0060 & 0.0743 \\
FF5       & 1.3368 & 0.23 & 0.0048 & 0.0597 & 0.0050 & 0.0615 \\
FF6       & 1.3854 & 0.21 & 0.0055 & 0.0682 & 0.0057 & 0.0705 \\
\midrule
\addlinespace[0.4em]

\multicolumn{7}{l}{\textit{Change in recommendation}} \\[3pt]
CAPM      & 2.4101 & 0.02 & 0.0058 & 0.0717 & 0.0062 & 0.0761 \\
FF3       & 2.2623 & 0.02 & 0.0061 & 0.0752 & 0.0065 & 0.0798 \\
Carhart4  & 2.4376 & 0.01 & 0.0069 & 0.0851 & 0.0072 & 0.0894 \\
FF5       & 2.1439 & 0.03 & 0.0059 & 0.0728 & 0.0063 & 0.0778 \\
FF6       & 2.2909 & 0.02 & 0.0066 & 0.0811 & 0.0069 & 0.0857 \\
\midrule
\addlinespace[0.4em]

\multicolumn{7}{l}{\textit{Change in Forecast and Accrual}} \\[3pt]
CAPM      & 1.5123 & 0.15 & 0.0046 & 0.0563 & 0.0048 & 0.0588 \\
FF3       & 1.5732 & 0.13 & 0.0049 & 0.0607 & 0.0051 & 0.0633 \\
Carhart4  & 1.5667 & 0.13 & 0.0056 & 0.0700 & 0.0059 & 0.0737 \\
FF5       & 1.5269 & 0.15 & 0.0048 & 0.0590 & 0.0050 & 0.0614 \\
FF6       & 1.5191 & 0.15 & 0.0054 & 0.0673 & 0.0057 & 0.0703 \\
\midrule
\addlinespace[0.4em]

\multicolumn{7}{l}{\textit{Long-vs-short EPS forecasts}} \\[3pt]
CAPM      & 2.5764 & 0.01 & 0.0042 & 0.0522 & 0.0048 & 0.0588 \\
FF3       & 2.4627 & 0.01 & 0.0045 & 0.0561 & 0.0050 & 0.0621 \\
Carhart4  & 2.8118 & 0.01 & 0.0054 & 0.0673 & 0.0059 & 0.0731 \\
FF5       & 2.3544 & 0.02 & 0.0044 & 0.0541 & 0.0048 & 0.0597 \\
FF6       & 2.7049 & 0.01 & 0.0050 & 0.0624 & 0.0055 & 0.0688 \\
\midrule
\addlinespace[0.4em]

\multicolumn{7}{l}{\textit{Analyst earnings per share}} \\[3pt]
CAPM      & 0.9325 & 0.51 & 0.0043 & 0.0527 & 0.0047 & 0.0587 \\
FF3       & 1.1561 & 0.33 & 0.0046 & 0.0566 & 0.0050 & 0.0625 \\
Carhart4  & 1.2215 & 0.29 & 0.0054 & 0.0673 & 0.0059 & 0.0729 \\
FF5       & 1.1588 & 0.33 & 0.0044 & 0.0541 & 0.0048 & 0.0597 \\
FF6       & 1.1697 & 0.33 & 0.0050 & 0.0624 & 0.0055 & 0.0681 \\
\midrule
\addlinespace[0.4em]

\multicolumn{7}{l}{\textit{EPS Forecast Dispersion}} \\[3pt]
CAPM      & 1.4355 & 0.18 & 0.0050 & 0.0618 & 0.0053 & 0.0662 \\
FF3       & 1.4461 & 0.18 & 0.0051 & 0.0635 & 0.0056 & 0.0693 \\
Carhart4  & 1.4664 & 0.17 & 0.0057 & 0.0706 & 0.0063 & 0.0791 \\
FF5       & 1.3880 & 0.20 & 0.0050 & 0.0617 & 0.0054 & 0.0670 \\
FF6       & 1.3785 & 0.21 & 0.0054 & 0.0675 & 0.0060 & 0.0749 \\
\midrule
\addlinespace[0.4em]

\multicolumn{7}{l}{\textit{Earnings forecast revisions}} \\[3pt]
CAPM      & 2.3250 & 0.02 & 0.0047 & 0.0585 & 0.0050 & 0.0616 \\
FF3       & 2.5170 & 0.01 & 0.0051 & 0.0631 & 0.0053 & 0.0659 \\
Carhart4  & 2.4525 & 0.01 & 0.0058 & 0.0724 & 0.0061 & 0.0761 \\
FF5       & 2.3764 & 0.02 & 0.0050 & 0.0614 & 0.0052 & 0.0638 \\
FF6       & 2.3054 & 0.02 & 0.0056 & 0.0695 & 0.0058 & 0.0724 \\
\midrule
\addlinespace[0.4em]

\multicolumn{7}{l}{\textit{Analyst Value}} \\[3pt]
CAPM      & 1.0717 & 0.39 & 0.0042 & 0.0520 & 0.0044 & 0.0547 \\
FF3       & 1.0577 & 0.41 & 0.0045 & 0.0555 & 0.0047 & 0.0580 \\
Carhart4  & 1.0581 & 0.41 & 0.0054 & 0.0671 & 0.0056 & 0.0689 \\
FF5       & 1.0187 & 0.44 & 0.0043 & 0.0526 & 0.0045 & 0.0551 \\
FF6       & 1.0238 & 0.43 & 0.0050 & 0.0621 & 0.0052 & 0.0639 \\
\midrule
\addlinespace[0.4em]

\multicolumn{7}{l}{\textit{Analyst Optimism}} \\[3pt]
CAPM      & 1.1976 & 0.31 & 0.0042 & 0.0521 & 0.0046 & 0.0573 \\
FF3       & 1.1288 & 0.35 & 0.0045 & 0.0556 & 0.0048 & 0.0600 \\
Carhart4  & 1.0821 & 0.39 & 0.0054 & 0.0672 & 0.0057 & 0.0706 \\
FF5       & 1.0605 & 0.40 & 0.0043 & 0.0527 & 0.0046 & 0.0573 \\
FF6       & 1.0211 & 0.44 & 0.0050 & 0.0623 & 0.0053 & 0.0658 \\
\midrule
\addlinespace[0.4em]

\end{longtable}
\end{small}

\clearpage

\begin{figure}
    \centering
    \begin{tikzpicture}[
      >=Latex,
      node distance=1.2cm and 1.6cm,
      every node/.style={font=\small},
      neuron/.style={circle, draw, fill=white,thick, minimum size=5mm, inner sep=0pt},
      layer box/.style={rounded corners=6pt, draw=red, very thick, dashed},
      subbox/.style={rounded corners=4pt, draw, thick},
      textlbl/.style={font=\small, align=center},
      conn/.style={-Latex, line width=0.9pt},
      connfaint/.style={draw=gray!70, line width=0.35pt, draw opacity=0.5},
      connsample/.style={draw, line width=0.35pt, draw opacity=0.8}
    ]
    
    \coordinate (col1) at (3,0);   
    \coordinate (col2) at (9,0);   
    
    \def\ys{0.63}
    
    \foreach \k in {1,...,10}{
      \node[neuron] (h1-\k) at ($(col1)+(-1.8,{(5.5-\k)*\ys})$) {};
    }
    
    \foreach \k in {1,...,7}{
      \node[neuron] (h2-\k) at ($(col1)+(0.00,{(4-\k)*\ys})$) {};
    }
    
    \foreach \k in {1,...,5}{
      \node[neuron] (c-\k)  at ($(col1)+(1.8,{(3-\k)*\ys})$) {};
    }
    
    \node[subbox, fit=(c-1)(c-5), inner sep=4pt] (consvecbox) {};
    
    \begin{scope}[on background layer]
      \foreach \a in {1,...,10}
        \foreach \b in {1,...,7}
          \draw[connfaint] (h1-\a) -- (h2-\b);
    \end{scope}
    
    \foreach \a in {1,3,5,7,9}
      \foreach \b in {1,4,7}
        \draw[connsample] (h1-\a) -- (h2-\b);

    \begin{scope}[on background layer]
      \foreach \a in {1,...,7}
        \foreach \b in {1,...,5}
          \draw[connfaint] (h2-\a) -- (c-\b);
    \end{scope}
    
    \foreach \a in {1,3,5,7}
      \foreach \b in {1,4}
        \draw[connsample] (h2-\a) -- (c-\b);

    \begin{pgfonlayer}{background}
        \node[layer box, fit=(h1-1)(h1-10)(h2-1)(h2-7)(consvecbox), minimum height = 70mm, inner sep=8pt,
            label={[textlbl]90:Consensus Module \\ $f(\phi)$}] (consbox) {};
    \end{pgfonlayer}

    \node[textlbl, anchor=east] (predlabel) at ($(col1)+(-3,0)$) {Predictors \\ $\mathcal{I}^f_{i,t}, \mathcal{I}^m_t$};
    \draw[conn] (predlabel.east) -- (consbox.west);
    
    \foreach \k in {1,...,5}{
      \node[neuron] (q\k) at ($(col2)+(-0.7,{(3-\k)*\ys})$) {};
    }
    \node[neuron] (r1) at ($(col2)+( 0.2, 0.0)$) {};
    \node[subbox, fit=(q1)(q5), inner sep=4pt] (approxsvecbox) {};

     \foreach \k in {1,...,5}{
        \draw[connsample] (q\k) -- (r1);
    }

    \begin{pgfonlayer}{background}
        \node[layer box, fit=(q1)(q5)(r1), minimum height = 70mm, inner sep=12pt, label={[textlbl]90:Prediction Module \\ $g(\theta)$}] (predbox2) {};
    \end{pgfonlayer}
    
    \node[textlbl, anchor=south] (conslabel) at ($(consbox.east)!0.5!(predbox2.west)$) {Consensus \\ $\hat{C}_{i,t}$};

    \draw[conn] (consbox.east) -- (predbox2.west);
    
    \node[textlbl, anchor=east] (outputlabel) at ($(col2)+(+4.3,0)$) {Excess Return \\ $\mathbb{E}_t\left[R_{i,t+h}\right]$};
    \draw[conn] (predbox2.east) -- (outputlabel.west);
    
    
    \node[circle, draw, thick, minimum size=5.5mm, below=3.0cm of outputlabel] (plus) {$+$};

    \node[subbox, draw=none] (lrbox) at ($(outputlabel.south)!0.5!(plus.north)$) {$\mathcal{L}_R(\phi,\theta)$};
    \draw[conn] (lrbox.south) -- (plus.north);
    \draw[conn] (outputlabel.south) -- (lrbox.north);
    
    \node[subbox, draw=none] (lcbox) at (conslabel |- plus) {$\lambda \, \langle\textbf{1}, \mathcal{L}_C(\phi)\rangle$};
    \draw[conn] (conslabel.south) -- (lcbox.north);
    \draw[conn] (lcbox.east) -- (plus.west);

    \node[subbox, draw=none, right=0.8cm of plus] (loss) {$\mathcal{L}$};
    \draw[conn] (plus.east) -- (loss.west);
    
    \end{tikzpicture}

    \caption{Architecture of the CB-APM. \\
    The model is composed of two modules, the consensus module $f(\phi)$ (left) and the prediction module $g(\theta)$ (right). The consensus module compresses firm-specific predictors $\mathcal{I}^f_{i,t}$ and macroeconomic variables $\mathcal{I}^m_t$ into a lower-dimensional consensus vector $\hat{C}_{i,t}$ through a feedforward neural network. This bottleneck enforces interpretability by design, as each coordinate of $\hat{C}_{i,t}$ is treated as a consensus concept. The prediction module then maps these consensus variables into expected excess returns $\mathbb{E}_t\left[R_{i,t+h}\right]$ using a linear layer. The return loss $\mathcal{L}_R(\phi,\theta)$ and the consensus loss $\mathcal{L}_C(\phi)$ are optimized jointly using the weighted sum $\mathcal{L}=\mathcal{L}_R+\lambda\, \langle\textbf{1}, \mathcal{L}_C\rangle$, ensuring that the consensus layer is both predictive of returns and interpretable.}
    \label{fig:model_architecture}
\end{figure}

\clearpage

\end{refsection}


\appendix

\begin{refsection}

\setcounter{page}{0}
\thispagestyle{empty}

\begingroup
\renewcommand{\thefootnote}{\fnsymbol{footnote}}
\begin{center}
    \LARGE
    Internet Appendices to \\
    ``Interpretable Deep Learning for Stock Returns:\\ A Consensus-Bottleneck Asset Pricing Model" \footnotemark[1]
\end{center}
\footnotetext[1]{\Acknowledgement}
\endgroup

\begin{center}
    \large
    Bong-Gyu Jang \qquad Younwoo Jeong \qquad Changeun Kim
\end{center}

\clearpage

\renewcommand{\thetable}{\thesection.\arabic{table}}
\renewcommand{\thefigure}{\thesection.\arabic{figure}}

\setcounter{table}{0}
\setcounter{figure}{0}

The Internet Appendix is organized as follows. Section~\ref{sec:interpret} reviews interpretable machine learning and situates CB-APM within this framework. Section~\ref{Appendix:preprocess} discusses data preprocessing procedures applied prior to estimation and evaluation, including macroeconomic state extraction via autoencoder. Section~\ref{Appendix:nn} provides details on the implementation and architectural choices of the neural network used throughout this study. Section~\ref{Appendix:additional} reports additional empirical results for robustness and supplementary insights. Section~\ref{Appendix:data} presents the list of variables comprising the dataset used in this study.

\section{Interpretable Artificial Intelligence} \label{sec:interpret}
In this section, we clarify the often-confused concept of explainability and interpretability, to prevent any potential misunderstanding. Although they both focus on understanding the nature of machine learning from a human perspective, the primary difference between these two fields with a long and intense history of research lies in their focus areas. 

\textit{Explainable AI}, also known as XAI, focuses on the reason \textit{why} the prediction of a model has been inferred, while interpretable AI is more interested in \textit{how} the model is trained to find the approximate mapping from the hypothesis set. In particular, XAI does not attempt to dissect the functioning of the black box model but rather accepts the opacity of such models as it is intended to be.  The fundamental assumption of XAI is that the result of such an opaque system should be strongly related to the input features, which, in the context of deep learning, is often the extent of our understanding. Therefore, researchers design an ad hoc statistical model, which is simpler compared to a model subject to explanation in most cases, to explain the relationships between the input and output of the estimated model. 

\textit{Interpretable AI}, in comparison, designs a model to be perceivable in human knowledge itself, without requiring further explanations. The most dominant and perhaps the most well-known example of an interpretable model is linear regression. Linear regressions are interpretable since their outputs can be directly represented as linear combinations of input features and coefficients. Conversely, \textcite{benitez1997artificial} argue that deep neural networks are considered non-interpretable since these models typically do not provide insight into how input features are transformed through hidden layers to produce outputs. 

Nevertheless the black box nature of deep neural networks, researchers may attempt to make such models understandable by introducing the concept of ``disentangled representations" \citep[See ][]{higgins2018towards, locatello2019challenging, locatello2019fairness}.  To understand the concept of disentanglement, we can consider the simple example of a na\"ive feedforward neural network, which consists of multiple hidden layers. The hidden layers are intermediate vector representations, often interpreted as features extracted from input data. Since they are trained with a downstream task-related objective in most cases, those layers are presumed to represent the essential information within the high-dimensional and noisy input data, well enough for the successful performance on that task. However, these features exist in an ``entangled" manner, meaning that each element of the feature representation is a mixture of multiple factors. This entanglement complicates the interpretability of the model, as it blurs the specific contributions of individual input features to the final output. In this context, ``disentanglement" refers to separating the underlying causal factors of the data and intermediate representations into distinct and non-overlapping representations. For instance, in fixed income markets, dozens of yields across maturities can be effectively summarized by three disentangled factors, level, slope, and curvature, as illustrated in Figure \ref{fig:disentangled}. Each of these dimensions captures an independent source of variation in the yield curve, and together they provide an interpretable low-dimensional representation that still preserves the essential structure of the data.\footnote{A clarification is needed to avoid conflating our framework with traditional factor models. \textcite{nelson1987parsimonious} impose level, slope, and curvature terms to parameterize the yield curve, while \textcite{litterman1991common} show via principal components that similar dimensions emerge empirically. Both approaches reduce high-dimensional yields to a few interpretable coordinates. By contrast, in CB-APM the consensus-bottleneck is not imposed ex ante but learned endogenously, closer in spirit to Litterman--Scheinkman than to Nelson--Siegel. In this sense, whereas Nelson--Siegel estimate factors to fit the curve, our model embeds interpretability within the network architecture itself, rendering part of the neural network transparent while retaining predictive capacity.}

\begin{figure}[htbp]
    \centering
    \begin{tikzpicture}[
      >=Latex,
      node distance=1.7cm and 1.6cm,
      every node/.style={font=\small},
      circle node/.style={circle, draw, thick, minimum size=10mm, inner sep=0pt},
      layer box/.style={rounded corners=4pt, draw, thick},
      concept/.style={circle node, very thick},
      conn/.style={-Latex, line width=0.7pt},
      lbl/.style={font=\small}
    ]
    
    \coordinate (col0) at (0,0);      
    \coordinate (col1) at (2.3,0);    
    \coordinate (col2) at (4.6,0);    
    \coordinate (col3) at (7.6,0);    
    \coordinate (col4) at (10.4,0);   
    
    \node[circle node] (i1) at ($(col0)+(0, 1.4)$) {};
    \node[circle node] (i2) at ($(col0)+(0, 0.0)$) {};
    \node[circle node] (i3) at ($(col0)+(0,-1.4)$) {};
    
    \node[layer box, fit=(i1)(i3), inner sep=8pt, label={[lbl]90:\shortstack{Input\\Layer}}] (ibox) {};
    
    \node[circle node] (h11) at ($(col1)+(0, 1.4)$) {};
    \node[circle node] (h12) at ($(col1)+(0, 0.0)$) {};
    \node[circle node] (h13) at ($(col1)+(0,-1.4)$) {};
    
    \node[circle node] (h21) at ($(col2)+(0, 1.4)$) {};
    \node[circle node] (h22) at ($(col2)+(0, 0.0)$) {};
    \node[circle node] (h23) at ($(col2)+(0,-1.4)$) {};
    
    \definecolor{conceptA}{RGB}{214,235,225} 
    \definecolor{conceptB}{RGB}{246,222,222} 
    \definecolor{conceptC}{RGB}{220,230,250} 
    
    \node[concept, fill=conceptA] (c1) at ($(col3)+(0, 1.4)$) {};
    \node[concept, fill=conceptB] (c2) at ($(col3)+(0, 0.0)$) {};
    \node[concept, fill=conceptC] (c3) at ($(col3)+(0,-1.4)$) {};

    \node[anchor=west] (lab1) at ($(c1)+(0.5,0)$) {Level};
    \node[anchor=west] (lab2) at ($(c2)+(0.5,0)$) {Slope};
    \node[anchor=west] (lab3) at ($(c3)+(0.5,0)$) {Curvature};
    
    \begin{pgfonlayer}{background}
      \node[layer box, fill=black!5, rounded corners,
            fit=(c1)(c3)(lab1)(lab2)(lab3), inner sep=12pt,
            label={[lbl]90:\shortstack{Disentangled\\Representation}}] (cbox) {};
    \end{pgfonlayer}
    
    \node[layer box, minimum width=15mm, minimum height=18mm, anchor=west] (outbox) at ($(col4)+(+7mm,0)$) {};
    \node[circle, draw, thick, minimum size=9mm, inner sep=0pt] (o1) at (outbox) {};
    \node[anchor=south, yshift=2mm, align=center] at (outbox.north) {Output\\Layer};
    
    \foreach \a in {i1,i2,i3}{
      \foreach \b in {h11,h12,h13}{
        \draw[conn] (\a) -- (\b);
      }
    }
    
    \foreach \a in {h11,h12,h13}{
      \foreach \b in {h21,h22,h23}{
        \draw[conn] (\a) -- (\b);
      }
    }
    
    \foreach \a in {h21,h22,h23}{
      \foreach \b in {c1,c2,c3}{
        \draw[conn] (\a) -- (\b);
      }
    }
    
    \draw[conn, line width=1pt] ($(cbox.east)+(0,0)$) -- ++(0.5,0) -- ($(outbox.west)+(0,0)$);
    
    \end{tikzpicture}
    \caption{Disentangled representations of neural network. \\
    This schematic illustrates how a high-dimensional input is compressed into a small set of latent representations that correspond to interpretable concepts. The example shown mirrors the yield curve decomposition in fixed income, where dozens of yields can be summarized by three factors: level, slope, and curvature \parencite{nelson1987parsimonious}. The disentangled representation isolates these dimensions, which are then mapped by the output layer into the final prediction, here defined as the yield curve.}
    \label{fig:disentangled}
\end{figure}

Applying this concept, fully interpretable neural networks are designed so that all aspects of their structure and function are understandable to humans. This means that every layer, neuron, and connection in the network has a clear and understandable purpose related to the task at hand. These fully interpretable models have been acknowledged for their transparency and the ease with which their decisions can be understood and trusted. However, a prevailing limitation of these models has been their generally lower predictive performance compared to their less interpretable counterparts due to the interpretability-accuracy trade-off, presented by \textcite{plate1999accuracy}. This trade-off has historically motivated researchers towards utilizing complex neural network-based models in machine learning studies, due to their superior predictive capabilities, despite their lack of interpretability. 

Recent advances in interpretable AI, however, are challenging the notion that interpretability must come at the cost of performance. Studies are now demonstrating that it is feasible to maintain the high performance of neural network structures while making them partially interpretable \footnote{Refer to the representative studies of \textcite{koh2020concept}, \textcite{chen2020concept} for discriminative models and \textcite{chen2016infogan}, \textcite{higgins2017beta} for probabilistic generative models.}. Partial interpretability takes an approach where a segment of the network is made interpretable, typically the layers closer to the input or output. The rest of the network may operate as a black box, allowing intervention on the learning process without significant loss in model performance. 

\section{Data Preprocessing} \label{Appendix:preprocess}

The quality and temporal consistency of input data are fundamental to the empirical validity of the CB-APM framework. Because our model relies on a rich set of firm-level and macroeconomic variables to approximate analysts' consensus and forecast future returns, ensuring that the data accurately reflect the information available to investors at each point in time is essential. Accordingly, this section outlines the complete preprocessing pipeline applied before model training. The procedures include (i) lagging variables to eliminate look-ahead bias, (ii) sampling and filtering firms and predictors to balance coverage and data quality, (iii) imputing missing observations through economically informed methods, and (iv) normalizing the data to harmonize variable scales. Together, these steps construct a temporally aligned, cross-sectionally coherent, and numerically stable dataset that serves as the foundation for the empirical analysis.

\subsection{Data lagging} 

Publicly available monthly asset pricing data are typically released with reporting delays, which can introduce look-ahead bias. This issue arises because the recorded date of the data often reflects when the metric was calculated rather than when it became available. Such discrepancies can mislead researchers about the actual availability of the data at a given time. Several papers point out this problem including \textcite{ChenZimmermann2021}, and we follow their recommended practices for handling such discrepancies from data delays.

Initially, we check the data frequency of firm-level characteristics as detailed in Table \ref{tab:firm-level-predictors}. Recognizing that consensus variables are provided irregularly, we conservatively assume an annual frequency. Then, quarterly data are lagged by three months and annual data by six months, respectively. All firm-level predictors are lagged prior to constructing the learning dataset to ensure temporal alignment between available information and subsequent returns. This approach ensures that the data utilized for predicting future returns are sufficiently historical, thereby minimizing the risk of inadvertently using information that would not have been available at the forecast horizon.

\subsection{Data Sampling}
\label{Appendix:data-sampling}

This appendix details the construction of the learning dataset employed in the CB-APM estimation. The preprocessing follows a systematic six-step procedure implemented in the \texttt{get\_data} function, which integrates firm-level predictors, macroeconomic variables, and risk-free rates into a unified panel suitable for model training. Each step ensures data quality, temporal consistency, and the preservation of meaningful cross-sectional information, as summarized below.

\subsubsection{Firm screening}

Because I/B/E/S analyst consensus variables are relatively sparse compared to other firm-level characteristics, filling in missing observations without preliminary filtering would yield an artificial dataset dominated by interpolated or substituted values. To avoid such distortion, firms without any valid analyst consensus data are excluded from the investable universe at the outset. Among the 17{,}743 stocks in the full \textcite{ChenZimmermann2021} dataset, we retain only 4{,}683 companies for which the complete set of analyst-related variables is available for a sufficiently long history to ensure stable estimation. After applying this screening, the resulting sample contains a total of 605{,}722 firm-month observations.

This exclusion criterion deliberately sacrifices sample size to ensure that the model learns directly from genuine analyst opinions rather than imputed proxies. Although previous studies in machine-learning-based asset pricing (e.g., \citealp{gu2020empirical}; \citealp{chen2024deep}) generally tolerate higher sparsity levels, the stricter sampling rule adopted here is essential for faithfully training the consensus module.

\subsubsection{Variable selection}

The original dataset from \textcite{ChenZimmermann2021} includes 161 firm-level predictors with significant long-short portfolio $t$-statistics exceeding 4 in absolute value. While these variables have been validated for statistical predictability, many suffer from low firm coverage and short sample histories. Such sparsity weakens the ability of the consensus module to capture variation across firms, since consensus approximation relies on observing cross-firm differences over comparable horizons.

Accordingly, we retain 114 predictors after applying the following criteria:
\begin{enumerate}
    \item variables with missing-value rates exceeding 20\% across the firm panel are removed;
    \item variables with insufficient historical coverage (sample starting year after January~1994) are excluded.
\end{enumerate}
The resulting set of firm-level characteristics provides a balanced trade-off between data completeness and information diversity, ensuring that each firm contributes a meaningful set of observations to both the consensus and return-prediction modules.

After all preprocessing steps, the final dataset comprises:
\begin{enumerate}
    \item 4{,}683 firms with nonmissing analyst consensus data,
    \item 114 firm-level predictors and 123 macroeconomic indicators (including 115 from FRED-MD and 8 from \citealp{welch2008comprehensive}),
    \item a total of 605{,}722 firm-month observations spanning January~1994 to December~2023.
\end{enumerate}
This refined panel forms the empirical foundation for all model estimation and evaluation procedures described in Section~\ref{sec:model}.

\subsection{Data imputation}

Although the majority of studies neglect the importance of data imputation methods and simply handle missing values by substituting a cross-sectional mean or median \parencite{green2017characteristics, gu2020empirical}, we adopt a distinct approach to ensure the integrity of the concept information set, which is crucial in bottleneck modeling.

For variables representing firm characteristics, we primarily employ the Last Observation Carried Forward (LOCF) method. This technique assumes that the most recent observation remains valid until updated information becomes available, thereby maintaining temporal continuity over short gaps. This approach also mirrors real-world information flow, as the last observation reflects the data actually observed by investors until the next public update. 

However, for variables that represent growth rates or changes in firm characteristics, \footnote{Variables marked with asterisks in Table~\ref{tab:firm-level-predictors} of Internet Appendix~\ref{Appendix:data}} we apply time-series mean imputation. This method captures the inherent continuity and trend in firm-specific dynamics, producing more realistic estimates than cross-sectional averaging. Relying repeatedly on the last observed value for growth-related factors could falsely imply persistence or monotonic trends, misrepresenting the inherently dynamic nature of such variables. 

When firms lack historical observations entirely, as in the case of newly listed firms or those undergoing restructuring, we revert to imputing missing values using the cross-sectional mean computed within the same month. Although less ideal, this fallback preserves dataset integrity without introducing excessive bias from outdated or anomalous firm histories.

For analyst consensus data, which include earnings forecasts and investment recommendations, we adopt a two-pronged strategy based on data availability. Because analyst estimates evolve gradually in response to changing fundamentals rather than shifting abruptly, we apply linear interpolation between adjacent data points to capture smooth temporal adjustments. When neither past nor future observations are available (e.g., for firms with sparse coverage), missing entries are filled using the cross-sectional mean of all firms in the same month. 

This multi-stage imputation strategy preserves both temporal coherence and cross-sectional comparability, ensuring that the constructed concept information set remains economically interpretable and suitable for the CB-APM framework.

\subsection{Data normalization}

In asset pricing research, just as in other fields that utilize numerical data, the presence of outliers can significantly distort model outputs, necessitating the standardization of data prior to model integration. Specifically, in cross-sectional asset pricing, the relative position of a metric within the spectrum of similar data points across firms is often more informative than the metric's absolute value. Consequently, aligning with methodologies employed in recent studies \parencite{kelly2019characteristics, gu2020empirical, freyberger2020dissecting, gu2021autoencoder}, we compute the rank percentage of firm-level data cross-sectionally, subsequently scaling these ranks to the interval of $[-1, 1]$. This transformation not only mitigates the influence of outliers but also facilitates more meaningful comparisons across firms. For macroeconomic data, min-max normalization is applied to ensure compatibility with the firm-level data scale. This method adjusts macroeconomic indicators to the same $[-1, 1]$ interval. To promote numerical stability during joint optimization and to align the scale of all model components, we also apply the same rank transformation to consensus variables, even though they primarily serve as target outputs.

\subsection{Non-linear macro embeddings} \label{Appendix:autoencoder}

To incorporate macroeconomic dynamics into the conditional expectation function \( \mathbb{E}_t[R_{i,t+h}] \) defined in equation~\eqref{eq:cond}, we encode the aggregate information set \( \mathcal{I}_t^m \) through an autoencoder-based representation. While macroeconomic variables are often dismissed in cross-sectional asset pricing due to their perceived homogeneity across firms, we argue that macro context exerts differentiated influence through sectoral dynamics, capital structure sensitivity, and behavioral channeling. However, the sheer volume and redundancy of macroeconomic indicators, particularly those sourced from databases such as FRED-MD, pose significant challenges for model training. Including hundreds of highly correlated variables not only increases the risk of overfitting but also dilutes the learning signal by overwhelming the model with noise and irrelevant information.

Moreover, many macroeconomic series track similar phenomena at varying lags, granularities, or levels of transformation (e.g., growth rates, differences, log-levels), creating unnecessary dimensionality without proportional gains in explanatory power. This redundancy hinders both the stability and interpretability of predictive models, especially those trained on firm-level data where macro variables are shared across the entire cross-section. Reducing this high-dimensional input into a compact, informative representation is thus not only computationally efficient but also essential for isolating the latent economic regimes that meaningfully affect asset returns.

Dimensionality reduction techniques have long been used in financial modeling to address such issues. Principal Component Analysis (PCA) has served as a standard tool for extracting latent factors from large panels of macroeconomic variables \parencite{ludvigson2007empirical}, while extensions such as Sparse PCA and Independent Component Analysis (ICA) have been applied to improve factor interpretability and reduce multicollinearity \parencite{fan2016projected, erichson2020sparse}. More recently, deep learning approaches, particularly autoencoders, have gained traction in the asset pricing literature for their ability to capture nonlinear interactions and extract economically meaningful latent structures from noisy, high-dimensional data \parencite{chen2024deep, gu2021autoencoder}. These methods have proven effective in modeling complex macro-financial dynamics that traditional linear techniques may fail to uncover.\footnote{Appendix~\ref{Appendix:vsPCA} demonstrates that replacing the autoencoder with a 32-factor PCA markedly weakens out-of-sample return predictability, despite both approaches delivering similar accuracy in reconstructing analysts' consensus. This divergence highlights the advantage of nonlinear compression in capturing macroeconomic structure relevant for pricing.}

To address these challenges, we enhance model performance by encoding the macroeconomic regime using an autoencoder, thereby providing structured, compact, and economically interpretable representations that condition firm-level predictions. Instead of feeding all 115 raw macroeconomic variables from the FRED-MD database directly into the model, we train an autoencoder to learn a lower-dimensional latent representation of the macroeconomic environment at each time step. Specifically, the encoder compresses high-dimensional macroeconomic inputs into a latent macroeconomic state vector \( \mathbf{z}_t \), which is subsequently concatenated with firm-level features and passed into the CB-APM architecture. During training, the decoder reconstructs the input variables, and the network is optimized to minimize the mean squared reconstruction error. After training, only the encoder is retained to generate macroeconomic embeddings for prediction.

Formally, let the macroeconomic input at time \( t \) be \( \mathbf{x}_t \in \mathbb{R}^D \), where \( D = 123 \). The encoder \( \mathcal{E}_\phi(\cdot) \) maps this input to a latent representation \( \mathbf{z}_t \in \mathbb{R}^d \):\footnote{Empirically, setting the latent dimension to \( d=32 \) yields the best out-of-sample performance (see Internet Appendix~\ref{Appendix:latent-dim}).}
\begin{equation*}
    \mathbf{z}_t
    =
    \mathcal{E}_{\phi}\!\left( \mathbf{x}_t \right).
\end{equation*}
The decoder \( \mathcal{D}_\theta(\cdot) \) reconstructs the input:
\begin{equation*}
    \hat{\mathbf{x}}_t
    =
    \mathcal{D}_{\theta}\!\left( \mathbf{z}_t \right),
\end{equation*}
and the model is trained to minimize the reconstruction loss:
\begin{equation*}
    \mathcal{L}_{\text{AE}}(\theta, \phi)
    =
    \frac{1}{T}
    \sum_{t=1}^{T}
    \left\|
    \mathbf{x}_t
    -
    \mathcal{D}_{\theta}\!\bigl(
    \mathcal{E}_{\phi}(\mathbf{x}_t)
    \bigr)
    \right\|_2^2.
\end{equation*}
After training, the encoder output \( \mathbf{z}_t \) is concatenated with each firm's feature vector \( \mathbf{x}_{i,t}^{\text{firm}} \) to form the model input:
\begin{equation*}
    \mathbf{x}_{i,t}^{\text{input}}
    =
    \left[
    \mathbf{x}_{i,t}^{\text{firm}}
    \,;\,
    \mathbf{z}_t
    \right].
\end{equation*}

Formally, the latent representation \( \mathbf{z}_t \) learned through the autoencoder serves as an empirical proxy for the macroeconomic information set \( \mathcal{I}_t^m \) introduced in equation~\eqref{eq:cond}. In this context, \( \mathbf{z}_t \) functions as a compressed, data-driven approximation of the macroeconomic state observable to investors at time \( t \). This design allows the CB-APM to integrate the high-dimensional macroeconomic information set into a tractable latent representation, ensuring that firm-level forecasts remain conditioned on a parsimonious yet informative depiction of the aggregate economic environment. By mapping \( \mathcal{I}_t^m \) into \( \mathbf{z}_t \), the model effectively operationalizes the theoretical information set within a learnable structure, thereby linking the empirical implementation of the macro encoder to the conditional expectation framework defined in equation~\eqref{eq:cond}.

As illustrated in Figure~\ref{fig:autoencoder_macro}, this framework visualizes the overall data pipeline of the CB-APM, depicting how macroeconomic inputs are encoded, compressed, and subsequently integrated with firm-level characteristics for return prediction. The figure serves as a conceptual representation clarifying the interaction between the macro autoencoder and the return-prediction module. The full architecture details of the autoencoder, including hidden-layer configurations and activation functions, are provided in Internet Appendix~\ref{Appendix:hyper}. The empirical findings underscore the importance of representing macroeconomic regimes in shaping cross-sectional return dynamics and highlight the utility of neural representation learning in extracting economically salient signals from high-dimensional macro data. At each expanding-window step, the autoencoder is trained only on macro data available up to the window end date, and the encoder is then used to compute \( \mathbf{z}_t \) for that window's validation and test months, thereby preventing look-ahead bias.

\begin{figure}[htbp]
    \centering
    \begin{tikzpicture}[
        node distance=12mm and 16mm,
        neuron/.style={circle, draw, minimum size=8mm},
        concat/.style={circle, draw, minimum size=6mm, fill=white, thick},
        box/.style={rectangle, draw, rounded corners, align=center, minimum height=1cm, minimum width=2.2cm},
        trapez/.style={draw, trapezium, trapezium angle=70, minimum height=2.5cm, minimum width=2.8cm, align=center},
        >=stealth
    ]
    \node[box] (macro) {\parbox{3.5cm}{\centering \textbf{Macroeconomic Inputs} \\ $\mathbf{x}_t$}};

    \node[trapez, right=1.0cm of macro, trapezium stretches=true, shape border rotate=270] (encoder) {\textbf{Encoder} \\ $f_\phi(\cdot)$};

    \node[neuron, right=0.5cm of encoder] (latent) {$\mathbf{z}_t$};

    \node[trapez, right=0.5cm of latent, trapezium stretches=true, shape border rotate=90] (decoder) {\textbf{Decoder} \\ $g_\theta(\cdot)$};

    \node[box, right=1.0cm of decoder] (recon)  {\parbox{3.5cm}{\centering \textbf{Reconstructed Inputs} \\ $\hat{\mathbf{x}}_t$}};

    \node[box, below=3cm of latent, xshift=-2.3cm] (firm) {\parbox{3.5cm}{\centering \textbf{Firm-Level Inputs} \\ $\mathbf{x}^{\text{firm}}_{i,t}$}};

    \node[concat, below=1.6cm of latent] (concat) {+};

    \node[box, right=2.0cm of concat] (cbapm) {\parbox{3.5cm}{\centering \textbf{CB-APM Model} \\ Return Prediction}};

    \draw[->] (macro) -- (encoder);
    \draw[->] (encoder) -- (latent);
    \draw[->] (latent) -- (decoder);
    \draw[->] (decoder) -- (recon);

    \draw[->, thick, dashed] (latent.south) -- (concat.north);

    \draw[->, thick, dashed] (firm.east) -| (concat.south);

    \draw[->, thick, dashed] (concat.east) -- (cbapm.west);

    \node[above=0.8cm of latent] {\scriptsize \parbox{3.5cm}{\centering Encoder output \\ retained for prediction}};
    
    \node[above=0.3cm of recon] {\scriptsize \parbox{3.5cm}{\centering Decoder used only \\ during training \\ (reconstruction error)}};
    
    \end{tikzpicture}
    \caption{Autoencoder-based macroeconomic embedding. \\
    The encoder narrows horizontally to compress high-dimensional macroeconomic inputs into a latent state $\mathbf{z}_t$, concatenated with firm-level features for return prediction. The decoder is used only during training for reconstruction loss.}
    \label{fig:autoencoder_macro}
\end{figure}

\subsection{The Expanding window approach} \label{Appendix:expanding}

To evaluate model performance under realistic and evolving market conditions, this study employs an expanding window as a sample splitting scheme. Unlike static train-validation-test splits, the expanding window approach incrementally grows the training dataset over time while keeping the validation and testing sets fixed in size. This dynamic design mirrors the constraints of real-world applications, where future regimes are unknown and models must generalize across economic environments without the benefit of hindsight. By gradually shifting the end point of the training set forward, the expanding window simulates a time-consistent learning process that naturally adapts to structural changes in the data. As a result, this framework offers both methodological rigor and practical relevance, allowing the model to be evaluated not only on statistical metrics but also on its robustness across different economic cycles.

Figure \ref{fig:expanding_window} illustrates the expanding window approach for dataset partitioning, with the arrow along the bottom denoting the timeline of the window. The validation dataset spans two years, while the testing dataset spans a single year. Starting from the training set from January 1994 to December 2010, each training window ends at December of a given year, and subsequently expands by one year for the next window. This process continues sequentially, ensuring that the testing datasets do not overlap in any window. Consequently, the complete testing set spans from January, 2013 to December, 2022. \footnote{The final year of the dataset (January--December 2023) is reserved solely for constructing annual stock returns, as computing these returns requires at least one full year of subsequent observations.}

\makeatletter
\newcommand*{\rom}[1]{\expandafter\@slowromancap\romannumeral #1@}
\makeatother

\begin{figure}[htbp]
\centering
\begin{tikzpicture}[
    >=latex,                                  
    x=0.8cm, y=0.8cm,                        
    every node/.style={font=\footnotesize}                        
]
\def\yTop{4.2}
\def\yMid{2.8}
\def\yBot{0.6}
\def\h{0.9}
\def\baseline{0}

\draw[->, line width=0.4pt]
  ({2005-2005},\baseline) -- ({2023-2005 + 0.8/0.52},\baseline) node[below]{time};

\foreach \yy in {2005,2011,2013,2014}{
  \draw[dashed,gray!70]
    ({\yy-2005},\baseline-0.1) -- ({\yy-2005},\yTop+\h);
}
\foreach \yy in {2015}{
  \draw[dashed,gray!70]
    ({\yy-2005},\baseline-0.1) -- ({\yy-2005},\yMid+\h);
}
\foreach \yy in {2023}{
  \draw[dashed,gray!70]
    ({\yy-2005},\baseline-0.1) -- ({\yy-2005},\yBot+\h);
}
\node[below] at ({2005-2005},\baseline) {1994};
\node[below] at ({2011-2005},\baseline) {2011};
\node[below] at ({2013-2005},\baseline) {2013};
\node[below] at ({2014-2005},\baseline) {2014};
\node[below] at ({2015-2005},\baseline) {2015};
\node[below] at ({2023-2005},\baseline) {2023};

\fill[gray!35] ({2005-2005},\yTop) rectangle ({2011-2005},\yTop+\h);
\fill[white]   ({2011-2005},\yTop) rectangle ({2013-2005},\yTop+\h);
\fill[gray!20] ({2013-2005},\yTop) rectangle ({2014-2005},\yTop+\h);

\node at ({2008-2005},\yTop+\h/2) {\rom{1}};
\node at ({2012-2005},\yTop+\h/2) {\rom{2}};
\node at ({2013.5-2005},\yTop+\h/2) {\rom{3}};

\fill[gray!35] ({2005-2005},\yMid) rectangle ({2012-2005},\yMid+\h);
\fill[white]   ({2012-2005},\yMid) rectangle ({2014-2005},\yMid+\h);
\fill[gray!20] ({2014-2005},\yMid) rectangle ({2015-2005},\yMid+\h);

\node[rotate=145, font=\large] at ({2017.5-2005}, {(\yMid+\yBot)/2 + \h/2 + 0.1}) {$\cdots$};

\fill[gray!35] ({2005-2005},\yBot) rectangle ({2020-2005},\yBot+\h);
\fill[white]   ({2020-2005},\yBot) rectangle ({2022-2005},\yBot+\h);
\fill[gray!35] ({2022-2005},\yBot) rectangle ({2023-2005},\yBot+\h);

\foreach \A/\B/\Y in {
  2005/2011/\yTop,
  2011/2013/\yTop,
  2013/2014/\yTop,
  2005/2012/\yMid,
  2012/2014/\yMid,
  2014/2015/\yMid,
  2005/2020/\yBot,
  2020/2022/\yBot,
  2022/2023/\yBot}{
  \draw[black!60,line width=0.3pt]
    ({\A-2005},\Y) rectangle ({\B-2005},\Y+\h);
}

\end{tikzpicture}
\caption{Expanding window evaluation. \\
This figure illustrates the expanding-window procedure used for model evaluation. At each iteration, the available data are divided into three subsets: \rom{1} (training set), \rom{2} (validation set), and \rom{3} (test set). The training set expands over time, while the validation and test sets are fixed in length at two years and one year, respectively.}
\label{fig:expanding_window}
\end{figure}

\setcounter{table}{0}
\setcounter{figure}{0}
\section{Implementing Neural Network for Asset Pricing} \label{Appendix:nn}

\subsection{Activation functions}
Rectified Linear Unit (ReLU) is frequently chosen in various machine learning applications due to its computational simplicity and efficiency, as also evidenced by its usage in empirical asset pricing research employing deep learning architectures such as \textcite{gu2020empirical} and \textcite{chen2024deep}. In contrast to activations such as SoftMax or Sigmoid functions, ReLU has demonstrated comparable performance while offering faster computational speed. Moreover, ReLU is particularly valued for its ability to address the gradient vanishing problem by consistently producing non-negative gradients for positive inputs. ReLU deactivates all negative inputs by setting the value to zero. This characteristic of ReLU occasionally leads to the ``dying ReLU" problem, where majority of layers become deactivated during training. This phenomenon occurs when the gradient's absolute value is high or when the bias is significantly negative, causing the activation to become zero and remain stuck in that state as their values are not updated for the rest of the training time. While this issue rarely arises in other applications and may even be considered a strength of ReLU due to its support for sparse learning, dying ReLU can pose a serious challenge in empirical asset pricing since higher learning rates and batch sizes are often employed to facilitate convergence to the global optimum. The problem can be mitigated by introducing a small amount of gradients on the negative side to prevent neurons from becoming completely inactive, achieved through the use of functions such as LeakyReLU, Exponential Linear Units \citep[ELU, ][]{clevert2016fast} or Continuously Differentiable Exponential Linear Units \citep[CELU, ][]{barron2017continuously}. 

GELU serves as a notable example of such activation functions, while it is also the most widely adopted activation function in recent deep learning architectures due to its additional benefits. Firstly, GELU has a zero mean and unit variance for inputs drawn from a Gaussian distribution, ensuring the stability of activations and gradients throughout the network. Secondly, the GELU function is smooth and non-monotonic, which contributes to its stability during training while enabling it to capture more complex patterns in the data.

Therefore, we utilize Gaussian Error Linear Units function (GELU) as non-linearity of the neural network. The mathematical formulation of GELU is given as below.

$$
    \mathrm{GELU}(x)
=
x \cdot \mathbb{P}(X \le x)
=
x \cdot \Phi(x), 
$$
where $\Phi(x)$ the cumulative distribution function for Gaussian distribution $X\sim N(0,\sigma^2)$. GELU was first introduced by \textcite{hendrycks2016gaussian} as an alternative of Rectified linear units (ReLU) \parencite{nair2010rectified}. GELU permits some small interval for negative inputs to propagate through subsequent layers. 

\subsection{Detailed model specification}
The mathematical form of the model architecture is given as follows. First, let $X$ denote the input layer, and $H^{(1)}, H^{(2)}, \ldots, H^{(n)}$ represent the hidden layers. The weight matrices connecting the layers are denoted as $W_{0}, W_{1}, \ldots, W_{n}^c, W_{n}^r$,  where $W_{0}$ connects the input layer to the first hidden layer, $W_{1}$ connects the first hidden layer to the second hidden layer, and so forth, up to $W_{n}^c$ connecting the $n$-th hidden layer to the output layer of the consensus module, and $W_{n}^r$ connecting the output layer of the consensus module to the output layer of the return module. Similarly, the bias vectors are represented as $b_0, b_1, \ldots, b_n^c, b_n^r$. The computations for the hidden layers are as follows,
\[
\begin{aligned}
H^{(1)}
&=
\mathrm{GELU}\!\left(
W_{0}\!\left( \mathcal{I}^f \oplus \mathcal{I}^m \right)
+ b_{0}
\right),
\\
H^{(2)}
&=
\mathrm{GELU}\!\left(
W_{1} H^{(1)}
+ b_{1}
\right),
\\
&\;\vdots
\\
H^{(n)}
&=
\mathrm{GELU}\!\left(
W_{n-1} H^{(n-1)}
+ b_{\,n-1}
\right).
\end{aligned}
\]
Then the output layer computation of the consensus module is given by,
\[
f\!\left( \mathcal{I}^f,\, \mathcal{I}^m \,;\, \phi \right)
=
W^{c}_{n} \, H^{(n)}
+
b^{c}_{n}.
\]
and the output layer computation of the return module is given by,
\[
g\!\left(
f\!\left( \mathcal{I}^f,\, \mathcal{I}^m \,;\, \phi \right)
\,;\, \theta
\right)
=
W^{r}_{n} \,
f\!\left( \mathcal{I}^f,\, \mathcal{I}^m \,;\, \phi \right)
+
b^{r}_{n}.
\]
Note that there are no activation layers between the consensus and return modules for interpretability. Therefore, when $\lambda=0$, the CB-APM functions as a simple feedforward network, with the number of hidden layers matching that of the consensus module. The learnable weights of CB-APM are initialized by adopting the He initialization proposed by \textcite{he2015delving}.

\subsection{Stabilized learning}
Stabilizing neural network training is essential for robust estimation, particularly when deploying deep learning models in high-stakes decision-making contexts.  The following sections detail the regularization methods and machine learning techniques we employ to enhance the stability of the learning process.\footnote{Detailed parameter settings are provided in Table \ref{tab:hyper-setting}.} Despite that this topic is related to experimental factors rather than based on theoretical backgrounds, it is still worthy to discuss since that not only providing this information is crucial for reproducing the results, but also ensures a clear understanding of deep learning techniques.  This is particularly important in fields like financial economics, where such concepts may not yet be widely understood or adopted. As such, this discussion is not only relevant but vital for integrating advanced computational methods into financial economic research effectively. 

\textbf{Early stopping} is a regularization technique commonly used in deep learning algorithms to prevent overfitting and improve generalization performance. During the training process, the model's performance on a validation set is monitored after each epoch. If the validation performance starts to decline or no longer improves, training is halted early, preventing the model from further fitting to noise in the training data. By stopping training before the model becomes overfitted, early stopping helps to achieve better generalization performance on unseen data. Early stopping is a simple yet effective method for improving the robustness and generalization ability of deep learning models, particularly in empirical asset pricing where the amount of training data is limited. 

\textbf{Adaptive Moment Estimation (ADAM)} is an optimization algorithm proposed by \textcite{kingma2017adam}, which is commonly used in practice of training deep learning models. It combines ideas from both momentum optimization and RMSprop \parencite{hinton2012neural}, making it well-suited for optimizing non-convex objective functions commonly encountered in neural network training. ADAM maintains the first moment and the second moment of the gradients as two separate moving averages. These moving averages are used to adaptively update the parameters of the model during training. ADAM automatically adjusts the learning rate for each parameter based on the magnitude of the gradients and the accumulated past gradients, allowing it to converge quickly and efficiently in practice.

\textbf{Learning rate scheduling} is a technique used to dynamically adjust the learning rate during training to improve optimization performance. Instead of using a fixed learning rate throughout the training process, learning rate scheduling gradually decreases the learning rate over time, allowing the model to fine-tune its parameters more effectively as training progresses. By annealing the learning rate, learning rate scheduling helps to prevent the optimization process from getting stuck in local minima, which happens surprisingly often as discussed in the next section. Although ADAM inherently adjusts the learning rate, combining it with learning rate scheduling further optimizes performance by refining the initial learning rate. For training CB-APM, we use the ReduceLROnPlateau scheduler provided in PyTorch  (See the official document provided by PyTorch for more details \textcite{ReduceLROnPlateau}), which monitors validation performance and reduces the learning rate by a specified factor. 

\textbf{Gradient clipping} is a technique introduced by \textcite{pascanu2013difficulty} to prevent the exploding gradient problem during the training of neural networks. This problem arises when the gradients become too large, causing numerical instability and hindering the convergence of the optimization algorithm. Gradient clipping limits the magnitude of the gradients to a predefined threshold. By capping the gradient values, gradient clipping helps stabilize the training process and improves the convergence of the model, allowing for more stable and efficient training of neural networks.

\textbf{Ensemble learning} combines the predictions of multiple individual models to improve overall performance. Such learning scheme is inspired by tree-based models such as Random Forests, where outputs of multiple tree estimators are aggregated to generate final prediction results. The most common approach is to compute an average of model outputs, which is also adopted for this work. Specifically, in the case of CB-APM, ensemble learning is applied to both consensuses and individual stock returns. This entails training the entire model multiple times, with the final approximations of analysts' consensus and future returns derived as the average of each model's output. 

\textbf{Layer normalization} is a technique designed by \textcite{ba2016layer} that is used in deep learning to normalize the activations of neurons within each layer of a neural network. Unlike batch normalization \parencite{ioffe2015batch}, which normalizes across the entire batch of data, layer normalization computes the mean and standard deviation of the inputs along the hidden layer for each individual training example. This normalization process ensures that the activations of neurons have a mean of zero and a standard deviation of one, which helps stabilize the training process and accelerates convergence. 

\textbf{Dropout} is a stochastic regularization technique introduced by \textcite{hinton2012improving}  that helps prevent overfitting in neural networks by randomly setting a proportion of neurons to zero during each training iteration. This dropout process effectively removes certain connections between neurons, forcing the network to learn more robust and generalizable features. During training, dropout is applied to the input and hidden layers of a neural network with a specified dropout probability, and each neuron in the selected layers is randomly dropped out with the specified probability. Dropout is applied independently to each training example, ensuring that different subsets of neurons are dropped out during each iteration. By randomly dropping out neurons, dropout prevents the network from relying too heavily on any individual neuron or feature, forcing it to learn more redundant representations. During inference, dropout is turned off, and the full network is used to make predictions. However, the weights of the network are usually scaled down by the dropout probability at inference time to account for the increased number of active neurons during training.

\subsection{Model hyperparameters} \label{Appendix:hyper}
A proper hyperparameter setting is well known to be a key for getting successful performance results in various machine learning applications \parencite{feurer2019hyperparameter}. The most common approach is a process called ``hyperparameter optimization", where optimal hyperparameters are chosen from a candidate set automatically by solving an optimization problem of either minimizing or maximizing validation metric. 
However, this kind of optimization approach can cause specific problems in cross-sectional asset pricing. 

In numerous applications involving regression problems, the mean squared error (MSE) is commonly selected as the objective function for hyperparameter tuning because it directly measures the model's predictive accuracy. Yet, in empirical asset pricing, achieving precise predictions of future returns is universally recognized as a pipe dream. Consequently, researchers in this field often use alternative metrics to evaluate the performance of asset pricing models. For instance, \textcite{kelly2024virtue} demonstrate that predictive models can exhibit negative $R^2$ values yet still deliver positive Sharpe ratios in long-short portfolios. This finding encourages the prioritization of portfolio performance metrics over traditional regression metrics. 

Because that reachable level of positive out-of-sample $R^2$ is nearly 0\%, which is significantly low compared to other prediction tasks, models can fall into the trap of converging towards the historical mean, a well-documented local optimum. \textcite{welch2008comprehensive} empirically show that simply taking an average of excess returns can beat regression models with predictive factors. Although the promising developments in employing various factors and modeling approaches over the years of research, such alternative solution can still take over the predictive accuracy of complex models, depending on the model hyperparameter settings and the chosen validation data window. Moreover, even in scenarios where using the historical mean as an expected return might appear statistically optimal, such results hold limited practical economic value, since we cannot apply the results to the investment strategies directly. 

Given these considerations, designing an appropriate hyperparameter optimization problem for asset pricing is important, which we leave it as an attractive and also challenging future research topic. In this paper, we rather summarize the hyperparameter ``choice" that produces reasonable results as practical guidelines for finding a rational setting. The list of all the hyperparameters under consideration is provided in Table \ref{tab:hyper-setting}. 

\setlength\LTleft{0pt}
\setlength\LTright{0pt}
\captionsetup{justification=raggedright,singlelinecheck=false}
\begin{small}
\begin{longtable}{@{\extracolsep{\fill}} lp{9cm}l}
    \caption{Hyperparameter settings for CB-APM and Autoencoder.}
    \label{tab:hyper-setting} \\
    \toprule
    \multicolumn{1}{l}{\small Hyperparameter} &
    \multicolumn{1}{l}{\small Description} &
    \multicolumn{1}{l}{\small Setting} \\
    \midrule
    \endfirsthead

    \caption[]{Hyperparameter settings for CB-APM and Autoencoder (cont'd).} \\
    \toprule
    \multicolumn{1}{l}{\small Hyperparameter} &
    \multicolumn{1}{l}{\small Description} &
    \multicolumn{1}{l}{\small Setting} \\
    \midrule
    \endhead

    \midrule
    \multicolumn{3}{r}{\textit{(cont'd on next page)}}\\
    \midrule
    \endfoot

    \bottomrule
    \endlastfoot

    \multicolumn{3}{l}{\small \textbf{Panel A: CB-APM}} \\
    \addlinespace[4pt]

    \multicolumn{3}{l}{\small \emph{Model}} \\
    \# hidden layers   & Number of hidden layers in consensus module & 2 \\
    \# nodes           & Nodes per hidden layer in consensus module  & 64, 32 \\
    Ensemble size      & Number of models used for ensembling        & 10 \\
    \addlinespace[6pt]

    \multicolumn{3}{l}{\small \emph{Learning}} \\
    Batch size            & Mini-batch size for stochastic optimization        & 5{,}000 \\
    Learning rate         & Initial step size (Adam optimizer)                 & 0.001 \\
    Weight decay          & $\ell_2$ penalty (Adam)                            & 0.005 \\
    \addlinespace[6pt]

    \multicolumn{3}{l}{\small \emph{Scheduling}} \\
    Scheduler patience    & Epochs without val. improvement before LR decay    & 2 \\
    Scheduler factor      & Multiplicative LR decay factor                     & 0.2 \\
    \addlinespace[6pt]

    \multicolumn{3}{l}{\small \emph{Regularization}} \\
    Early stopping patience & Epochs without val. improvement before stopping  & 5 \\
    Gradient clip value     & Max absolute gradient (global clipping)          & 1.0 \\
    Dropout probability     & Dropout probability per linear layer             & 0.5 \\
    \addlinespace[8pt]

    \multicolumn{3}{l}{\small \textbf{Panel B: Autoencoder}} \\
    \addlinespace[4pt]

    \multicolumn{3}{l}{\small \emph{Model}} \\
    \# hidden layers      & Hidden layers in encoder and decoder                     & 2 \\
    \# nodes              & Nodes per hidden layer in encoder and decoder            & 128, 64 \\
    Latent dimension      & Dimension of macro latent state $\mathbf{z}_t$           & 32 \\
    \addlinespace[6pt]

    \multicolumn{3}{l}{\small \emph{Learning}} \\
    Batch size            & Mini-batch size                                        & 1 \\
    Learning rate         & Initial step size (Adam optimizer)                     & 0.00005 \\
    \addlinespace[6pt]

    \multicolumn{3}{l}{\small \emph{Regularization}} \\
    Early stopping patience & Epochs without val. improvement before stopping      & 2500 \\
    Dropout probability     & Dropout probability per linear layer                 & 0.2 \\
\end{longtable}
\end{small}

Although various hyperparameters contribute to predictive performance, the CB-APM framework generally exhibits robustness to modest changes in most settings. By contrast, the choice of batch size is markedly more influential for model performance in our application. This distinction arises from two structural differences in the data used by each component. First, the CB-APM consensus module is trained on high-dimensional panel data with a large cross-sectional dimension (over 600{,}000 firm-monthly observations), making a relatively large batch size (5{,}000) computationally efficient while ensuring stable gradient estimates. In contrast, the autoencoder is trained on macroeconomic variables with a very limited number of monthly observations (fewer than 1{,}000 in total), which more closely resemble low-frequency time-series data. Regarding that the autoencoder does not explicitly exploit temporal dependence, the scarcity of observations motivates a batch size of 1, effectively adopting a stochastic gradient regime that maximizes the diversity of parameter updates. These settings reflect the interaction between data structure (panel versus macroeconomic series) and data availability, and they are critical for achieving stable training dynamics and avoiding overfitting in each model.

\setcounter{table}{0}
\setcounter{figure}{0}
\section{Additional Results} \label{Appendix:additional}
\subsection{Forecast horizon analysis} \label{Appendix:horizon}

\begin{sidewaystable}[htbp]
        \centering
        \small
        \renewcommand{\arraystretch}{1.3}
        \setlength{\tabcolsep}{6pt}
        \newcolumntype{Y}{>{\centering\arraybackslash}X}
        
        \caption{Out-of-Sample $R^2$ for Stock Return and Consensus Approximations. \\
        This table reports monthly $R^2 (\%)$ of monthly stock return estimation and analysts' consensus variable approximation over the entire evaluation sets for different $\lambda$ settings. }
        \label{tab:monthly}
        
        \begin{tabular}{
            S[table-format=1.3] 
            S[table-format=-2.2] 
            S[table-format=-2.2] 
            S[table-format=-2.2] 
            S[table-format=-2.2] 
            S[table-format=2.2]  
            S[table-format=2.2]  
            S[table-format=2.2]  
            S[table-format=2.2]  
            S[table-format=2.2]  
            S[table-format=2.2]  
            S[table-format=1.2]  
        }
            \toprule
            & \multicolumn{9}{c}{\small Consensus Variables} 
            & \multicolumn{2}{c}{\small Overall Results} \\
            \cmidrule(lr){2-10} \cmidrule(lr){11-12}
            {$\lambda$} & {\small \makecell{EPS \\ Forecast \\ Revision}} & {\small \makecell{Change \\ in \\ Rec-\\ommend-\\ation}} & {\small \makecell{Change \\in \\ Forecast \\ \& \\ Accrual}} & {\small \makecell{Long \\ vs \\ short \\ EPS \\ Forecasts}} & {\small \makecell{Analyst \\ Earnings \\ per \\ Share}} & {\small \makecell{EPS \\ Forecast \\ Dispersion}} & {\small \makecell{ Earnings \\ Forecast \\ Revisions}} & {\small \makecell{Analyst \\ Value}} & {\small \makecell{Analyst \\ Optimism}} & {\small \makecell{Consensus \\ Average}} & {\small \makecell{Stock \\ Returns}} \\
            \midrule
            0  &  {-} & {-} & {-} & {-} & {-} & {-} & {-} & {-} & {-} & {-} & 0.72 \\
            0.1  &  4.17 & -0.01 &  3.86 &  7.78 & 59.12 & 30.98 & 10.62 & 26.77 & 28.65 & 19.10  & 0.32 \\
            0.2  &  4.89 &  0.05 &  4.65 &  9.09 & 68.94 & 37.31 & 13.96 & 33.20 & 35.40 & 23.05  & 0.40 \\
            0.3  &  5.19 &  0.08 &  5.00 &  9.41 & 73.01 & 39.99 & 15.52 & 36.27 & 38.20 & 24.74  & 0.41 \\
            0.4  &  5.34 &  0.09 &  5.17 &  9.66 & 75.35 & 41.42 & 16.50 & 38.15 & 40.05 & 25.75  & 0.40 \\
            0.5  &  5.39 &  0.10 &  5.21 &  9.81 & 76.81 & 42.62 & 17.26 & 39.64 & 41.26 & 26.46  & 0.37 \\
            0.6  &  5.42 &  0.10 &  5.26 &  9.92 & 77.82 & 43.47 & 17.82 & 40.68 & 42.19 & 26.96  & 0.24 \\
            0.7  &  5.42 &  0.09 &  5.25 &  9.89 & 78.50 & 44.09 & 18.31 & 41.52 & 42.84 & 27.32  & 0.16 \\
            0.8  &  5.42 &  0.09 &  5.26 &  9.95 & 79.06 & 44.62 & 18.66 & 42.16 & 43.34 & 27.62  & 0.06 \\
            0.9  &  5.47 &  0.07 &  5.27 & 10.04 & 79.50 & 44.98 & 18.97 & 42.65 & 43.83 & 27.86  & 0.02 \\
            1.0  &  5.46 &  0.07 &  5.27 & 10.07 & 79.81 & 45.34 & 19.27 & 43.06 & 44.20 & 28.06  & -0.03 \\
            \bottomrule
        \end{tabular}
    \begin{flushleft}
    {\textit{Note:} Consensus average is calculated as the mean of consensus variable $R^2$.}
    \end{flushleft}
\end{sidewaystable}

\begin{sidewaystable}[htbp]
        \centering
        \small
        \renewcommand{\arraystretch}{1.3}
        \setlength{\tabcolsep}{6pt}
        \newcolumntype{Y}{>{\centering\arraybackslash}X}
        \caption{Out-of-Sample $R^2$ for Stock Return and Consensus Approximations. \\
        This table reports monthly $R^2 (\%)$ of quarterly stock return estimation and analysts' consensus variable approximation over the entire evaluation sets for different $\lambda$ settings.}
        \label{tab:quarterly}

        \begin{tabular}{
            S[table-format=1.3] 
            S[table-format=-2.2] 
            S[table-format=-2.2] 
            S[table-format=-2.2] 
            S[table-format=-2.2] 
            S[table-format=2.2]  
            S[table-format=2.2]  
            S[table-format=2.2]  
            S[table-format=2.2]  
            S[table-format=2.2]  
            S[table-format=2.2]  
            S[table-format=1.2]  
        }
            \toprule
            & \multicolumn{9}{c}{\small Consensus Variables} 
            & \multicolumn{2}{c}{\small Overall Results} \\
            \cmidrule(lr){2-10} \cmidrule(lr){11-12}
            {$\lambda$} & {\small \makecell{EPS \\ Forecast \\ Revision}} & {\small \makecell{Change \\ in \\ Rec-\\ommend-\\ation}} & {\small \makecell{Change \\in \\ Forecast \\ \& \\ Accrual}} & {\small \makecell{Long \\ vs \\ short \\ EPS \\ Forecasts}} & {\small \makecell{Analyst \\ Earnings \\ per \\ Share}} & {\small \makecell{EPS \\ Forecast \\ Dispersion}} & {\small \makecell{ Earnings \\ Forecast \\ Revisions}} & {\small \makecell{Analyst \\ Value}} & {\small \makecell{Analyst \\ Optimism}} & {\small \makecell{Consensus \\ Average}} & {\small \makecell{Stock \\ Returns}} \\
            \midrule
            0  & {-} & {-} & {-} & {-} & {-} & {-} & {-} & {-} & {-} & {-} & 3.37 \\
            0.1  &  2.73 & -0.05 &   2.51 &  5.22 & 43.70 & 22.34 &  7.44 & 18.61 & 20.37 & 13.65    & 3.68 \\
            0.2  &  4.12 & -0.06 &   3.78 &  8.03 & 62.85 & 33.52 & 12.02 & 28.67 & 30.80 & 20.41    & 3.49 \\
            0.3  &  4.60 & -0.02 &   4.36 &  8.90 & 68.32 & 36.79 & 14.21 & 32.33 & 34.58 & 22.67    & 3.35 \\
            0.4  &  4.83 & -0.01 &   4.64 &  9.33 & 71.16 & 38.54 & 15.44 & 34.53 & 36.82 & 23.92    & 3.39 \\
            0.5  &  5.03 &  0.01 &   4.90 &  9.65 & 73.37 & 39.88 & 16.29 & 36.36 & 38.43 & 24.88    & 3.66 \\
            0.6  &  5.20 &  0.02 &   5.08 &  9.88 & 74.87 & 40.95 & 16.91 & 37.73 & 39.61 & 25.58    & 3.86 \\
            0.7  &  5.30 &  0.05 &   5.21 &  9.99 & 75.87 & 41.70 & 17.43 & 38.71 & 40.53 & 26.09    & 4.01 \\
            0.8  &  5.36 &  0.07 &   5.28 & 10.05 & 76.64 & 42.26 & 17.85 & 39.52 & 41.21 & 26.47    & 3.97 \\
            0.9  &  5.44 &  0.06 &   5.34 & 10.12 & 77.37 & 42.80 & 18.20 & 40.36 & 41.99 & 26.85    & 3.84 \\
            1.0  &  5.48 &  0.07 &   5.39 & 10.17 & 77.94 & 43.35 & 18.52 & 41.05 & 42.51 & 27.16    & 3.95 \\
            \bottomrule
        \end{tabular}
    \begin{flushleft}
    {\textit{Note:} Results are reported for a sampled subset of $\lambda$ settings due to redundancy.}
    \end{flushleft}
\end{sidewaystable}

\begin{sidewaystable}[htbp]
        \centering
        \small
        \renewcommand{\arraystretch}{1.3}
        \setlength{\tabcolsep}{6pt}
        \newcolumntype{Y}{>{\centering\arraybackslash}X}
        
        \caption{Out-of-Sample $R^2$ for Stock Return and Consensus Approximations. \\
        This table reports monthly $R^2 (\%)$ of semiannual stock return estimation and analysts' consensus variable approximation over the entire evaluation sets for different $\lambda$ settings. }
        \label{tab:semiannual}

        \begin{tabular}{
            S[table-format=1.3] 
            S[table-format=-2.2] 
            S[table-format=-2.2] 
            S[table-format=-2.2] 
            S[table-format=-2.2] 
            S[table-format=2.2]  
            S[table-format=2.2]  
            S[table-format=2.2]  
            S[table-format=2.2]  
            S[table-format=2.2]  
            S[table-format=2.2]  
            S[table-format=1.2]  
        }
            \toprule
            & \multicolumn{9}{c}{\small Consensus Variables} 
            & \multicolumn{2}{c}{\small Overall Results} \\
            \cmidrule(lr){2-10} \cmidrule(lr){11-12}
            {$\lambda$} & {\small \makecell{EPS \\ Forecast \\ Revision}} & {\small \makecell{Change \\ in \\ Rec-\\ommend-\\ation}} & {\small \makecell{Change \\in \\ Forecast \\ \& \\ Accrual}} & {\small \makecell{Long \\ vs \\ short \\ EPS \\ Forecasts}} & {\small \makecell{Analyst \\ Earnings \\ per \\ Share}} & {\small \makecell{EPS \\ Forecast \\ Dispersion}} & {\small \makecell{ Earnings \\ Forecast \\ Revisions}} & {\small \makecell{Analyst \\ Value}} & {\small \makecell{Analyst \\ Optimism}} & {\small \makecell{Consensus \\ Average}} & {\small \makecell{Stock \\ Returns}} \\
            \midrule
            0  & {-} & {-} & {-} & {-} & {-} & {-} & {-} & {-} & {-} & {-} & 4.34 \\
            0.1  &  1.77 & -0.12 &   1.95 &  4.02 & 29.53 & 15.52 &   6.17 &  12.53 & 13.59 &  9.44    & 6.87 \\
            0.2  &  3.12 & -0.10 &   3.09 &  6.33 & 50.97 & 27.36 &   9.42 &  22.19 & 24.40 & 16.31    & 6.91 \\
            0.3  &  3.89 & -0.08 &   3.67 &  7.64 & 60.95 & 32.78 &  11.98 &  27.60 & 29.94 & 19.82    & 6.56 \\
            0.4  &  4.34 & -0.08 &   4.09 &  8.51 & 66.32 & 35.91 &  13.73 &  30.92 & 33.20 & 21.88    & 6.31 \\
            0.5  &  4.60 & -0.05 &   4.38 &  8.94 & 69.33 & 37.58 &  14.82 &  33.12 & 35.20 & 23.10    & 6.19 \\
            0.6  &  4.79 & -0.07 &   4.62 &  9.17 & 71.53 & 38.80 &  15.69 &  34.89 & 36.77 & 24.02    & 6.26 \\
            0.7  &  4.95 & -0.05 &   4.83 &  9.34 & 73.20 & 39.84 &  16.34 &  36.21 & 38.05 & 24.75    & 6.18 \\
            0.8  &  5.11 & -0.05 &   4.99 &  9.53 & 74.42 & 40.56 &  16.90 &  37.27 & 39.08 & 25.31    & 5.99 \\
            0.9  &  5.21 & -0.04 &   5.10 &  9.66 & 75.42 & 41.23 &  17.31 &  38.20 & 39.92 & 25.78    & 5.93 \\
            1.0  &  5.32 & -0.02 &   5.18 &  9.76 & 76.15 & 41.67 &  17.75 &  38.89 & 40.56 & 26.14    & 5.95 \\
            \bottomrule
        \end{tabular}
    \begin{flushleft}
    {\textit{Note:} Results are reported for a sampled subset of $\lambda$ settings due to redundancy.}
    \end{flushleft}
\end{sidewaystable}

\begin{figure}[htbp]
    \centering
    \begin{subfigure}{\columnwidth}
        \centering
        \includegraphics[width=\columnwidth]{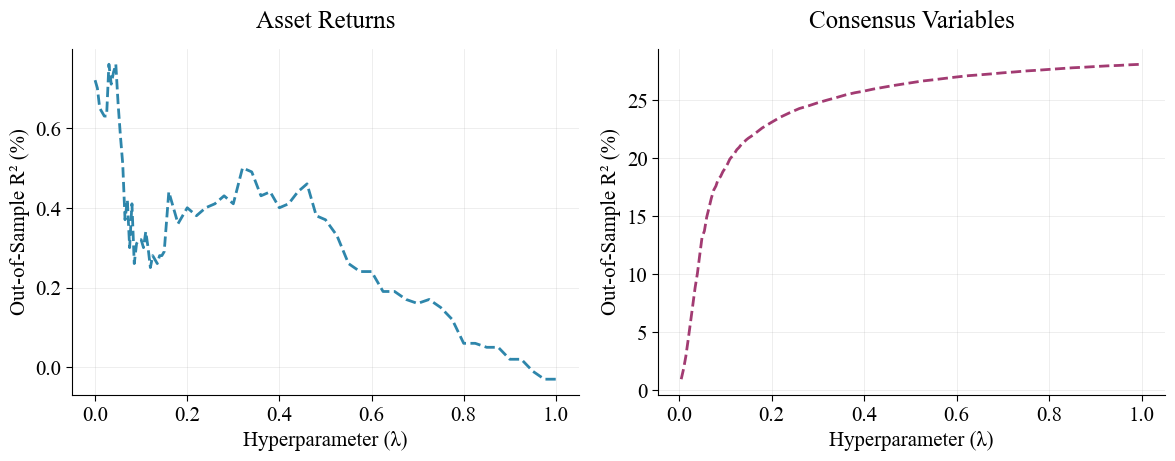}
        \captionsetup{justification=centering}
        \caption{Monthly horizon}
        \label{fig:r2-1month}
    \end{subfigure}
    
    \begin{subfigure}{\columnwidth}
        \centering
        \includegraphics[width=\columnwidth]{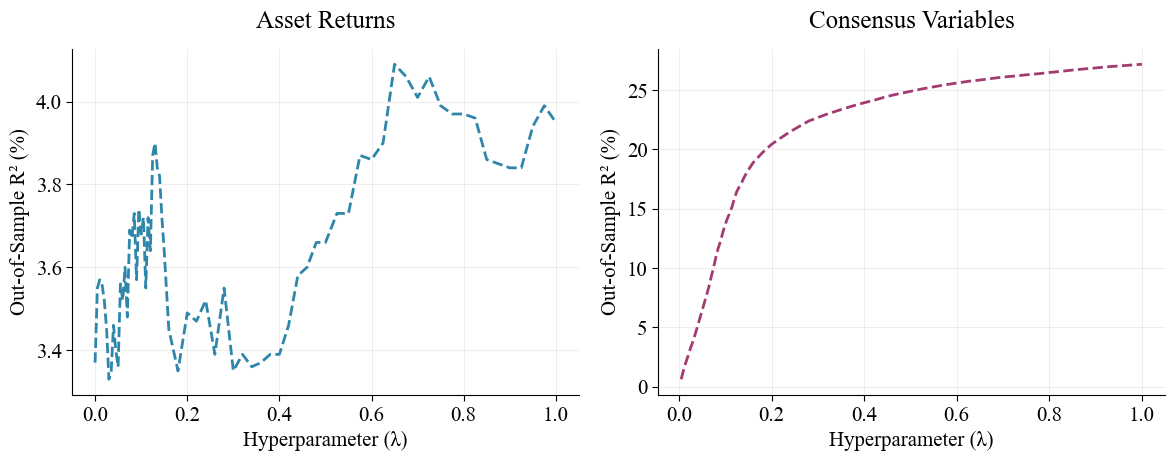}
        \captionsetup{justification=centering}
        \caption{Quarterly horizon}
        \label{fig:r2-3month}
    \end{subfigure}
    
    \begin{subfigure}{\columnwidth}
        \centering
        \includegraphics[width=\columnwidth]{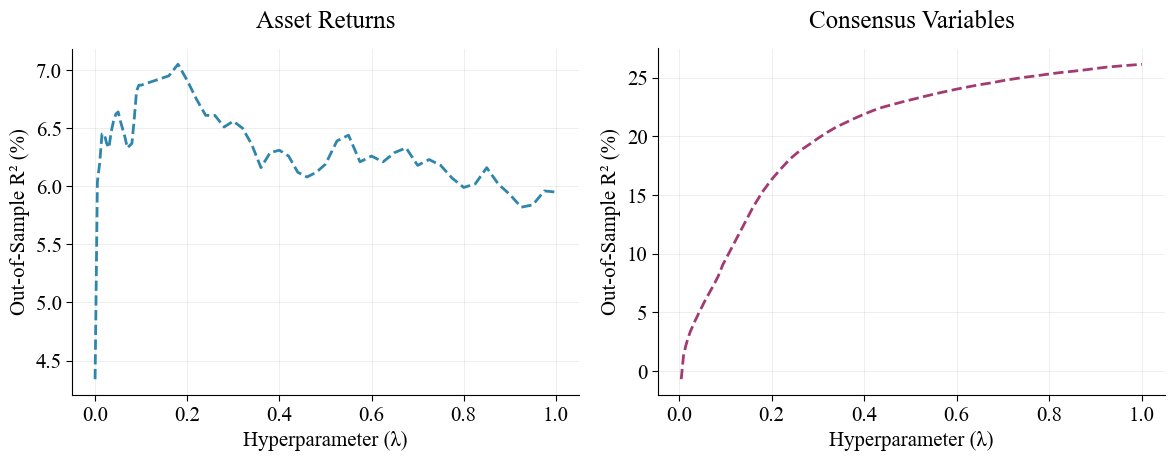}
        \captionsetup{justification=centering}
        \caption{Semiannual horizon}
        \label{fig:r2-6month}
    \end{subfigure}
    
    \caption{Out-of-sample $R^2$ of return predictions and consensus approximations. \\
    Panels (a), (b), and (c) present results for the monthly, quarterly, and semiannual returns respectively. }
    \label{fig:r2-horizon}
\end{figure}

To complement the primary analysis of annual return prediction, we extend our empirical evaluation of CB-APM to alternative horizons, including monthly, quarterly, and semiannual forecasts. This exercise serves two purposes, first, to investigate the model's robustness across varying temporal horizons, and second, to examine whether the observed interpretability-accuracy amplification at long horizons persists in shorter-term predictions.

Tables \ref{tab:monthly}--\ref{tab:semiannual} and Figures \ref{fig:r2-1month}--\ref{fig:r2-6month} present the out-of-sample $R^2$ results and their period-by-period decomposition for these horizons. For shorter horizons such as one month, return predictability remains marginal, with $R^2$ values close to zero and even slightly negative at higher values of $\lambda$, consistent with the well-documented difficulty of forecasting near-term returns. In this regime, increasing $\lambda$ intensifies the interpretability-accuracy trade-off: while consensus approximation improves monotonically, return $R^2$ declines, suggesting that allocating more weight to consensus modeling diverts representational capacity away from short-horizon return-specific signals.

In contrast, quarterly and semiannual horizons exhibit intermediate behavior between the monthly and annual cases. For these horizons, the inclusion of consensus learning yields positive predictive gains without incurring the sharp performance penalty seen in monthly forecasting. Notably, the semiannual horizon begins to display a pattern closer to that of the annual horizon, with joint optimization reinforcing both consensus approximation and return predictability.

These results collectively underscore the horizon-dependent effectiveness of CB-APM. At longer horizons (semiannual and annual), the integration of consensus learning acts as an economically grounded regularizer, anchoring predictions to persistent, macro-fundamental drivers that dominate long-term returns. By contrast, for near-term horizons dominated by transitory noise and market microstructure effects, interpretability constraints impose structural rigidity that impairs predictive accuracy. This divergence aligns with the theoretical intuition that analysts' consensus reflects slow-moving fundamentals, making it more complementary to long-horizon forecasting than to short-term return prediction.

From a practical standpoint, this evidence suggests that CB-APM is particularly well-suited for medium- to long-term investment horizons, where its interpretable architecture not only improves accuracy but also aligns predictions with economically meaningful signals. Conversely, for short-term horizons, where price dynamics are less tied to fundamentals, purely data-driven models may retain an edge in capturing relatively high-frequency fluctuations.

\subsection{Properties of the joint optimization} \label{Appendix:optim}

Given that CB-APM is trained using a joint loss function as defined in equation \eqref{eq:loss}, a weighted sum of return prediction loss and consensus approximation loss, it is essential to verify that the model is learning in line with its design. While out-of-sample $R^2$ is the primary metric for evaluating forecasting performance, it does not reveal how the model balances its dual objectives during training or whether the intended interaction between predictive accuracy and interpretability materializes. In particular, because CB-APM explicitly incorporates a hyperparameter $\lambda$ to control the trade-off between these two objectives, examining the in-sample MSE dynamics is crucial for understanding how different $\lambda$ settings shape the model's optimization behavior. This analysis is especially important in our context, as results for the annual forecasting horizon (Section \ref{sec:cross-section}) suggest that CB-APM may improve both interpretability and performance simultaneously, deviating from the classical interpretability-accuracy trade-off often documented in machine learning applications, as described in \textcite{koh2020concept}.

\begin{figure}[htbp]
    \centering
    \begin{subfigure}{\columnwidth}
        \centering
        \includegraphics[width=\columnwidth]{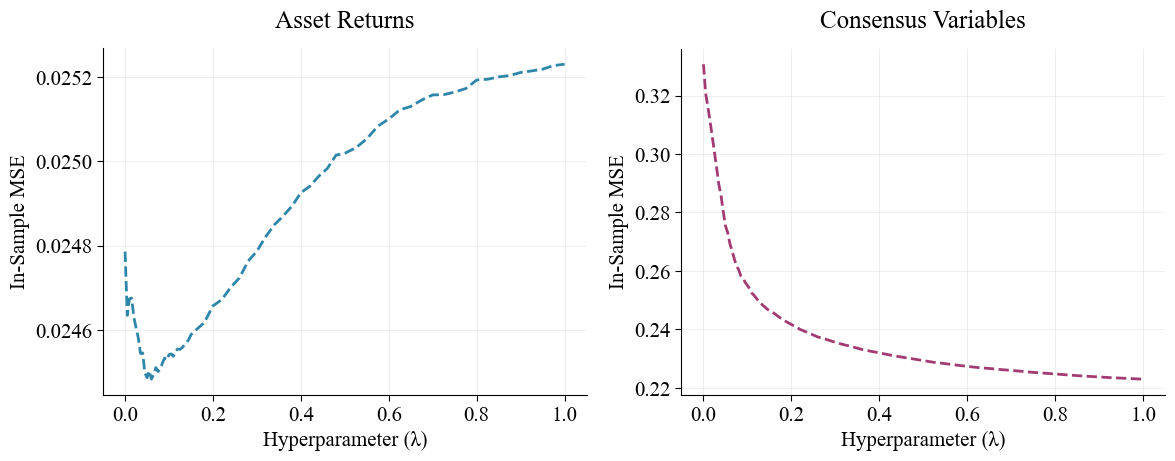}
        \caption{Monthly horizon}
        \label{fig:mse_is_1month}
    \end{subfigure}
    
    \vspace{0.5cm} 
    
    \begin{subfigure}{\columnwidth}
        \centering
        \includegraphics[width=\columnwidth]{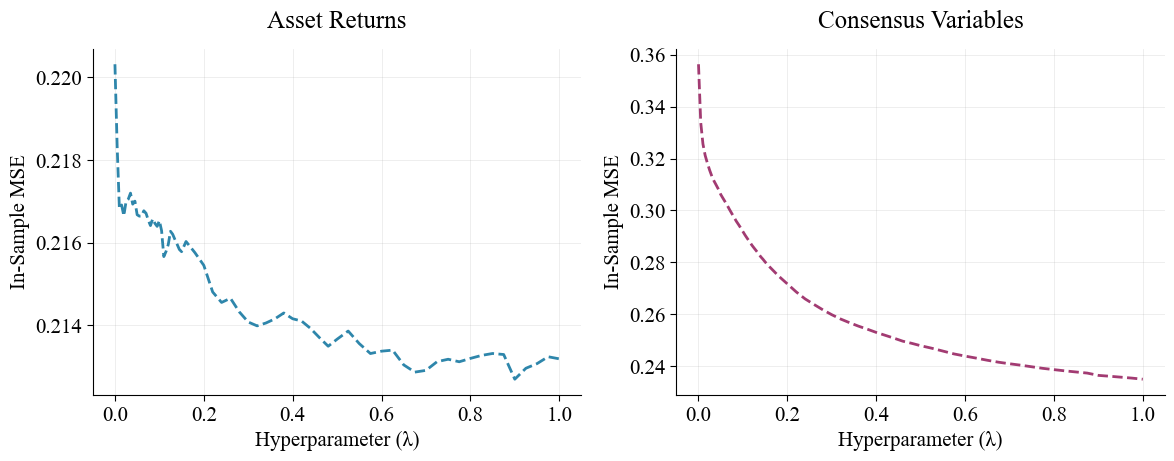}
        \caption{Annual horizon}
        \label{fig:mse_is_12month}
    \end{subfigure}
    
    \caption{In-sample MSE of return and consensus approximations. \\
    This figure plots the in-sample MSE of stock return (left) and consensus approximation (right) for different $\lambda$ settings. Panel (a) reports results for the monthly return, and Panel (b) for the annual return. These plots illustrate how forecasting horizons and $\lambda$ values govern the trade-off between predictive accuracy and consensus reconstruction in the joint loss function.}

    \label{fig:in-sample-mse}
\end{figure}

Figure \ref{fig:in-sample-mse} illustrates the in-sample MSE dynamics of the CB-APM under varying values of the hyperparameter $\lambda$, separately for monthly and annual forecasting horizons. Each panel presents two curves: the left axis depicts the in-sample MSE of stock return predictions, while the right axis reports the average in-sample MSE for consensus variable approximation. The first panel corresponds to the monthly horizon (Figure \ref{fig:mse_is_1month}), and the second panel presents the results for the annual horizon (Figure \ref{fig:mse_is_12month}).

The observed patterns reveal a striking divergence between the two horizons. For monthly returns, the stock return MSE exhibits a U-shaped trajectory: it initially decreases slightly for small values of $\lambda$ but increases steadily thereafter, suggesting a trade-off between return prediction accuracy and consensus approximation performance. This pattern aligns with the theoretical role of $\lambda$ in the loss function in equation \eqref{eq:loss}, which explicitly prioritizes consensus approximation as its value increases. Placing greater weight on $\mathcal{L}_C$ directs more representational capacity of the network toward modeling analyst consensus at the expense of direct return prediction. This is consistent with the interpretability-accuracy trade-off widely documented in the interpretable machine learning literature (e.g., \citealp{rudin2019stop}), where models constrained to capture auxiliary structure or explanatory variables tend to sacrifice marginal predictive performance in favor of enhanced interpretability or alignment with economic reasoning.

In sharp contrast, the annual horizon exhibits what we term an interpretability-accuracy amplification effect. Here, increasing $\lambda$ monotonically reduces the in-sample return MSE, even as the consensus approximation error steadily improves. Rather than trading off predictive accuracy for interpretability, joint learning of consensus variables appears to reinforce the return prediction objective at longer horizons. This result is particularly noteworthy in the context of financial forecasting, where long-horizon returns are notoriously noisy and difficult to predict using traditional methods. The amplification effect implies that, for CB-APM, jointly learning analyst expectations, serving as a structured, economically meaningful regularizer, can improve the model's capacity to extract signal for long-horizon returns.

This divergence between short- and long-horizon dynamics underscores an important methodological implication of interpretable neural networks in finance. For short-horizon return prediction, forcing the model to align with analyst consensus imposes additional structure that constrains flexibility, thereby introducing a predictable accuracy penalty. However, at longer horizons, the alignment between professional analyst forecasts and fundamental asset value drivers becomes more pronounced, such that the inclusion of consensus loss improves return prediction by anchoring the learning process on more persistent, macroeconomically relevant signals. This finding suggests that interpretable architectures such as CB-APM may be particularly well-suited for applications where the economic rationale underlying predictions is inherently long-term, a domain where conventional ``black-box" approaches often fail to yield stable or economically meaningful forecasts.

From a broader perspective, these results provide empirical evidence that interpretability and predictive accuracy in financial neural networks need not be inherently conflicting objectives. Their relationship depends critically on the forecast horizon and on the economic structure embedded in the auxiliary interpretable signals: at short horizons the two objectives compete for representational capacity, whereas at longer horizons economically grounded auxiliary targets appear to regularize the return-prediction task rather than constrain it. The amplification effect documented in annual return forecasts illustrates this mechanism and, more broadly, suggests that interpretability constraints anchored in economically meaningful concepts can contribute positively to out-of-sample performance. This perspective opens avenues for research on financially grounded, interpretable deep learning models that leverage economically motivated auxiliary tasks to improve both transparency and forecasting efficacy.

\subsection{Structure of the learned macroeconomic representations} \label{Appendix:autoencoder-representation}

While CB-APM achieves interpretability primarily through its consensus-bottleneck, it also relies on macroeconomic embeddings learned by an autoencoder as part of its input structure. Because these embeddings are learned in an unsupervised manner and directly influence return prediction, it is critical to empirically verify that they capture meaningful economic structure rather than spurious patterns. Thus, this section focuses on analyzing the autoencoder's latent representation through visualization and dimensionality reduction techniques. This analysis does not aim to provide instance-level explanations of model predictions but instead validates that the latent macroeconomic state aligns with established business cycle dynamics. In doing so, we complement CB-APM's built-in interpretability with evidence that its macroeconomic component operates transparently and in an economically coherent manner.

\begin{figure}[htbp]
    \centering
    \includegraphics[width=12cm]{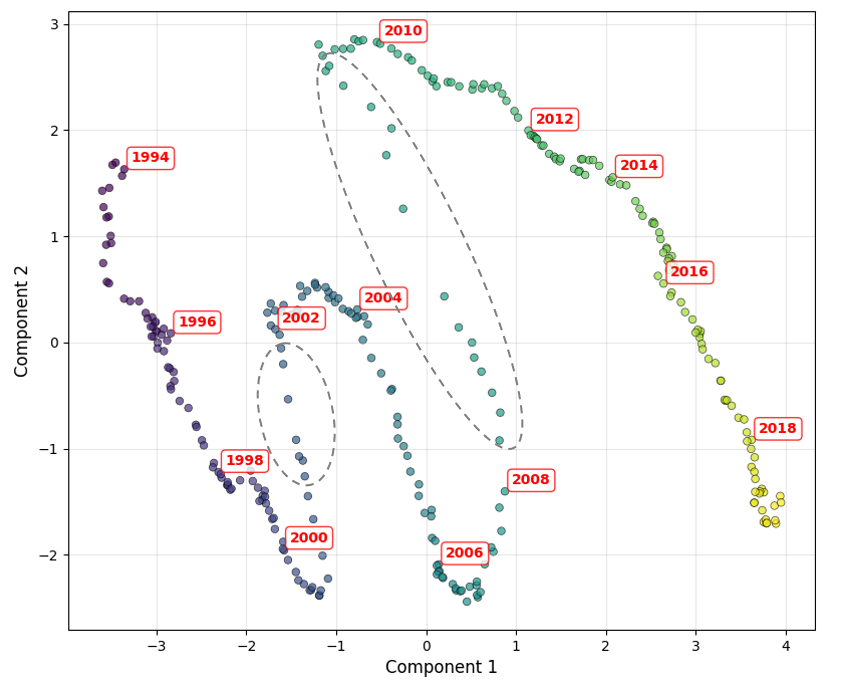}
    \caption{PCA projection of autoencoder latent state variables. \\ 
    PCA projection of in-sample 32-dimensional autoencoder latent state vectors into two dimensions, colored by month and annotated with red labels for January of select years. The grey dashed ovals mark NBER recession periods (2001Q1--Q4 and 2007Q4--2009Q2).}
    \label{fig:latent-vector}
\end{figure}

Figure \ref{fig:latent-vector} illustrates the two-dimensional principal component projection of the 32-dimensional latent state vectors produced by the macroeconomic autoencoder, color-coded by month and annotated with January observations for selected years. This visualization highlights how the autoencoder successfully encodes macroeconomic conditions into a smooth, low-dimensional manifold that evolves coherently over time. The trajectory of the latent vectors follows a clear temporal progression, demonstrating that the learned embedding captures the gradual transitions and structural shifts in the U.S. macroeconomic environment across the sample period.

A notable feature of this representation is its ability to distinguish major economic regimes. The grey dashed ovals in Figure \ref{fig:latent-vector} correspond to periods classified as recessions by the National Bureau of Economic Research (NBER); the early 2000s recession (2001Q1--Q4) and the Global Financial Crisis (2007Q4--2009Q2). During these intervals, the latent vectors exhibit marked departures from their preceding trajectories, forming clusters that are distinct from surrounding expansionary phases. This pattern indicates that the autoencoder embedding effectively internalizes macroeconomic shocks and regime shifts, producing representations that align with well-established business cycle chronologies without direct supervision from recession labels.

Beyond capturing these discrete regime shifts, the latent trajectory also reflects continuous macroeconomic evolution during non-recessionary periods. The progression from the early 1990s through the late 2010s shows a gradual unfolding in the latent space, with local curvature corresponding to cyclical fluctuations and persistent structural changes, such as those associated with the post-2008 recovery and subsequent expansion. This smooth temporal ordering suggests that the latent factors not only encode discrete downturns but also represent broader secular dynamics in economic conditions, including shifts in growth, inflation, and monetary policy regimes.

Recent deep factor models make it clear that neural networks can extract a parsimonious set of latent factors from high‐dimensional financial and macroeconomic data; these latent variables then drive improved asset pricing and predictive performance (see, for example, \citealp{feng2018deep, gu2021autoencoder, chen2024deep}). Our macroeconomic autoencoder performs a closely related function for aggregate time‐series data: it distills hundreds of macro indicators into a smooth latent trajectory that aligns with well‐known business‐cycle chronologies and regime shifts. While many prior studies emphasize quantitative performance metrics and offer only limited visual exploration of their latent factors, the clear temporal patterns in Figure~\ref{fig:latent-vector} demonstrate that CB‐APM's autoencoder uncovers economically meaningful state dynamics from complex data.

These findings validate the autoencoder's role in distilling high-dimensional macroeconomic data into an economically meaningful latent state. By learning unsupervised representations that exhibit both temporal coherence and sensitivity to regime changes, the model effectively embeds the prevailing macroeconomic environment into a compact form that can be integrated into return prediction. This latent structure provides a powerful mechanism for conditioning asset pricing on macroeconomic context: it transmits shared, time-varying information to the cross-section of firms while mitigating redundancy and noise inherent in raw macroeconomic predictors. Importantly, this approach aligns with our broader CB-APM framework by ensuring that firm-level predictions are informed by a parsimonious yet rich representation of the macro-financial backdrop.

\subsection{The explanatory fidelity of filtered expectations} \label{Appendix:linear_regression}

Once the consensus mapping is estimated, the subsequent linear relation
\[
\hat{R}_{i,t+h} = a + \boldsymbol{b}^\top \hat{C}_{i,t},
\]
can be analyzed using standard asset pricing tools. This formulation allows us to assess whether the CB-APM–implied consensus captures economically meaningful variation in expected returns.

To formalize this connection, we implement pooled panel regressions that relate future returns to either raw analyst consensus or CB-APM–implied consensus. These regressions are not intended as structural pricing tests; rather, they serve as diagnostic tools to compare the explanatory content of the learned consensus representations against the original analyst signals.

Table~\ref{tab:ols-consensus} reports pooled OLS regressions that relate future annual stock returns to either raw analyst consensus variables or CB-APM-implied consensus at \((\lambda = 1)\). We estimate the following pooled panel regression:

\begin{equation}
\label{eq:ols}
R_{i,t+h} = a^{(C)} + \boldsymbol{b}^{(C)\top} C_{i,t} + \varepsilon^{(C)}_{i,t+h}
\quad \text{or} \quad
R_{i,t+h} = a^{(\hat{C})} + \boldsymbol{b}^{(\hat{C})\top} \hat{C}_{i,t} + \varepsilon^{(\hat{C})}_{i,t+h}.
\end{equation}
For the analysis, we set $h=12$ to focus on annual return predictability, thereby aligning the return horizon with prior empirical studies discussed in the preceding sections. The model is estimated on the stacked cross-section of firm-month \((i,t)\) observations to obtain a time-invariant coefficient vector \(\widehat{\boldsymbol{b}}\). Inference is based on heteroskedasticity-robust covariance estimation tailored to the dependent variable’s structure. To handle an overlapping long-horizon return, we compute Driscoll--Kraay (kernel HAC) standard errors with a Bartlett kernel and an eleven-month bandwidth, which are robust to heteroskedasticity, cross-sectional dependence, and the serial correlation induced by overlapping observations \parencite{driscoll1998consistent, newey1986simple, hodrick1992dividend}.

Panel~A reports coefficient estimates, \(t\)-statistics, and a variable-level fit measure; Panel~B summarizes the intercept and overall adjusted \(R^2\). The comparison isolates the incremental explanatory content obtained when analyst information is first synthesized by CB-APM’s consensus-bottleneck and then mapped linearly into expected returns.

\begin{table}[htbp]
            \centering
            \small
            \renewcommand{\arraystretch}{1.3}
            \setlength{\tabcolsep}{6pt}

             \caption{OLS regressions with raw versus model-inferred consensus. \\
            This table reports pooled OLS regressions in which the dependent variable is the \emph{annual} stock return \(R_{i,t+12}\). We compare specifications that use raw analyst consensus variables to those that use CB-APM-inferred consensus estimates at ($\lambda=1$), evaluated on the longest training set from the expanding-window procedure. The CB-APM consensus corresponds to the averaged output of an ensemble of models. Panel~A reports coefficient estimates, $t$-statistics, and predictor-level $R^2$ for each variable, while Panel~B summarizes the intercept, its standard error, and the overall in-sample adjusted $R^2$.}
            \label{tab:ols-consensus}

            \textbf{Panel A. Coefficients and t-statistics} \\
            \vspace{0.2cm}
            \begin{tabular*}{\textwidth}{@{\extracolsep{\fill}} 
                l
                S[table-format=-.4]
                S[table-format=-.2]@{\hspace{-1.7em}}l
                S[table-format=-.2] 
                S[table-format=-.4]
                S[table-format=-.2]@{\hspace{-1.7em}}l
            }
                \toprule
                \multicolumn{1}{c}{} 
                & \multicolumn{3}{c}{\small Raw Consensus} 
                & \multicolumn{4}{c}{\small Approximated Consensus} \\
                \cmidrule(lr){2-4} \cmidrule(lr){5-8}
                {\small Variable} & {\small Coef.} & {\small t-stat.} & & {\small $R^2$ (\%)} & {\small Coef.} & {\small t-stat.} & \\
                \midrule
        EPS forecast revision & -0.0049 & -1.31 & \st{} & 4.97 & 0.1164 & 0.38 & \st{} \\
        Change in recommendation & 0.0307 & 11.54 & \st{***} & -0.16 & -3.9080 & -6.67 & \st{***} \\
        Change in Forecast and Accrual & 0.0398 & 8.57 & \st{***} & 4.62 & 0.3781 & 1.53 & \st{} \\
        Long-vs-short EPS forecasts & 0.0002 & 0.04 & \st{} & 9.12 & -0.0940 & -0.87 & \st{} \\
        Analyst earnings per share & -0.0240 & -1.63 & \st{} & 71.43 & 0.3174 & 4.46 & \st{***} \\
        EPS Forecast Dispersion & 0.0076 & 0.59 & \st{} & 39.06 & -0.6263 & -4.87 & \st{***} \\
        Earnings forecast revisions & -0.0068 & -0.62 & \st{} & 16.18 & -0.3328 & -1.82 & \st{*} \\
        Analyst Value & 0.0146 & 0.88 & \st{} & 35.45 & -0.1364 & -2.75 & \st{***} \\
        Analyst Optimism & 0.0216 & 2.37 & \st{**} & 37.24 & 0.1250 & 1.55 & \st{} \\
                \bottomrule
            \end{tabular*}

            \vspace{0.6cm}
        
    \textbf{Panel B. Summary statistics} \\
    \vspace{0.2cm}
    \begin{tabular*}{\textwidth}{@{\extracolsep{\fill}}l S[table-format=-2.4] S[table-format=-2.4]}
        \toprule
        & {\small Raw Consensus} & {\small Approximated Consensus} \\
        \midrule
        Intercept & -0.0064 & 0.0005 \\
        SE of Intercept & 0.0321 & 0.0159 \\
        In-Sample $adj\text{-}R^2$ (\%) & 0.40 & 8.35 \\
        \bottomrule
    \end{tabular*}
    
    \begin{flushleft}
    \textit{Note:} *** significance at the 1\% level; ** significance at the 5\% level; * significance at the 10\% level.
    Standard errors are computed using the Driscoll--Kraay (kernel HAC) estimator with a Bartlett kernel and an eleven-month bandwidth, robust to heteroskedasticity, cross-sectional dependence, and the serial correlation induced by overlapping returns.
    \end{flushleft}
\end{table}

The pooled OLS regression using raw analyst consensus variables yields limited explanatory power, with an adjusted \(R^2\) of just 0.40\%. Most predictors exhibit weak statistical significance; for example, \textit{EPS forecast revision} and \textit{Earnings forecast revisions} produce \(t\)-statistics of \(-1.31\) and \(-0.62\), respectively, with neither variable exhibiting meaningful predictive content. Although \textit{Change in recommendation} (\(t=11.54\)) and \textit{Change in Forecast and Accrual} (\(t=8.57\)) are statistically significant at the 1\% level, their estimated effects are modest in magnitude, and the overall model fit remains poor.

By contrast, the regression using CB-APM-inferred consensus achieves a substantially higher adjusted \(R^2\) of 8.35\%, representing more than a twentyfold improvement in explanatory power. A few key coefficients also reverse in sign relative to their raw counterparts. For instance, the coefficient on \textit{Change in recommendation} shifts from \(+0.0307\) (\(t=11.54\)) to \(-3.9080\) (\(t=-6.67\)), while \textit{EPS Forecast Dispersion} turns from \(+0.0076\) (\(t=0.59\)) to \(-0.6263\) (\(t=-4.87\)). In addition, \textit{Analyst earnings per share} becomes strongly positive and significant (\(t=4.46\)), whereas \textit{Analyst Value} becomes significantly negative (\(t=-2.75\)). These shifts suggest that CB-APM extracts transformed representations that encode economically distinct pricing content beyond the raw analyst signals.

While the CB-APM-inferred consensus variables deliver substantial improvements in explanatory power, caution is warranted in interpreting their individual coefficients. As shown in Table~\ref{tab:ols-consensus}, consensus variables with low predictor-level approximation \(R^2\) values often exhibit coefficient patterns that diverge from those estimated using raw consensus inputs. For example, \textit{Change in recommendation}, which has one of the lowest approximation \(R^2\) values, exhibits a pronounced sign reversal, while \textit{Change in Forecast and Accrual} weakens substantially in magnitude. This pattern highlights that the CB-APM approximations are not one-to-one reconstructions of analyst beliefs but rather encode transformed features with distinct pricing implications.

Consequently, interpretability must be grounded in a dual-lens framework. The consensus-level approximation \(R^2\), previously reported in Table~\ref{tab:annual}, reflects the degree to which a model-inferred variable aligns with its human-interpretable counterpart, whereas the coefficient estimate from the return regression captures the economic relevance of that signal. Coefficients associated with well-approximated variables are more directly interpretable as refinements of analyst expectations, whereas those tied to poorly reconstructed signals likely reflect alternative representations or re-weightings learned by the model. Thus, proper interpretation requires joint consideration of both approximation fidelity and return sensitivity rather than treating coefficients in isolation.

Importantly, this improvement follows directly from the design of the framework, which trains the approximated consensus layer under a joint objective that simultaneously targets return prediction and consensus reconstruction. By doing so, the model synthesizes information from a wide set of firm-level characteristics and macroeconomic variables into consensus features that retain risk-relevant content while reducing noise. Although the reported \(t\)-statistics primarily capture in-sample explanatory strength and are not intended for direct investment use due to inherent look-ahead bias, they underscore that the approximated consensus simultaneously explains both realized analyst consensus and future returns. This dual property makes the learned consensus features a rich source of information that merits closer examination beyond forecasting alone.

Taken together, the results indicate that CB-APM's consensus module extracts signals that are both more informative and more economically meaningful than raw analyst inputs. Although the reported regressions are estimated in-sample and do not directly measure out-of-sample predictive accuracy, they nonetheless support the model’s central objective: to learn interpretable latent representations that jointly capture analyst expectations and priced return variation. Proper interpretation of these results requires concurrent evaluation of (i) approximation \(R^2\), which gauges the alignment between model-inferred signals and observable analyst variables, and (ii) coefficient sign and magnitude, which reflect the economic relevance of each signal for cross-sectional return prediction.

\subsection{Further robustness checks}
\subsubsection{Sensitivity of autoencoder performance to latent dimensionality} \label{Appendix:latent-dim}

An additional robustness check examines the sensitivity of CB-APM's performance to the choice of latent dimension in the autoencoder used for macroeconomic feature compression. While the main body of the paper reports results based on a 32-dimensional latent representation, we also experimented with smaller latent spaces of 8 and 16 dimensions, and larger latent spaces of 64 dimensions. The motivation for this analysis is straightforward. Too small latent space may discard valuable information embedded in macroeconomic predictors, while too large a latent space risks retaining noise and reducing the regularization benefits of dimensionality reduction, as discussed in \textcite{hinton2006reducing}.

\begin{sidewaystable}[htbp]
  \centering
  \small
  \renewcommand{\arraystretch}{1.3}
  \setlength{\tabcolsep}{6pt}
  \newcolumntype{Y}{>{\centering\arraybackslash}X}

  \caption{Out-of-Sample $R^2$ for Stock Return and Consensus Approximations for Different Autoencoder Dimensions. \\
  This table reports monthly $R^2 (\%)$ of annual stock return estimation and analysts' consensus variable approximation over the entire evaluation sets for different $\lambda$ settings. }
  \label{tab:autoencoder-dim-compare}

  \begin{tabular}{
    c
    S[table-format=1.3]
    S[table-format=-2.2]
    S[table-format=-2.2]
    S[table-format=-2.2]
    S[table-format=-2.2]
    S[table-format=2.2]
    S[table-format=2.2]
    S[table-format=2.2]
    S[table-format=2.2]
    S[table-format=2.2]
    S[table-format=2.2]
    S[table-format=1.2]
  }
    \toprule
    & & \multicolumn{9}{c}{\small Consensus Variables}
    & \multicolumn{2}{c}{\small Overall Results} \\
    \cmidrule(lr){3-11} \cmidrule(lr){12-13}
    {\small \makecell{Auto-\\encoder \\ dim}} & {$\lambda$}
    & {\small \makecell{EPS \\ Forecast \\ Revision}}
    & {\small \makecell{Change \\ in \\ Rec-\\ommend-\\ation}}
    & {\small \makecell{Change \\in \\ Forecast \\ \& \\ Accrual}}
    & {\small \makecell{Long \\ vs \\ short \\ EPS \\ Forecasts}}
    & {\small \makecell{Analyst \\ Earnings \\ per \\ Share}}
    & {\small \makecell{EPS \\ Forecast \\ Dispersion}}
    & {\small \makecell{ Earnings \\ Forecast \\ Revisions}}
    & {\small \makecell{Analyst \\ Value}}
    & {\small \makecell{Analyst \\ Optimism}}
    & {\small \makecell{Consensus \\ Average}}
    & {\small \makecell{Stock \\ Returns}} \\
    \midrule

    \multirow{4}{*}{$D=8$}
      & 0   & {-}   & {-}   & {-}   & {-}   & {-}   & {-}   & {-}   & {-}   & {-}   & {-}   & -5.82 \\
      & 0.3 &  -1.36 & -2.55 &  0.58 &  2.91 & 43.51 & 27.05 &  10.05 & 14.94 & 19.66 &  12.75 & -7.86 \\
      & 0.6 &  1.3 & -1.84 &  1.84 &  3.66 & 59.24 & 34.15 &  13.91 & 24.17 & 30.66 & 18.57 & -9.71 \\
      & 0.9 &  2.25 & -1.5 &  2.67 &  5.00 & 66.51 & 37.09 & 15.71 & 29.26 & 35.59 & 21.40 & -9.70 \\ 
    \addlinespace \hline

    \multirow{4}{*}{$D=16$}
      & 0   & {-}   & {-}   & {-}   & {-}   & {-}   & {-}   & {-}   & {-}   & {-}   & {-}   & 0.29 \\
      & 0.3 &  3.05 & -0.3 &  2.81 &  5.5 & 51.36 & 28.07 & 9.85 & 21.95 & 23.72 & 16.22 & 3.70 \\
      & 0.6 &  4.24 & -0.24 &  3.98 & 7.54 & 64.22 & 35.10 & 13.73 & 30.07 & 31.91 & 21.17 & 3.40 \\
      & 0.9 &  4.79 & -0.17 &  4.52 &  8.59 & 69.76 & 38.23 & 15.67 & 34.10 & 36.10 & 23.51 & 2.73 \\ 
    \addlinespace \hline

    \multirow{4}{*}{$D=32$}
      & 0   & {-}   & {-}   & {-}   & {-}   & {-}   & {-}   & {-}   & {-}   & {-}   & {-}   & 4.34 \\
      & 0.3 & 3.21 & -0.25 & 2.92 & 5.91 & 50.96 & 27.25 & 9.81 & 22.22 & 24.97 & 16.33 & 10.46 \\
      & 0.6   & 4.3 & -0.19 & 3.9 & 7.92 & 64.31 & 34.97 & 13.71 & 30.27 & 32.70 & 21.32 & 9.9 \\
      & 0.9   & 4.85 & -0.17 & 4.5 & 8.89 & 70.39 & 38.44 & 15.73 & 34.65         & 36.53 & 23.76 & 9.51    \\
    \addlinespace \hline

    \multirow{4}{*}{$D=64$}
      & 0   & {-}   & {-}   & {-}   & {-}   & {-}   & {-}   & {-}   & {-}   & {-}   & {-}   & 1.29 \\
      & 0.3 & 3.81 & -0.22 & 3.54 & 7.17 & 58.10 & 32.13 & 12.85 & 27.17 & 27.71 & 19.14 & 4.44 \\
      & 0.6   & 5.02 & -0.15 & 4.55 & 9.09 & 69.20 & 38.66 & 16.33 & 34.46 & 35.16 & 23.59 & 3.98 \\
      & 0.9   & 5.49 & -0.11 & 5.03 & 9.73 & 73.97 & 41.05 & 17.92 & 37.69         & 39.02 & 25.53 & 3.29    \\
    \addlinespace


    \bottomrule
  \end{tabular}

  \begin{flushleft}
    {\textit{Note:} Results are reported for a sampled subset of $\lambda$ settings due to redundancy.}
  \end{flushleft}
\end{sidewaystable}

The comparative results across latent dimensionalities highlight a clear information–bottleneck trade-off in the macroeconomic autoencoder. Increasing the latent dimension generally improves the model’s ability to reconstruct analysts' consensus variables: approximation $R^{2}$ values rise monotonically for most consensus categories as $D$ increases from 8 to 64, reflecting the greater capacity of higher-dimensional embeddings to capture the underlying macroeconomic structure. However, these gains come with diminishing marginal benefits and introduce the risk of over-parameterization. Very small latent spaces (e.g., $D=8$) underfit the macroeconomic state, leading to weaker consensus approximation and substantially lower return $R^{2}$. Conversely, very large embeddings (e.g., $D=64$) improve consensus reconstruction but begin to attenuate the regularization benefits of compression, slightly weakening return predictability in line with the classical bias–variance trade-off in autoencoder architectures \parencite{hinton2006reducing}. The 32-dimensional specification achieves a favorable balance where it captures most of the consensus-relevant macroeconomic variation while maintaining sufficient regularization for stable long-horizon return forecasting. For this reason, the main empirical analysis adopts $D=32$ as the benchmark latent dimensionality.

\subsubsection{Comparison of state variables from principal components and autoencoder} \label{Appendix:vsPCA}

\begin{figure}[htbp]

    \centering
    \includegraphics[width=\columnwidth]{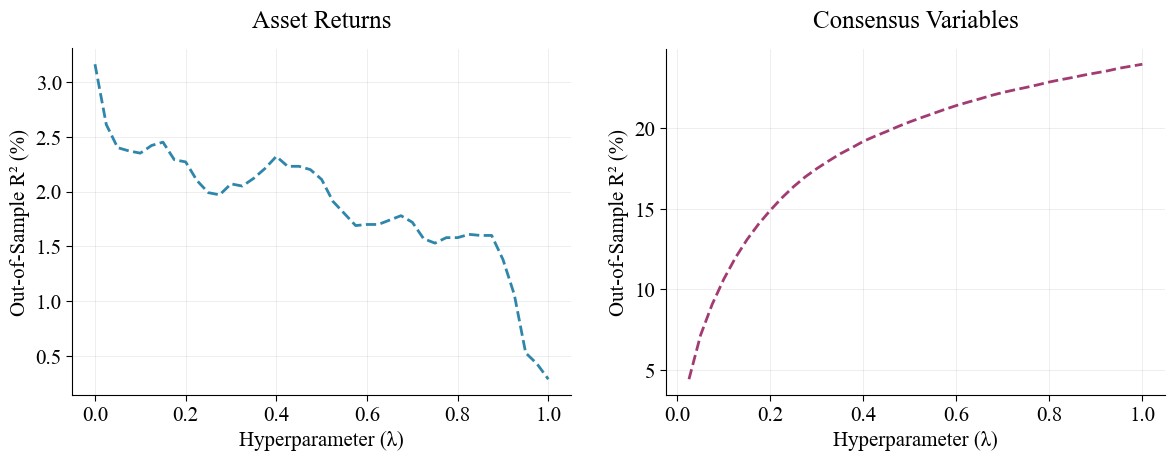}

    \caption{Out-of-sample $R^2$ of return predictions and consensus approximations after compressing state variables to 32 dimensions from PCA}

    \label{fig:r2_pca32dim}
\end{figure}
We next examine how the choice of macroeconomic feature compression method affects predictability. In particular, we compare autoencoder based compression with PCA, both reduced to 32 dimensions. As shown in Figure \ref{fig:r2_pca32dim}, PCA-based compression produces a pronounced decline in return predictability as $\lambda$ increases. This stands in sharp contrast to the autoencoder based compression, for which out-of-sample performance peaks around $\lambda = 0.4$ and remains substantially higher overall. However, the two approaches deliver broadly similar performance for the consensus variable approximation. Taken together, these results indicate that, in our setting, the autoencoder provides a more effective representation of high-dimensional macroeconomic information for return prediction, even though both methods are comparable for consensus approximation. A plausible explanation is that PCA, being a linear and variance-based method, treats all input variables symmetrically and focuses solely on capturing overall variance, whereas the autoencoder can learn nonlinear transformations that emphasize features most relevant for the prediction task, thereby yielding more informative embeddings for returns.

\subsubsection{Portfolio turnover and real-world implementability}
\label{Appendix:TC}

Because the CB-APM long--short portfolios exhibit relatively high turnover, an important practical consideration is whether the documented out-of-sample performance remains economically meaningful once realistic trading frictions are introduced. High-turnover strategies typically face nontrivial execution costs, and it is therefore natural to examine whether the model’s profitability persists after accounting for these frictions. To address this concern, we conduct a transaction-cost robustness analysis that adjusts returns according to
\[
R_t^{\mathrm{net}}
=
R_t^{\mathrm{gross}}
-
c \cdot \mathrm{TO}_t ,
\]
where $c \in \{25,50,75\}$ basis points denotes the proportional transaction-cost rate. The term $\mathrm{TO}_t$ represents the period-$t$ one-way turnover implied by the portfolio’s rebalancing rule and corresponds to the per-period rebalancing component of the turnover expression defined in the main text (Equation~\eqref{eq:turnover}). That is, $\mathrm{TO}_t$ measures the absolute adjustment in portfolio weights required to move from drifted holdings to the target weights at $t+1$. The transaction-cost adjustment therefore applies directly to the same notion of turnover used to construct the turnover statistics reported earlier.

This structure follows standard execution-cost decompositions emphasizing effective bid--ask spreads and market impact as primary sources of trading frictions (e.g., \citealp{bessembinder2003trade}; \citealp{frazzini2012trading}). We evaluate representative hyperparameter values $\lambda \in \{0, 0.3, 0.5, 0.7, 1.0\}$ under four cost scenarios (0, 25, 50, 75 bps). Table~\ref{tab:tc-robustness} summarizes the resulting performance measures.

Across all specifications, incorporating transaction costs reduces mean returns, cumulative log returns, and annualized Sharpe ratios in a monotonic and economically plausible manner. Importantly, however, the cross-sectional ordering of performance across $\lambda$ values remains essentially unchanged: the hyperparameter configurations that perform best in a frictionless environment continue to do so after transaction costs are applied. This stability indicates that the superior performance of CB-APM is driven by its predictive structure rather than by the absence of trading frictions.

The economic implications of the cost adjustment differ across models. The benchmark case of $\lambda = 0$, corresponding to an unconstrained neural network without the consensus-bottleneck restriction, already delivers the weakest frictionless Sharpe ratio among the specifications. Once transaction costs are incorporated, its performance converges toward that of a passive S\&P~500 buy-and-hold portfolio, especially under the 75~bps cost assumption, where the Sharpe ratios of the two become nearly indistinguishable. This pattern suggests that a plain neural network—despite having the lowest turnover among the models—does not generate sufficiently strong or persistent cross-sectional signals to overcome even moderate levels of trading frictions.

In contrast, higher-$\lambda$ specifications retain economically meaningful performance even under conservative transaction-cost assumptions. Their Sharpe ratios remain above one at 75~bps, indicating that the consensus-bottleneck architecture produces predictive signals with sufficient strength to remain profitable after accounting for realistic execution costs. These results underscore that the economic value of CB-APM arises not from frictionless idealizations but from its ability to extract stable, priced structure in the cross-section of returns.

Overall, the transaction-cost analysis confirms that the main findings of the paper are not artifacts of assuming frictionless trading. Although the main text reports frictionless results for comparability with the empirical asset pricing literature, the cost-adjusted evidence demonstrates that CB-APM's performance advantages are robust to trading frictions and remain relevant for real-world portfolio implementation.

\begin{table}[htbp]
\centering
\small
\renewcommand{\arraystretch}{1.25}
\setlength{\tabcolsep}{4.8pt}

\caption{Robustness of CB-APM Long--Short Portfolio Performance to Transaction Costs.\\
This table reports portfolio performance for representative hyperparameter values 
($\lambda \in \{0, 0.3, 0.5, 0.7, 1.0\}$) under four proportional transaction-cost assumptions: 
0, 25, 50, and 75~bps.}
\label{tab:tc-robustness}

\begin{tabular*}{\textwidth}{@{\extracolsep{\fill}}
l
S[table-format=1.3]
S[table-format=1.3]
S[table-format=1.3]
S[table-format=1.3]
S[table-format=1.3]
S[table-format=1.3]
S[table-format=1.3]
S[table-format=1.3]
S[table-format=2.1]
}
\toprule
 & \multicolumn{4}{c}{\small Mean Return}
 & \multicolumn{4}{c}{\small Sharpe Ratio}
 & {\small Turnover} \\
\cmidrule(lr){2-5} \cmidrule(lr){6-9}
{\small $\lambda$}
& {\small 0 bps} & {\small 25 bps} & {\small 50 bps} & {\small 75 bps}
& {\small 0 bps} & {\small 25 bps} & {\small 50 bps} & {\small 75 bps}
& {} \\
\midrule

0.0 
& 0.0153 & 0.0138 & 0.0124 & 0.0109
& 1.0997 & 0.9865 & 0.8752 & 0.7658
& 58.3 \\

0.3 
& 0.0220 & 0.0204 & 0.0189 & 0.0174
& 1.4375 & 1.3253 & 1.2152 & 1.1071
& 60.9 \\

0.5 
& 0.0211 & 0.0196 & 0.0181 & 0.0166
& 1.3169 & 1.2110 & 1.1071 & 1.0050
& 60.7 \\

0.7 
& 0.0219 & 0.0204 & 0.0189 & 0.0175
& 1.3535 & 1.2487 & 1.1459 & 1.0450
& 60.3 \\

1.0 
& 0.0223 & 0.0208 & 0.0192 & 0.0177
& 1.3766 & 1.2706 & 1.1665 & 1.0644
& 60.8 \\

\bottomrule
\end{tabular*}

\begin{flushleft}
\textit{Note:}
Transaction costs are applied as 
$r_t^{\mathrm{net}} = r_t^{\mathrm{gross}} - c \cdot \mathrm{TO}_t$,
where $\mathrm{TO}_t$ denotes one-way portfolio turnover. Turnover values do not vary with cost assumptions. 
\end{flushleft}
\end{table}

\subsubsection{Pricing benchmark portfolios with CB-APM-implied factors} \label{Appendix:benchmark-portfolio}

We assess whether CB-APM-implied factor-mimicking portfolios can price standard benchmark portfolios. It is important to point out that the consensus representations are not designed to approximate the span of the stochastic discount factor (SDF), but instead summarize belief-based heterogeneity that is complementary to the traditional factor space.

We consider three sets of standard test portfolios widely used in empirical asset pricing: the Fama--French 25 portfolios sorted on size and book-to-market ratio ($5\times 5$), the 25 portfolios sorted on size and momentum, and the 30 value-weighted industry portfolios. We evaluate whether the CB-APM factor-mimicking portfolios can jointly price the benchmark 25-- and 30--portfolio test assets.

\begin{table}[htbp]
    \centering
    \footnotesize
    \renewcommand{\arraystretch}{1.1}
    \setlength{\tabcolsep}{6pt}

    \caption{GRS tests for CB-APM-mimicking portfolios versus standard factor models. \\
    This table reports in-sample GRS tests of mean--variance efficiency for competing factor models.The test assets include (i) the 25 size--book-to-market portfolios, (ii) the 25 size--momentum portfolios, and (iii) the 30 value-weighted industry portfolios. CB-APM factors are constructed by sorting stocks into deciles on each consensus dimension and taking the value-weighted return spread between the top and bottom deciles, yielding tradable factor-mimicking portfolios.Different values of $\lambda$ correspond to CB-APM models trained under varying strengths of the consensus-bottleneck.Each panel reports the GRS $F$-statistic, $p$-value, mean absolute and RMS pricing errors (monthly and annualized), and the number of factors ($K$). All statistics use monthly excess returns.}

    \label{tab:grs-traditional}

\textbf{Panel A. 25 portfolios formed on size and book-to-market ratio} \\
\vspace{0.2cm}
\begin{tabular*}{\textwidth}{@{\extracolsep{\fill}}
    l
    S[table-format=1.3]
    S[table-format=1.2]
    S[table-format=1.4]
    S[table-format=1.4]
    S[table-format=1.4]
    S[table-format=1.4]
    S[table-format=1.0]
}
    \toprule
    {\small Factor Model} & {\small GRS $F$} & {\small $p$-value} &
    {\small Mean$|{\alpha}|$ (M)} & {\small Mean$|{\alpha}|$ (A)} &
    {\small RMS$\alpha$ (M)} & {\small RMS$\alpha$ (A)} &
    {\small $K$} \\
    \midrule
    CB-APM ($\lambda$=0.1) & 4.091 & 0.00 & 0.0075 & 0.0946 & 0.0077 & 0.0969 & 9 \\
    CB-APM ($\lambda$=0.5) & 4.069 & 0.00 & 0.0074 & 0.0925 & 0.0076 & 0.0947 & 9 \\
    CB-APM ($\lambda$=1.0) & 4.042 & 0.00 & 0.0102 & 0.1292 & 0.0103 & 0.1307 & 9 \\
    CAPM & 4.196 & 0.00 & 0.0016 & 0.0196 & 0.0021 & 0.0252 & 1 \\
    FF3 & 4.392 & 0.00 & 0.0014 & 0.0163 & 0.0018 & 0.0216 & 3 \\
    Carhart4 & 4.000 & 0.00 & 0.0013 & 0.0152 & 0.0016 & 0.0195 & 4 \\
    FF5 & 3.620 & 0.00 & 0.0013 & 0.0153 & 0.0016 & 0.0197 & 5 \\
    FF6 & 3.410 & 0.00 & 0.0012 & 0.0143 & 0.0015 & 0.0180 & 6 \\
    \bottomrule
\end{tabular*}

\vspace{0.2cm}

\textbf{Panel B. 25 portfolios formed on size and momentum} \\
\vspace{0.2cm}
\begin{tabular*}{\textwidth}{@{\extracolsep{\fill}}
    l
    S[table-format=1.3]
    S[table-format=1.3]
    S[table-format=1.4]
    S[table-format=1.4]
    S[table-format=1.4]
    S[table-format=1.4]
    S[table-format=1.0]
}
    \toprule
    {\small Factor Model} & {\small GRS $F$} & {\small $p$-value} &
    {\small Mean$|{\alpha}|$ (M)} & {\small Mean$|{\alpha}|$ (A)} &
    {\small RMS$\alpha$ (M)} & {\small RMS$\alpha$ (A)} &
    {\small $K$} \\
    \midrule
    CB-APM ($\lambda$=0.1) & 2.389 & 0.00 & 0.0075 & 0.0942 & 0.0078 & 0.0987 & 9 \\
    CB-APM ($\lambda$=0.5) & 2.676 & 0.00 & 0.0073 & 0.0922 & 0.0078 & 0.0988 & 9 \\
    CB-APM ($\lambda$=1.0) & 3.115 & 0.00 & 0.0104 & 0.1327 & 0.0106 & 0.1356 & 9 \\
    CAPM & 2.270 & 0.00 & 0.0028 & 0.0329 & 0.0035 & 0.0411 & 1 \\
    FF3 & 2.329 & 0.00 & 0.0028 & 0.0331 & 0.0036 & 0.0424 & 3 \\
    Carhart4 & 2.135 & 0.00 & 0.0016 & 0.0193 & 0.0019 & 0.0228 & 4 \\
    FF5 & 2.070 & 0.00 & 0.0022 & 0.0257 & 0.0028 & 0.0332 & 5 \\
    FF6 & 1.970 & 0.00 & 0.0010 & 0.0127 & 0.0015 & 0.0179 & 6 \\
    \bottomrule
\end{tabular*}

\vspace{0.2cm}

\textbf{Panel C. 30 industry portfolios} \\
\vspace{0.2cm}
\begin{tabular*}{\textwidth}{@{\extracolsep{\fill}}
    l
    S[table-format=1.3]
    S[table-format=1.3]
    S[table-format=1.4]
    S[table-format=1.4]
    S[table-format=1.4]
    S[table-format=1.4]
    S[table-format=1.0]
}
    \toprule
    {\small Factor Model} & {\small GRS $F$} & {\small $p$-value} &
    {\small Mean$|{\alpha}|$ (M)} & {\small Mean$|{\alpha}|$ (A)} &
    {\small RMS$\alpha$ (M)} & {\small RMS$\alpha$ (A)} &
    {\small $K$} \\
    \midrule
    CB-APM ($\lambda$=0.1) & 1.517 & 0.05 & 0.0068 & 0.0849 & 0.0072 & 0.0905 & 9 \\
    CB-APM ($\lambda$=0.5) & 1.461 & 0.06 & 0.0066 & 0.0827 & 0.0070 & 0.0883 & 9 \\
    CB-APM ($\lambda$=1.0) & 1.644 & 0.02 & 0.0092 & 0.1166 & 0.0095 & 0.1201 & 9 \\
    CAPM & 1.233 & 0.19 & 0.0023 & 0.0278 & 0.0031 & 0.0368 & 1 \\
    FF3 & 1.572 & 0.03 & 0.0026 & 0.0305 & 0.0032 & 0.0379 & 3 \\
    Carhart4 & 1.583 & 0.03 & 0.0024 & 0.0283 & 0.0029 & 0.0347 & 4 \\
    FF5 & 1.774 & 0.01 & 0.0030 & 0.0350 & 0.0039 & 0.0452 & 5 \\
    FF6 & 1.666 & 0.02 & 0.0026 & 0.0311 & 0.0035 & 0.0405 & 6 \\
    \bottomrule
\end{tabular*}

\begin{flushleft}
\textit{Note:} 
The GRS $F$-statistic tests the null hypothesis that all pricing errors ($\boldsymbol{\alpha}$) are jointly zero. Mean$|{\alpha}|$ and RMS$\alpha$ denote mean absolute and root-mean-squared pricing errors, reported in monthly (M) and annualized (A) terms. $p$-values are rounded to two decimal places; values below $0.005$ appear as $0.00$.
\end{flushleft}
\end{table}

Panels~A--C of Table~\ref{tab:grs-traditional} show that CB-APM factors partially capture systematic return variation across benchmark portfolios but do not match the pricing performance of established factor models. Pricing errors remain modest but are consistently larger than those of FF5 and FF6. These results indicate that the consensus-based factors provide incremental but incomplete coverage of the traditional factor space.

\subsection{Ablation studies} \label{Appendix:ablation}

To better understand the mechanisms that drive the performance of CB-APM, we conduct a series of ablation studies. An ablation study refers to systematically removing or modifying key model components to evaluate their incremental contribution to predictive accuracy and interpretability. This approach is widely adopted in machine learning research to clarify the role of specific architectural choices,\footnote{See \textcite{gao2019res2net} and \textcite{devlin2019bert} for representative examples in the deep learning literature.} and has recently been extended to interpretable models such as concept-bottleneck architectures \parencite{koh2020concept}. In empirical asset pricing, where models often involve high-dimensional predictors and complex nonlinear interactions, ablation studies provide a transparent way to disentangle whether observed performance gains stem from meaningful economic mechanisms or from generic model flexibility.

In the context of CB-APM, ablation studies allow us to assess the value of two key design features. First, we evaluate whether dimensionality reduction of macroeconomic predictors via an autoencoder provides genuine improvements in signal extraction compared to using the raw, redundant set of macroeconomic variables. Second, we examine the role of joint optimization of consensus approximation and return prediction. \footnote{\emph{To be done and reported in the next version of the paper.}} In particular, we consider both extreme cases: when the model ignores consensus learning altogether ($\lambda=0$), and when it focuses exclusively on consensus approximation without return prediction ($\lambda \to \infty$). These tests enable us to evaluate whether the consensus-bottleneck provides unique value beyond replicating analysts' forecasts, and whether simultaneous optimization is critical for linking consensus formation to expected returns.

\subsubsection{Effect of macroeconomic feature compression} \label{Appendix:ablation-autoencoder}
\begin{figure}[htbp]
    \centering
    \includegraphics[width=\columnwidth]{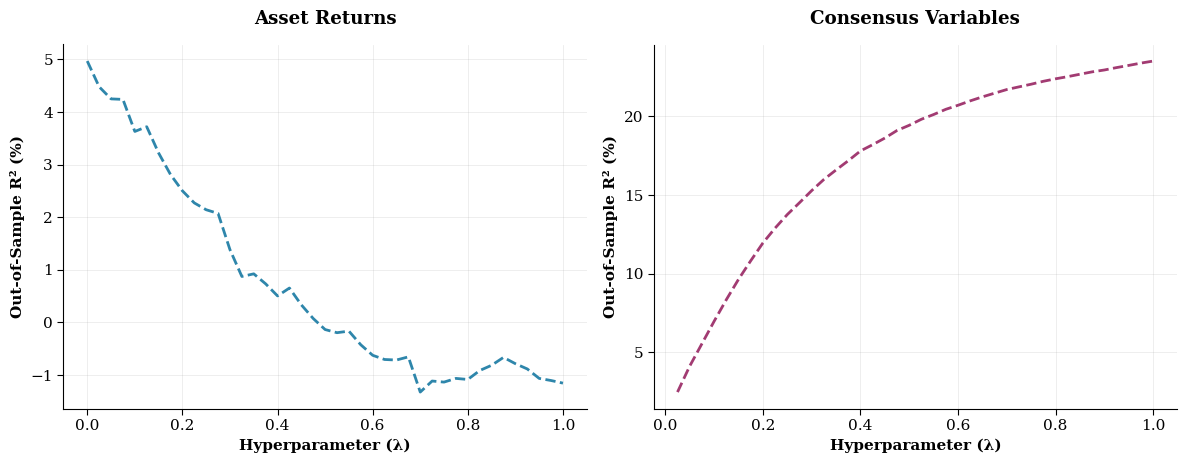}
    \caption{Out-of-Sample $R^2$ without Macroeconomic State Embeddings. \\
    This figure reports monthly $R^2$ of annual stock return estimation (left) and average $R^2$ of analysts' consensus variable approximation (right) when macroeconomic state variables are not embedded via the autoencoder.}

    \label{fig:no-autoencoder}
\end{figure}

To evaluate the contribution of macroeconomic state embeddings to CB-APM's performance, we conduct an ablation study by re-estimating the model without the autoencoder component. Figure \ref{fig:no-autoencoder} reports the out-of-sample $R^2$ for annual stock return prediction (left) and consensus variable approximation (right) across varying values of the hyperparameter $\lambda$.

The results show that excluding the autoencoder leads to a sharp deterioration in return predictability. Without macroeconomic embeddings, the out-of-sample $R^2$ for annual returns declines steadily with increasing $\lambda$, ultimately falling below zero for moderate-to-high values of the regularization parameter. This pattern contrasts starkly with the baseline CB-APM, where joint learning with macroeconomic state variables amplifies long-horizon predictive performance. These findings highlight the critical role of macroeconomic context in anchoring the consensus-bottleneck and enhancing its informativeness for return forecasting.

Importantly, consensus approximation remains largely unaffected in this ablated model, as shown in the right panel of Figure \ref{fig:no-autoencoder}. While the model continues to reconstruct analysts' consensus variables with reasonable accuracy, the absence of macroeconomic embeddings severs an important informational channel linking consensus to return-relevant fundamentals. This divergence underscores the complementary function of the autoencoder that by distilling high-dimensional macroeconomic signals into latent state variables, it enriches the consensus layer with persistent economy-wide information, thereby mitigating noise in firm-level predictors and improving the model's capacity to extract long-horizon risk premiums.

\subsubsection{Avoiding look-ahead of AI analysts} \label{Appendix:ablation-consensus}

A further component analysis evaluates the role of joint optimization in CB-APM, where the model simultaneously learns to approximate analyst consensus and to predict future returns. While the main body of the paper focused on the case $\lambda = 0$, where the model collapses to a pure return prediction architecture, it is equally informative to consider the opposite extreme. When $\lambda \to \infty$, the model is trained solely to replicate contemporary consensus variables without any direct return forecasting objective. This setting allows us to assess whether the architectural design is effective in extracting meaningful consensus representations from firm and macro characteristics.

Table~\ref{tab:consensus_only_r2} reports annual out-of-sample $R^2$ for this consensus-only specification, separately for a range of consensus-based targets and for returns. Consensus-related $R^2$ measures such as \emph{EPS forecast revision}, \emph{Earnings forecast revisions}, and \emph{Analyst Value} remain economically sizable and relatively stable over time, while \emph{Analyst earnings per share} stays in a narrow band around $80-86\%$ and \emph{EPS Forecast Dispersion} between roughly $41\%$ and $52\%$. The composite \emph{Consensus average} fluctuates only modestly between $27.52\%$ and $31.72\%$ across 2014--2023, with a full-sample value of $30.30\%$, indicating that the model recovers a stable consensus structure even without any return signal.

Importantly, the full-sample consensus average ($30.30\%$) is close in magnitude to the out-of-sample consensus-approximation performance obtained under the empirical baseline of $\lambda = 1$, where the model jointly learns consensus and returns. Across the various return forecasting horizons considered in the main analysis, the consensus $R^2$ under $\lambda = 1$ typically lies in the $24\%$--$28\%$ range. The similarity of these values demonstrates that consensus-approximation accuracy does not improve markedly beyond the interpretability constraint used in the empirical specification. In other words, the consensus-learning component of CB-APM effectively converges by the time $\lambda = 1$ is reached, and further increasing the weight on consensus approximation yields only marginal gains.

By comparing this consensus-only specification with the baseline joint optimization, we can more clearly identify the incremental role that consensus learning plays in shaping return predictions. The central insight from this analysis is that analysts' consensus variables are themselves highly learnable from the same firm-level characteristics and macroeconomic information that the asset pricing literature already employs for return prediction. This establishes that the consensus-bottleneck is not an artificial architectural constraint, but an empirically legitimate representation: it extracts a predictable, economically interpretable signal embedded in observable characteristics.

\begin{sidewaystable}[t]
    \centering
    \renewcommand{\arraystretch}{1.3}
    \caption{Annual out-of-sample $R^2$ (\%) under consensus-only training ($\lambda \to \infty$). \\
    This table reports annual out-of-sample $R^2$ (\%) for a consensus-only specification of CB-APM, in which the model is trained exclusively to approximate analysts' consensus variables with no return-prediction objective. For each year from 2014 to 2023, the table presents the predictive $R^2$ for a broad set of consensus variables. The final column reports full-sample $R^2$ values computed by concatenating realized targets across all test periods. The bottom row (“Consensus average”) summarizes the cross-variable average $R^2$ for each year. These results quantify the model’s ability to recover stable and economically meaningful consensus structure in the absence of any return-based loss component.}

    \label{tab:consensus_only_r2}
    \begin{tabular}{lrrrrrrrrrrr}
        \toprule
        Year & 2014 & 2015 & 2016 & 2017 & 2018 & 2019 & 2020 & 2021 & 2022 & 2023 & Full sample \\
        \midrule
        EPS forecast revision          & 7.29  & 6.38  & 7.48  & 5.78  & 6.32  & 6.00  & 5.87  & 7.37  & 5.95  & 7.56  & 6.60 \\
        Change in recommendation       & -0.08 & 0.13  & 0.00  & 0.04  & -0.02 & 0.12  & 0.11  & -0.04 & 0.21  & 0.01  & 0.05 \\
        Change in Forecast and Accrual & 7.22  & 6.43  & 7.38  & 7.20  & 7.83  & 6.35  & 7.87  & 1.66  & 4.79  & 6.85  & 6.45 \\
        Long-vs-short EPS forecasts    & 19.92 & 17.50 & 17.46 & 15.16 & 10.01 & 11.13 & 4.89  & 6.16  & 2.97  & 6.49  & 11.60 \\
        Analyst earnings per share     & 82.10 & 84.17 & 83.63 & 83.44 & 85.70 & 83.50 & 83.86 & 79.76 & 76.38 & 79.51 & 82.36 \\
        EPS Forecast Dispersion        & 50.46 & 50.54 & 49.48 & 51.68 & 51.96 & 48.19 & 49.84 & 44.01 & 43.72 & 41.34 & 48.37 \\
        Earnings forecast revisions    & 21.77 & 20.44 & 27.48 & 22.92 & 23.77 & 17.47 & 21.11 & 28.08 & 24.31 & 25.65 & 23.22 \\
        Analyst Value                  & 39.06 & 42.41 & 44.39 & 42.88 & 52.02 & 49.24 & 52.07 & 51.42 & 42.17 & 50.42 & 46.40 \\
        Analyst Optimism               & 49.03 & 48.52 & 48.14 & 48.62 & 45.57 & 47.72 & 49.84 & 47.36 & 47.17 & 44.03 & 47.68 \\
        Consensus average              & 30.75 & 30.72 & 31.72 & 30.86 & 31.46 & 29.97 & 30.61 & 29.53 & 27.52 & 29.10 & 30.30 \\
        \bottomrule
    \end{tabular}
\end{sidewaystable}

\setcounter{table}{0}
\setcounter{figure}{0}

\begin{landscape}
\section{Detailed Data Description} \label{Appendix:data}
\subsection{Firm-level predictors}
\small
\setlength\LTleft{0pt}
\setlength\LTright{0pt}
\begin{longtable}{@{\extracolsep{\fill}} 
    l                   
    l                   
    l                   
    l                   
    l                   
    l                   
    l                   
}
    \caption{Descriptions of firm-level predictors from \textcite{ChenZimmermann2021}.}
    \label{tab:firm-level-predictors} \\
    \toprule
    \multicolumn{1}{l}{\small No.} & 
    \multicolumn{1}{l}{\small Acronym} & 
    \multicolumn{1}{l}{\small Firm-level Predictor} & 
    \multicolumn{1}{l}{\small Authors} & 
    \multicolumn{1}{l}{\small Year} & 
    \multicolumn{1}{l}{\small Journal} & 
    \multicolumn{1}{l}{\small Frequency} \\
    \midrule
    \endfirsthead

    \caption[]{Descriptions of firm-level predictors from \textcite{ChenZimmermann2021} (cont'd).} \\
    \toprule
    \multicolumn{1}{l}{\small No.} & 
    \multicolumn{1}{l}{\small Acronym} & 
    \multicolumn{1}{l}{\small Firm-level Predictor} & 
    \multicolumn{1}{l}{\small Authors} & 
    \multicolumn{1}{l}{\small Year} & 
    \multicolumn{1}{l}{\small Journal} & 
    \multicolumn{1}{l}{\small Frequency} \\
    \midrule
    \endhead

    \bottomrule
    \endfoot

    \bottomrule
    \multicolumn{7}{l}{\footnotesize \textit{Note:} Predictors marked with * are inherently defined as change or growth rates.}
    \endlastfoot
    
    1 & AbnormalAccruals & Abnormal Accruals & Xie & 2001 & AR & Annual \\
    2 & Accruals & Accruals & Sloan & 1996 & AR & Annual \\
    3 & AM & Total assets to market & Fama and French & 1992 & JF & Monthly \\
    4 & AnnouncementReturn & Earnings announcement return & Chan, Jegadeesh and Lakonishok & 1996 & JF & Quarterly \\
    5 & AssetGrowth* & Asset growth & Cooper, Gulen and Schill & 2008 & JF & Annual \\
    6 & BetaLiquidityPS & Pastor-Stambaugh liquidity beta & Pastor and Stambaugh & 2003 & JPE & Monthly \\
    7 & betaVIX & Systematic volatility & Ang et al. & 2006 & JF & Monthly \\
    8 & BM & Book to market & Stattman & 1980 & Other & Annual \\
    9 & BMdec & Book to market using December ME & Fama and French & 1992 & JPM & Half \\
    10 & BookLeverage & Book leverage (annual) & Fama and French & 1992 & JF & Annual \\
    11 & BPEBM & Leverage component of BM & Penman, Richardson and Tuna & 2007 & JAR & Monthly \\
    12 & Cash & Cash to assets & Palazzo & 2012 & JFE & Quarterly \\
    13 & CashProd & Cash Productivity & Chandrashekar and Rao & 2009 & WP & Monthly \\
    14 & CBOperProf & Cash-based operating profitability & Ball et al. & 2016 & JFE & Annual \\
    15 & CF & Cash flow to market & Lakonishok, Shleifer, Vishny & 1994 & JF & Monthly \\
    16 & cfp & Operating Cash flows to price & Desai, Rajgopal, Venkatachalam & 2004 & AR & Monthly \\
    17 & ChEQ* & Growth in book equity & Lockwood and Prombutr & 2010 & JFR & Annual \\
    18 & ChInv* & Inventory Growth & Thomas and Zhang & 2002 & RAS & Annual \\
    19 & ChInvIA & Change in capital inv (ind adj) & Abarbanell and Bushee & 1998 & AR & Monthly \\
    20 & ChNNCOA & Change in Net Noncurrent Op Assets & Soliman & 2008 & AR & Annual \\
    21 & ChNWC & Change in Net Working Capital & Soliman & 2008 & AR & Annual \\
    22 & ChTax & Change in Taxes & Thomas and Zhang & 2011 & JAR & Quarterly \\
    23 & ConvDebt & Convertible debt indicator & Valta & 2016 & JFQA & Annual \\
    24 & CoskewACX & Coskewness using daily returns & Ang, Chen and Xing & 2006 & RFS & Monthly \\
    25 & DelBreadth & Breadth of ownership & Chen, Hong and Stein & 2002 & JFE & Quarterly \\
    26 & DelCOA & Change in current operating assets & Richardson et al. & 2005 & JAE & Annual \\
    27 & DelCOL & Change in current operating liabilities & Richardson et al. & 2005 & JAE & Annual \\
    28 & DelEqu & Change in equity to assets & Richardson et al. & 2005 & JAE & Annual \\
    29 & DelFINL & Change in financial liabilities & Richardson et al. & 2005 & JAE & Annual \\
    30 & DelLTI & Change in long-term investment & Richardson et al. & 2005 & JAE & Annual \\
    31 & DelNetFin & Change in net financial assets & Richardson et al. & 2005 & JAE & Annual \\
    32 & DivInit & Dividend Initiation & Michaely, Thaler and Womack & 1995 & JF & Annual \\
    33 & DivOmit & Dividend Omission & Michaely, Thaler and Womack & 1995 & JF & Annual \\
    34 & dNoa & change in net operating assets & Hirshleifer, Hou, Teoh, Zhang & 2004 & JAE & Annual \\
    35 & DolVol & Past trading volume & Brennan, Chordia, Subra & 1998 & JFE & Monthly \\
    36 & EarningsConsistency & Earnings consistency & Alwathainani & 2009 & BAR & Annual \\
    37 & EarningsStreak & Earnings surprise streak & Loh and Warachka & 2012 & MS & Quarterly \\
    38 & EarningsSurprise & Earnings Surprise & Foster, Olsen and Shevlin & 1984 & AR & Quarterly \\
    39 & EBM & Enterprise component of BM & Penman, Richardson and Tuna & 2007 & JAR & Monthly \\
    40 & EntMult & Enterprise Multiple & Loughran and Wellman & 2011 & JFQA & Monthly \\
    41 & EP & Earnings-to-Price Ratio & Basu & 1977 & JF & Monthly \\
    42 & EquityDuration & Equity Duration & Dechow, Sloan and Soliman & 2004 & RAS & Annual \\
    43 & ExchSwitch & Exchange Switch & Dharan and Ikenberry & 1995 & JF & Monthly \\
    44 & grcapx & Change in capex (two years) & Anderson and Garcia-Feijoo & 2006 & JF & Annual \\
    45 & grcapx3y & Change in capex (three years) & Anderson and Garcia-Feijoo & 2006 & JF & Annual \\
    46 & Herf & Industry concentration (sales) & Hou and Robinson & 2006 & JF & Monthly \\
    47 & HerfBE & Industry concentration (equity) & Hou and Robinson & 2006 & JF & Monthly \\
    48 & hire* & Employment growth & Bazdresch, Belo and Lin & 2014 & JPE & Annual \\
    49 & IdioVol3F & Idiosyncratic risk (3 factor) & Ang et al. & 2006 & JF & Monthly \\
    50 & IdioVolAHT & Idiosyncratic risk (AHT) & Ali, Hwang, and Trombley & 2003 & JFE & Monthly \\
    51 & Illiquidity & Amihud's illiquidity & Amihud & 2002 & JFM & Monthly \\
    52 & IndIPO & Initial Public Offerings & Ritter & 1991 & JF & Monthly \\
    53 & IndMom & Industry Momentum & Grinblatt and Moskowitz & 1999 & JF & Monthly \\
    54 & IntMom & Intermediate Momentum & Novy-Marx & 2012 & JFE & Monthly \\
    55 & Investment & Investment to revenue & Titman, Wei and Xie & 2004 & JFQA & Monthly \\
    56 & InvestPPEInv & change in ppe and inv/assets & Lyandres, Sun and Zhang & 2008 & RFS & Annual \\
    57 & iomom\_cust & Customers momentum & Menzly and Ozbas & 2010 & JF & Monthly \\
    58 & iomom\_supp & Suppliers momentum & Menzly and Ozbas & 2010 & JF & Monthly \\
    59 & Leverage & Market leverage & Bhandari & 1988 & JF & Monthly \\
    60 & LRreversal & Long-run reversal & De Bondt and Thaler & 1985 & JF & Monthly \\
    61 & MaxRet & Maximum return over month & Bali, Cakici, and Whitelaw & 2011 & JFE & Monthly \\
    62 & Mom12m & Momentum (12 month) & Jegadeesh and Titman & 1993 & JF & Monthly \\
    63 & Mom12mOffSeason & Momentum without the seasonal part & Heston and Sadka & 2008 & JFE & Monthly \\
    64 & Mom6m & Momentum (6 month) & Jegadeesh and Titman & 1993 & JF & Monthly \\
    65 & Mom6mJunk & Junk Stock Momentum & Avramov et al & 2007 & JF & Monthly \\
    66 & MomOffSeason & Off season long-term reversal & Heston and Sadka & 2008 & JFE & Monthly \\
    67 & MomSeason & Return seasonality years 2 to 5 & Heston and Sadka & 2008 & JFE & Monthly \\
    68 & MomSeasonShort & Return seasonality last year & Heston and Sadka & 2008 & JFE & Monthly \\
    69 & NetDebtFinance & Net debt financing & Bradshaw, Richardson, Sloan & 2006 & JAE & Annual \\
    70 & NetEquityFinance & Net equity financing & Bradshaw, Richardson, Sloan & 2006 & JAE & Annual \\
    71 & NOA & Net Operating Assets & Hirshleifer et al. & 2004 & JAE & Annual \\
    72 & OPLeverage & Operating leverage & Novy-Marx & 2011 & ROF & Annual \\
    73 & Price & Price & Blume and Husic & 1973 & JF & Monthly \\
    74 & PriceDelayRsq & Price delay r square & Hou and Moskowitz & 2005 & RFS & Annual \\
    75 & RDIPO & IPO and no RD spending & Gou, Lev and Shi & 2006 & JBFA & Annual \\
    76 & RDS & Real dirty surplus & Landsman et al. & 2011 & AR & Annual \\
    77 & RealizedVol & Realized (Total) Volatility & Ang et al. & 2006 & JF & Monthly \\
    78 & ResidualMomentum & Momentum based on FF3 residuals & Blitz, Huij and Martens & 2011 & JEmpFin & Monthly \\
    79 & ReturnSkew & Return skewness & Bali, Engle and Murray & 2015 & Book & Monthly \\
    80 & ReturnSkew3F & Idiosyncratic skewness (3F model) & Bali, Engle and Murray & 2015 & Book & Monthly \\
    81 & RevenueSurprise & Revenue Surprise & Jegadeesh and Livnat & 2006 & JAE & Quarterly \\
    82 & roaq & Return on assets (qtrly) & Balakrishnan, Bartov and Faurel & 2010 & JAE & Quarterly \\
    83 & ShareIss1Y & Share issuance (1 year) & Pontiff and Woodgate & 2008 & JF & Annual \\
    84 & ShareVol & Share Volume & Datar, Naik and Radcliffe & 1998 & JFM & Monthly \\
    85 & Size & Size & Banz & 1981 & JFE & Monthly \\
    86 & STreversal & Short term reversal & Jegadeesh & 1990 & JF & Monthly \\
    87 & Tax & Taxable income to income & Lev and Nissim & 2004 & AR & Annual \\
    88 & TotalAccruals & Total accruals & Richardson et al. & 2005 & JAE & Annual \\
    89 & TrendFactor & Trend Factor & Han, Zhou, Zhu & 2016 & JFE & Monthly \\
    90 & VolSD & Volume Variance & Chordia, Subra, Anshuman & 2001 & JFE & Monthly \\
    91 & XFIN & Net external financing & Bradshaw, Richardson, Sloan & 2006 & JAE & Annual \\
    92 & zerotrade & Days with zero trades & Liu & 2006 & JFE & Monthly \\
    93 & zerotradeAlt1 & Days with zero trades & Liu & 2006 & JFE & Monthly \\
    94 & zerotradeAlt12 & Days with zero trades & Liu & 2006 & JFE & Monthly \\
    95 & Beta & CAPM beta & Fama and MacBeth & 1973 & JPE & Monthly \\
    96 & BetaFP & Frazzini-Pedersen Beta & Frazzini and Pedersen & 2014 & JFE & Monthly \\
    97 & BidAskSpread & Bid-ask spread & Amihud and Mendelsohn & 1986 & JFE & Monthly \\
    98 & Coskewness & Coskewness & Harvey and Siddique & 2000 & JF & Monthly \\
    99 & DebtIssuance & Debt Issuance & Spiess and Affleck-Graves & 1999 & JFE & Annual \\
    100 & FirmAge & Firm age based on CRSP & Barry and Brown & 1984 & JFE & Monthly \\
    101 & GrLTNOA* & Growth in long term operating assets & Fairfield, Whisenant and Yohn & 2003 & AR & Annual \\
    102 & HerfAsset & Industry concentration (assets) & Hou and Robinson & 2006 & JF & Monthly \\
    103 & High52 & 52 week high & George and Hwang & 2004 & JF & Monthly \\
    104 & MRreversal & Medium-run reversal & De Bondt and Thaler & 1985 & JF & Monthly \\
    105 & NumEarnIncrease & Earnings streak length & Loh and Warachka & 2012 & MS & Quarterly \\
    106 & PriceDelaySlope & Price delay coeff & Hou and Moskowitz & 2005 & RFS & Annual \\
    107 & PriceDelayTstat & Price delay SE adjusted & Hou and Moskowitz & 2005 & RFS & Biennial \\
    108 & RoE & net income / book equity & Haugen and Baker & 1996 & JFE & Annual \\
    109 & ShareRepurchase & Share repurchases & Ikenberry, Lakonishok, Vermaelen & 1995 & JFE & Annual \\
    110 & SP & Sales-to-price & Barbee, Mukherji and Raines & 1996 & FAJ & Monthly \\
    111 & Spinoff & Spinoffs & Cusatis, Miles and Woolridge & 1993 & JFE & Monthly \\
    112 & VarCF & Cash-flow  to price variance & Haugen and Baker & 1996 & JFE & Monthly \\
    113 & VolMkt & Volume to market equity & Haugen and Baker & 1996 & JFE & Monthly \\
    114 & VolumeTrend & Volume Trend & Haugen and Baker & 1996 & JFE & Monthly
\end{longtable}
\end{landscape}

\begin{landscape}

\subsection{Macroeconomic predictors}

\setlength\LTleft{0pt}
\setlength\LTright{0pt}
\begin{small}
\begin{longtable}{@{\extracolsep{\fill}} 
    l                   
    l                   
    l                   
    l                   
}
    \caption{Descriptions of macroeconomic predictors from \textcite{welch2008comprehensive}.}
    \label{tab:macro-predictors} \\
    \toprule
    \multicolumn{1}{l}{\small No.} & 
    \multicolumn{1}{l}{\small Acronym} & 
    \multicolumn{1}{l}{\small Macroeconomic Predictor} & 
    \multicolumn{1}{l}{\small Description} \\
    \midrule
    \endfirsthead

    \caption[]{Descriptions of macroeconomic predictors from \textcite{welch2008comprehensive} (Cont'd).} \\
    \toprule
    \multicolumn{1}{l}{\small No.} & 
    \multicolumn{1}{l}{\small Acronym} & 
    \multicolumn{1}{l}{\small Macroeconomic Predictor} & 
    \multicolumn{1}{l}{\small Description} \\
    \midrule
    \endhead

    \bottomrule
    \endfoot

    \bottomrule
    \endlastfoot
    1 & dp & Dividend-price ratio & The difference between the log of dividends and the log of prices \\
    2 & ep & Earnings-price ratio & The difference between the log of earnings and the log of prices \\
    3 & bm & Book-to-market ratio & The ratio of book value to market value for the Dow Jones Industrial Average \\
    4 & ntis & Net equity expansion & \begin{tabular}[c]{@{}l@{}}The ratio of 12-month moving sums of net issues by NYSE listed stocks divided \\ by the total end-of-year market capitalization of NYSE stocks\end{tabular} \\
    5 & tbl & Treasury-bill rate & The 3-Month Treasury Bill: Secondary Market Rate \\
    6 & tms & Term spread & \begin{tabular}[c]{@{}l@{}}The difference between the long term yield on government bonds and the\\Treasury-bill\end{tabular} \\
    7 & dfy & Default yield spread & The difference between BAA and AAA-rated corporate bond yields \\
    8 & svar & Stock variance & Sum of squared daily returns on the S\&P 500
\end{longtable}
\end{small}
\end{landscape}

\begin{landscape}

\small
\setlength\LTleft{0pt}
\setlength\LTright{0pt}
\begin{longtable}{@{\extracolsep{\fill}} 
    l    
    l    
    l    
    p{8cm} 
}
    \caption{Descriptions of macroeconomic predictors from FRED-MD \parencite{mccracken2016fred}.}
    \label{tab:macro-fred} \\
    \toprule
    \multicolumn{1}{l}{\small No.} &
    \multicolumn{1}{l}{\small Group} &
    \multicolumn{1}{l}{\small Acronym} &
    \multicolumn{1}{l}{\small Macroeconomic Predictor} \\
    \midrule
    \endfirsthead

    \caption[]{Descriptions of macroeconomic predictors from FRED-MD \parencite{mccracken2016fred} (cont'd).} \\
    \toprule
    \multicolumn{1}{l}{\small No.} &
    \multicolumn{1}{l}{\small Group} &
    \multicolumn{1}{l}{\small Acronym} &
    \multicolumn{1}{l}{\small Macroeconomic Predictor} \\
    \midrule
    \endhead

    \bottomrule
    \endfoot

    \bottomrule
    \endlastfoot

    1 & Output and Income & RPI & Real Personal Income \\
    2 & Output and Income & W875RX1 & Real personal income ex transfer receipts \\
    3 & Output and Income & INDPRO & IP Index \\
    4 & Output and Income & IPFPNSS & IP: Final Products and Nonindustrial Supplies \\
    5 & Output and Income & IPFINAL & IP: Final Products (Market Group) \\
    6 & Output and Income & IPCONGD & IP: Consumer Goods \\
    7 & Output and Income & IPDCONGD & IP: Durable Consumer Goods \\
    8 & Output and Income & IPNCONGD & IP: Nondurable Consumer Goods \\
    9 & Output and Income & IPBUSEQ & IP: Business Equipment \\
    10 & Output and Income & IPMAT & IP: Materials \\
    11 & Output and Income & IPDMAT & IP: Durable Materials \\
    12 & Output and Income & IPNMAT & IP: Nondurable Materials \\
    13 & Output and Income & IPMANSICS & IP: Manufacturing (SIC) \\
    14 & Output and Income & IPFUELS & IP: Fuels \\
    15 & Output and Income & CUMFNS & Capacity Utilization:  Manufacturing \\
    16 & Labor Market & HWI & Help-Wanted Index for United States \\
    17 & Labor Market & HWIURATIO & Ratio of Help Wanted/No.  Unemployed \\
    18 & Labor Market & CLF16OV & Civilian Labor Force \\
    19 & Labor Market & CE16OV & Civilian Employment \\
    20 & Labor Market & UNRATE & Civilian Unemployment Rate \\
    21 & Labor Market & UEMPMEAN & Average Duration of Unemployment (Weeks) \\
    22 & Labor Market & UEMPLT5 & Civilians Unemployed - Less Than 5 Weeks \\
    23 & Labor Market & UEMP5TO14 & Civilians Unemployed for 5-14 Weeks \\
    24 & Labor Market & UEMP15OV & Civilians Unemployed - 15 Weeks and Over \\
    25 & Labor Market & UEMP15T26 & Civilians Unemployed for 15-26 Weeks \\
    26 & Labor Market & UEMP27OV & Civilians Unemployed for 27 Weeks and Over \\
    27 & Labor Market & CLAIMSx & Initial Claims \\
    28 & Labor Market & PAYEMS & All Employees:  Total nonfarm \\
    29 & Labor Market & USGOOD & All Employees:  Goods-Producing Industries \\
    30 & Labor Market & CES1021000001 & All Employees:  Mining and Logging:  Mining \\
    31 & Labor Market & USCONS & All Employees:  Construction \\
    32 & Labor Market & MANEMP & All Employees:  Manufacturing \\
    33 & Labor Market & DMANEMP & All Employees:  Durable goods \\
    34 & Labor Market & NDMANEMP & All Employees:  Nondurable goods \\
    35 & Labor Market & SRVPRD & All Employees:  Service-Providing Industries \\
    36 & Labor Market & USTPU & All Employees:  Trade, Transportation, and Utilities \\
    37 & Labor Market & USWTRADE & All Employees:  Wholesale Trade \\
    38 & Labor Market & USTRADE & All Employees:  Retail Trade \\
    39 & Labor Market & USFIRE & All Employees:  Financial Activities \\
    40 & Labor Market & USGOVT & All Employees:  Government \\
    41 & Labor Market & CES0600000007 & Avg Weekly Hours :  Goods-Producing \\
    42 & Labor Market & AWOTMAN & Avg Weekly Overtime Hours :  Manufacturing \\
    43 & Labor Market & AWHMAN & Avg Weekly Hours :  Manufacturing \\
    44 & Labor Market & CES0600000008 & Avg Hourly Earnings :  Goods-Producing \\
    45 & Labor Market & CES2000000008 & Avg Hourly Earnings :  Construction \\
    46 & Labor Market & CES3000000008 & Avg Hourly Earnings :  Manufacturing \\
    47 & Housing & HOUST & Housing Starts:  Total New Privately Owned \\
    48 & Housing & HOUSTNE & Housing Starts, Northeast \\
    49 & Housing & HOUSTMW & Housing Starts, Midwest \\
    50 & Housing & HOUSTS & Housing Starts, South \\
    51 & Housing & HOUSTW & Housing Starts, West \\
    52 & Consumption, Orders, and Inventories & DPCERA3M086SBEA & Real personal consumption expenditures \\
    53 & Consumption, Orders, and Inventories & CMRMTSPLx & Real Manu.  and Trade Industries Sales \\
    54 & Consumption, Orders, and Inventories & RETAILx & Retail and Food Services Sales \\
    55 & Consumption, Orders, and Inventories & AMDMNOx & New Orders for Durable Goods \\
    56 & Consumption, Orders, and Inventories & AMDMUOx & Unfilled Orders for Durable Goods \\
    57 & Consumption, Orders, and Inventories & BUSINVx & Total Business Inventories \\
    58 & Consumption, Orders, and Inventories & ISRATIOx & Total Business:  Inventories to Sales Ratio \\
    59 & Money and Credit & M1SL & M1 Money Stock \\
    60 & Money and Credit & M2SL & M2 Money Stock \\
    61 & Money and Credit & M2REAL & Real M2 Money Stock \\
    62 & Money and Credit & BOGMBASE & Monetary Base \\
    63 & Money and Credit & TOTRESNS & Total Reserves of Depository Institutions \\
    64 & Money and Credit & NONBORRES & Reserves Of Depository Institutions \\
    65 & Money and Credit & BUSLOANS & Commercial and Industrial Loans \\
    66 & Money and Credit & REALLN & Real Estate Loans at All Commercial Banks \\
    67 & Money and Credit & NONREVSL & Total Nonrevolving Credit \\
    68 & Money and Credit & CONSPI & Nonrevolving consumer credit to Personal Income \\
    69 & Money and Credit & DTCOLNVHFNM & Consumer Motor Vehicle Loans Outstanding \\
    70 & Money and Credit & DTCTHFNM & Total Consumer Loans and Leases Outstanding \\
    71 & Money and Credit & INVEST & Securities in Bank Credit at All Commercial Banks \\
    72 & Interest and Exchange Rates & FEDFUNDS & Effective Federal Funds Rate \\
    73 & Interest and Exchange Rates & CP3Mx & 3-Month AA Financial Commercial Paper Rate \\
    74 & Interest and Exchange Rates & TB3MS & 3-Month Treasury Bill: \\
    75 & Interest and Exchange Rates & TB6MS & 6-Month Treasury Bill: \\
    76 & Interest and Exchange Rates & GS1 & 1-Year Treasury Rate \\
    77 & Interest and Exchange Rates & GS5 & 5-Year Treasury Rate \\
    78 & Interest and Exchange Rates & GS10 & 10-Year Treasury Rate \\
    79 & Interest and Exchange Rates & AAA & Moody Seasoned Aaa Corporate Bond Yield \\
    80 & Interest and Exchange Rates & BAA & Moody Seasoned Baa Corporate Bond Yield \\
    81 & Interest and Exchange Rates & COMPAPFFx & 3-Month Commercial Paper Minus FEDFUNDS \\
    82 & Interest and Exchange Rates & TB3SMFFM & 3-Month Treasury C Minus FEDFUNDS \\
    83 & Interest and Exchange Rates & TB6SMFFM & 6-Month Treasury C Minus FEDFUNDS \\
    84 & Interest and Exchange Rates & T1YFFM & 1-Year Treasury C Minus FEDFUNDS \\
    85 & Interest and Exchange Rates & T5YFFM & 5-Year Treasury C Minus FEDFUNDS \\
    86 & Interest and Exchange Rates & T10YFFM & 10-Year Treasury C Minus FEDFUNDS \\
    87 & Interest and Exchange Rates & AAAFFM & Moody's Aaa Corporate Bond Minus FEDFUNDS \\
    88 & Interest and Exchange Rates & BAAFFM & Moody's Baa Corporate Bond Minus FEDFUNDS \\
    89 & Interest and Exchange Rates & EXSZUSx & Switzerland / U.S. Foreign Exchange Rate \\
    90 & Interest and Exchange Rates & EXJPUSx & Japan / U.S. Foreign Exchange Rate \\
    91 & Interest and Exchange Rates & EXUSUKx & U.S. / U.K. Foreign Exchange Rate \\
    92 & Interest and Exchange Rates & EXCAUSx & Canada / U.S. Foreign Exchange Rate \\
    93 & Prices & WPSFD49207 & PPI: Finished Goods \\
    94 & Prices & WPSFD49502 & PPI: Finished Consumer Goods \\
    95 & Prices & WPSID61 & PPI: Intermediate Materials \\
    96 & Prices & WPSID62 & PPI: Crude Materials \\
    97 & Prices & OILPRICEx & Crude Oil, spliced WTI and Cushing \\
    98 & Prices & PPICMM & PPI: Metals and metal products: \\
    99 & Prices & CPIAUCSL & CPI : All Items \\
    100 & Prices & CPIAPPSL & CPI : Apparel \\
    101 & Prices & CPITRNSL & CPI : Transportation \\
    102 & Prices & CPIMEDSL & CPI : Medical Care \\
    103 & Prices & CUSR0000SAC & CPI : Commodities \\
    104 & Prices & CUSR0000SAD & CPI : Durables \\
    105 & Prices & CUSR0000SAS & CPI : Services \\
    106 & Prices & CPIULFSL & CPI : All Items Less Food \\
    107 & Prices & CUSR0000SA0L2 & CPI : All items less shelter \\
    108 & Prices & CUSR0000SA0L5 & CPI : All items less medical care \\
    109 & Prices & PCEPI & Personal Cons.  Expend.:  Chain Index \\
    110 & Prices & DDURRG3M086SBEA & Personal Cons.  Exp:  Durable goods \\
    111 & Prices & DNDGRG3M086SBEA & Personal Cons.  Exp:  Nondurable goods \\
    112 & Prices & DSERRG3M086SBEA & Personal Cons.  Exp:  Services \\
    113 & Stock Market & S\&P 500 & S\&P500 Common Stock Price Index: Composite \\
    114 & Stock Market & S\&P div yield & S\&P500 Composite Common Stock: Dividend Yield \\
    115 & Stock Market & S\&P PE ratio & S\&P500 Composite Common Stock: Price-Earnings Ratio
\end{longtable}
\end{landscape}

\begin{landscape}

\subsection{Analysts' consensus variables}

\small
\setlength\LTleft{0pt}
\setlength\LTright{0pt}
\begin{longtable}{@{\extracolsep{\fill}} 
    l    
    l    
    p{5cm} 
    p{4cm} 
    l    
    l    
}
    \caption{Descriptions of analysts' consensus variables from \textcite{ChenZimmermann2021}.}
    \label{tab:analyst_var} \\
    \toprule
    \multicolumn{1}{l}{\small No.} &
    \multicolumn{1}{l}{\small Acronym} &
    \multicolumn{1}{l}{\small Analyst Consensus} &
    \multicolumn{1}{l}{\small Authors} &
    \multicolumn{1}{l}{\small Year} &
    \multicolumn{1}{l}{\small Journal} \\
    \midrule
    \endfirsthead

    \caption[]{Descriptions of analysts' consensus variables from \textcite{ChenZimmermann2021} (cont'd).} \\
    \toprule
    \multicolumn{1}{l}{\small No.} &
    \multicolumn{1}{l}{\small Acronym} &
    \multicolumn{1}{l}{\small Analyst Consensus} &
    \multicolumn{1}{l}{\small Authors} &
    \multicolumn{1}{l}{\small Year} &
    \multicolumn{1}{l}{\small Journal} \\
    \midrule
    \endhead

    \bottomrule
    \endfoot

    \bottomrule
    \endlastfoot
    
    1 & AnalystRevision & EPS forecast revision & Hawkins, Chamberlin, Daniel & 1984 & FAJ \\
    2 & ChangeInRecommendation & Change in recommendation & Jegadeesh et al. & 2004 & JF \\
    3 & ChForecastAccrual & Change in Forecast and Accrual & Barth and Hutton & 2004 & RAS \\
    4 & EarningsForecastDisparity & Long-vs-short EPS forecasts & Da and Warachka & 2011 & JFE \\
    5 & FEPS & Analyst earnings per share & Cen, Wei, and Zhang & 2006 & WP \\
    6 & ForecastDispersion & EPS Forecast Dispersion & Diether, Malloy and Scherbina & 2002 & JF \\
    7 & REV6 & Earnings forecast revisions & Chan, Jegadeesh and Lakonishok & 1996 & JF \\
    8 & AnalystValue & Analyst Value & Frankel and Lee & 1998 & JAE \\
    9 & AOP & Analyst Optimism & Frankel and Lee & 1998 & JAE
    
\end{longtable}
\end{landscape}

{
\singlespacing
\printbibliography[title={Internet Appendix References}]

@inproceedings{koh2020concept,
  title={Concept bottleneck models},
  author={Koh, Pang Wei and Nguyen, Thao and Tang, Yew Siang and Mussmann, Stephen and Pierson, Emma and Kim, Been and Liang, Percy},
  booktitle={International conference on machine learning},
  pages={5338--5348},
  year={2020},
  organization={PMLR}
}

@article{gu2020empirical,
  title={Empirical asset pricing via machine learning},
  author={Gu, Shihao and Kelly, Bryan and Xiu, Dacheng},
  journal={The Review of Financial Studies},
  volume={33},
  number={5},
  pages={2223--2273},
  year={2020},
  publisher={Oxford University Press}
}

@article{leippold2022machine,
  title={Machine learning in the Chinese stock market},
  author={Leippold, Markus and Wang, Qian and Zhou, Wenyu},
  journal={Journal of Financial Economics},
  volume={145},
  number={2},
  pages={64--82},
  year={2022},
  publisher={Elsevier}
}

@article{jaquart2021short,
  title={Short-term bitcoin market prediction via machine learning},
  author={Jaquart, Patrick and Dann, David and Weinhardt, Christof},
  journal={The journal of finance and data science},
  volume={7},
  pages={45--66},
  year={2021},
  publisher={Elsevier}
}

@article{fang2024ascertaining,
  title={Ascertaining price formation in cryptocurrency markets with machine learning},
  author={Fang, Fan and Chung, Waichung and Ventre, Carmine and Basios, Michail and Kanthan, Leslie and Li, Lingbo and Wu, Fan},
  journal={The European Journal of Finance},
  volume={30},
  number={1},
  pages={78--100},
  year={2024},
  publisher={Taylor \& Francis}
}

@article{green2017characteristics,
  title={The characteristics that provide independent information about average US monthly stock returns},
  author={Green, Jeremiah and Hand, John RM and Zhang, X Frank},
  journal={The Review of Financial Studies},
  volume={30},
  number={12},
  pages={4389--4436},
  year={2017},
  publisher={Oxford University Press}
}

@article{welch2008comprehensive,
  title={A comprehensive look at the empirical performance of equity premium prediction},
  author={Welch, Ivo and Goyal, Amit},
  journal={The Review of Financial Studies},
  volume={21},
  number={4},
  pages={1455--1508},
  year={2008},
  publisher={Society for Financial Studies}
}

@article{mccracken2016fred,
  title={FRED-MD: A monthly database for macroeconomic research},
  author={McCracken, Michael W and Ng, Serena},
  journal={Journal of Business \& Economic Statistics},
  volume={34},
  number={4},
  pages={574--589},
  year={2016},
  publisher={Taylor \& Francis}
}

@article{bianchi2021bond,
  title={Bond risk premiums with machine learning},
  author={Bianchi, Daniele and B{\"u}chner, Matthias and Tamoni, Andrea},
  journal={The Review of Financial Studies},
  volume={34},
  number={2},
  pages={1046--1089},
  year={2021},
  publisher={Oxford University Press}
}

@article{kelly2019characteristics,
  title={Characteristics are covariances: A unified model of risk and return},
  author={Kelly, Bryan T and Pruitt, Seth and Su, Yinan},
  journal={Journal of Financial Economics},
  volume={134},
  number={3},
  pages={501--524},
  year={2019},
  publisher={Elsevier}
}

@article{feng2018deep,
  title={Deep learning in characteristics-sorted factor models},
  author={Feng, Guanhao and He, Jingyu and Polson, Nicholas G and Xu, Jianeng},
  journal={Journal of Financial and Quantitative Analysis},
  pages={1--36},
  year={2018},
  publisher={Cambridge University Press}
}

@article{gu2021autoencoder,
  title={Autoencoder asset pricing models},
  author={Gu, Shihao and Kelly, Bryan and Xiu, Dacheng},
  journal={Journal of Econometrics},
  volume={222},
  number={1},
  pages={429--450},
  year={2021},
  publisher={Elsevier}
}

@article{chen2024deep,
  title={Deep learning in asset pricing},
  author={Chen, Luyang and Pelger, Markus and Zhu, Jason},
  journal={Management Science},
  volume={70},
  number={2},
  pages={714--750},
  year={2024},
  publisher={INFORMS}
}

@article{bybee2023narrative,
  title={Narrative asset pricing: Interpretable systematic risk factors from news text},
  author={Bybee, Leland and Kelly, Bryan and Su, Yinan},
  journal={The Review of Financial Studies},
  volume={36},
  number={12},
  pages={4759--4787},
  year={2023},
  publisher={Oxford University Press}
}

@article{rudin2022interpretable,
  title={Interpretable machine learning: Fundamental principles and 10 grand challenges},
  author={Rudin, Cynthia and Chen, Chaofan and Chen, Zhi and Huang, Haiyang and Semenova, Lesia and Zhong, Chudi},
  journal={Statistic Surveys},
  volume={16},
  pages={1--85},
  year={2022},
  publisher={The American Statistical Association, the Bernoulli Society, the Institute~…}
}

@article{rudin2019stop,
  title={Stop explaining black box machine learning models for high stakes decisions and use interpretable models instead},
  author={Rudin, Cynthia},
  journal={Nature machine intelligence},
  volume={1},
  number={5},
  pages={206--215},
  year={2019},
  publisher={Nature Publishing Group UK London}
}

@article{higgins2018towards,
  title={Towards a definition of disentangled representations},
  author={Higgins, Irina and Amos, David and Pfau, David and Racaniere, Sebastien and Matthey, Loic and Rezende, Danilo and Lerchner, Alexander},
  journal={arXiv preprint arXiv:1812.02230},
  year={2018}
}

@article{cochrane2008dog,
  title={The dog that did not bark: A defense of return predictability},
  author={Cochrane, John H},
  journal={The Review of Financial Studies},
  volume={21},
  number={4},
  pages={1533--1575},
  year={2008},
  publisher={Society for Financial Studies}
}

@article{ang2007stock,
  title={Stock return predictability: Is it there?},
  author={Ang, Andrew and Bekaert, Geert},
  journal={The Review of Financial Studies},
  volume={20},
  number={3},
  pages={651--707},
  year={2007},
  publisher={Oxford University Press}
}

@article{campbell2008predicting,
  title={Predicting excess stock returns out of sample: Can anything beat the historical average?},
  author={Campbell, John Y and Thompson, Samuel B},
  journal={The Review of Financial Studies},
  volume={21},
  number={4},
  pages={1509--1531},
  year={2008},
  publisher={Society for Financial Studies}
}

@article{fama1993common,
  title={Common risk factors in the returns on stocks and bonds},
  author={Fama, Eugene F and French, Kenneth R},
  journal={Journal of financial economics},
  volume={33},
  number={1},
  pages={3--56},
  year={1993},
  publisher={Elsevier}
}

@article{carhart1997persistence,
  title={On persistence in mutual fund performance},
  author={Carhart, Mark M},
  journal={The Journal of finance},
  volume={52},
  number={1},
  pages={57--82},
  year={1997},
  publisher={Wiley Online Library}
}

@article{fama2015five,
  title={A five-factor asset pricing model},
  author={Fama, Eugene F and French, Kenneth R},
  journal={Journal of financial economics},
  volume={116},
  number={1},
  pages={1--22},
  year={2015},
  publisher={Elsevier}
}

@article{cochrane2011presidential,
  title={Presidential address: Discount rates},
  author={Cochrane, John H},
  journal={The Journal of finance},
  volume={66},
  number={4},
  pages={1047--1108},
  year={2011},
  publisher={Wiley Online Library}
}

@article{freyberger2020dissecting,
  title={Dissecting characteristics nonparametrically},
  author={Freyberger, Joachim and Neuhierl, Andreas and Weber, Michael},
  journal={The Review of Financial Studies},
  volume={33},
  number={5},
  pages={2326--2377},
  year={2020},
  publisher={Oxford University Press}
}

@article{feng2020taming,
  title={Taming the factor zoo: A test of new factors},
  author={Feng, Guanhao and Giglio, Stefano and Xiu, Dacheng},
  journal={The Journal of Finance},
  volume={75},
  number={3},
  pages={1327--1370},
  year={2020},
  publisher={Wiley Online Library}
}

@article{harvey2016and,
  title={… and the cross-section of expected returns},
  author={Harvey, Campbell R and Liu, Yan and Zhu, Heqing},
  journal={The Review of Financial Studies},
  volume={29},
  number={1},
  pages={5--68},
  year={2016},
  publisher={Oxford University Press}
}

@article{green2013supraview,
  title={The supraview of return predictive signals},
  author={Green, Jeremiah and Hand, John RM and Zhang, X Frank},
  journal={Review of Accounting Studies},
  volume={18},
  pages={692--730},
  year={2013},
  publisher={Springer}
}

@article{jensen2023there,
  title={Is there a replication crisis in finance?},
  author={Jensen, Theis Ingerslev and Kelly, Bryan and Pedersen, Lasse Heje},
  journal={The Journal of Finance},
  volume={78},
  number={5},
  pages={2465--2518},
  year={2023},
  publisher={Wiley Online Library}
}

@article{he2017intermediary,
  title={Intermediary asset pricing: New evidence from many asset classes},
  author={He, Zhiguo and Kelly, Bryan and Manela, Asaf},
  journal={Journal of Financial Economics},
  volume={126},
  number={1},
  pages={1--35},
  year={2017},
  publisher={Elsevier}
}

@article{hou2015digesting,
  title={Digesting anomalies: An investment approach},
  author={Hou, Kewei and Xue, Chen and Zhang, Lu},
  journal={The Review of Financial Studies},
  volume={28},
  number={3},
  pages={650--705},
  year={2015},
  publisher={Oxford University Press}
}

@article{ChenZimmermann2021,
  title={Open Source Cross-Sectional Asset Pricing},
  author={Chen, Andrew Y. and Tom Zimmermann},
  journal={Critical Finance Review},
  year={2022},
  volume={27},
  number={2},
  pages={207--264} 
}

@article{kelly2024virtue,
  title={The virtue of complexity in return prediction},
  author={Kelly, Bryan and Malamud, Semyon and Zhou, Kangying},
  journal={The Journal of Finance},
  volume={79},
  number={1},
  pages={459--503},
  year={2024},
  publisher={Wiley Online Library}
}

@article{plate1999accuracy,
  title={Accuracy versus interpretability in flexible modeling: Implementing a tradeoff using gaussian process models},
  author={Plate, Tony A},
  journal={Behaviormetrika},
  volume={26},
  pages={29--50},
  year={1999},
  publisher={Springer}
}

@article{lovell1986tests,
  title={Tests of the rational expectations hypothesis},
  author={Lovell, Michael C},
  journal={The American Economic Review},
  volume={76},
  number={1},
  pages={110--124},
  year={1986},
  publisher={JSTOR}
}

@article{feurer2019hyperparameter,
  title={Hyperparameter optimization},
  author={Feurer, Matthias and Hutter, Frank},
  journal={Automated machine learning: Methods, systems, challenges},
  pages={3--33},
  year={2019},
  publisher={Springer International Publishing}
}

@article{benitez1997artificial,
  title={Are artificial neural networks black boxes?},
  author={Ben{\'\i}tez, Jos{\'e} Manuel and Castro, Juan Luis and Requena, Ignacio},
  journal={IEEE Transactions on neural networks},
  volume={8},
  number={5},
  pages={1156--1164},
  year={1997},
  publisher={IEEE}
}

@inproceedings{locatello2019challenging,
  title={Challenging common assumptions in the unsupervised learning of disentangled representations},
  author={Locatello, Francesco and Bauer, Stefan and Lucic, Mario and Raetsch, Gunnar and Gelly, Sylvain and Sch{\"o}lkopf, Bernhard and Bachem, Olivier},
  booktitle={international conference on machine learning},
  pages={4114--4124},
  year={2019},
  organization={PMLR}
}

@article{locatello2019fairness,
  title={On the fairness of disentangled representations},
  author={Locatello, Francesco and Abbati, Gabriele and Rainforth, Thomas and Bauer, Stefan and Sch{\"o}lkopf, Bernhard and Bachem, Olivier},
  journal={Advances in neural information processing systems},
  volume={32},
  year={2019}
}

@article{chen2020concept,
  title={Concept whitening for interpretable image recognition},
  author={Chen, Zhi and Bei, Yijie and Rudin, Cynthia},
  journal={Nature Machine Intelligence},
  volume={2},
  number={12},
  pages={772--782},
  year={2020},
  publisher={Nature Publishing Group UK London}
}

@article{chen2016infogan,
  title={Infogan: Interpretable representation learning by information maximizing generative adversarial nets},
  author={Chen, Xi and Duan, Yan and Houthooft, Rein and Schulman, John and Sutskever, Ilya and Abbeel, Pieter},
  journal={Advances in neural information processing systems},
  volume={29},
  year={2016}
}

@article{higgins2017beta,
  title={beta-vae: Learning basic visual concepts with a constrained variational framework.},
  author={Higgins, Irina and Matthey, Loic and Pal, Arka and Burgess, Christopher P and Glorot, Xavier and Botvinick, Matthew M and Mohamed, Shakir and Lerchner, Alexander},
  journal={ICLR (Poster)},
  volume={3},
  year={2017}
}

@article{lim2001rationality,
  title={Rationality and analysts' forecast bias},
  author={Lim, Terence},
  journal={The journal of Finance},
  volume={56},
  number={1},
  pages={369--385},
  year={2001},
  publisher={Wiley Online Library}
}

@article{jegadeesh2004analyzing,
  title={Analyzing the analysts: When do recommendations add value?},
  author={Jegadeesh, Narasimhan and Kim, Joonghyuk and Krische, Susan D and Lee, Charles MC},
  journal={The journal of finance},
  volume={59},
  number={3},
  pages={1083--1124},
  year={2004},
  publisher={Wiley Online Library}
}

@article{diether2002differences,
  title={Differences of opinion and the cross section of stock returns},
  author={Diether, Karl B and Malloy, Christopher J and Scherbina, Anna},
  journal={The journal of finance},
  volume={57},
  number={5},
  pages={2113--2141},
  year={2002},
  publisher={Wiley Online Library}
}

@article{sorescu2006cross,
  title={The cross section of analyst recommendations},
  author={Sorescu, Sorin and Subrahmanyam, Avanidhar},
  journal={Journal of Financial and Quantitative Analysis},
  volume={41},
  number={1},
  pages={139--168},
  year={2006},
  publisher={Cambridge University Press}
}

@misc{ba2016layer,
      title={Layer Normalization}, 
      author={Jimmy Lei Ba and Jamie Ryan Kiros and Geoffrey E. Hinton},
      year={2016},
      eprint={1607.06450},
      archivePrefix={arXiv},
      primaryClass={stat.ML}
}

@misc{hinton2012improving,
      title={Improving neural networks by preventing co-adaptation of feature detectors}, 
      author={Geoffrey E. Hinton and Nitish Srivastava and Alex Krizhevsky and Ilya Sutskever and Ruslan R. Salakhutdinov},
      year={2012},
      eprint={1207.0580},
      archivePrefix={arXiv},
      primaryClass={cs.NE}
}

@inproceedings{pascanu2013difficulty,
  title={On the difficulty of training recurrent neural networks},
  author={Pascanu, Razvan and Mikolov, Tomas and Bengio, Yoshua},
  booktitle={International conference on machine learning},
  pages={1310--1318},
  year={2013},
  organization={Pmlr}
}

@misc{kingma2017adam,
      title={Adam: A Method for Stochastic Optimization}, 
      author={Diederik P. Kingma and Jimmy Ba},
      year={2017},
      eprint={1412.6980},
      archivePrefix={arXiv},
      primaryClass={cs.LG}
}

@article{muth1961rational,
  title={Rational expectations and the theory of price movements},
  author={Muth, John F},
  journal={Econometrica: journal of the Econometric Society},
  pages={315--335},
  year={1961},
  publisher={JSTOR}
}

@article{daniel1997evidence,
  title={Evidence on the characteristics of cross sectional variation in stock returns},
  author={Daniel, Kent and Titman, Sheridan},
  journal={the Journal of Finance},
  volume={52},
  number={1},
  pages={1--33},
  year={1997},
  publisher={Wiley Online Library}
}

@article{hinton2012neural,
  title={Neural networks for machine learning lecture 6a overview of mini-batch gradient descent},
  author={Hinton, Geoffrey and Srivastava, Nitish and Swersky, Kevin},
  journal={Cited on},
  volume={14},
  number={8},
  pages={2},
  year={2012}
}

@misc{ReduceLROnPlateau,
  title = {{PyTorch}: ReduceLROnPlateau - PyTorch 2.3.0 documentation.},
  howpublished = {\url{https://pytorch.org/docs/stable/generated/torch.optim.lr_scheduler.ReduceLROnPlateau.html}}
}

@inproceedings{ioffe2015batch,
  title={Batch normalization: Accelerating deep network training by reducing internal covariate shift},
  author={Ioffe, Sergey and Szegedy, Christian},
  booktitle={International conference on machine learning},
  pages={448--456},
  year={2015},
  organization={pmlr}
}

@article{hendrycks2016gaussian,
  title={Gaussian error linear units (gelus)},
  author={Hendrycks, Dan and Gimpel, Kevin},
  journal={arXiv preprint arXiv:1606.08415},
  year={2016}
}

@inproceedings{nair2010rectified,
  title={Rectified linear units improve restricted boltzmann machines},
  author={Nair, Vinod and Hinton, Geoffrey E},
  booktitle={Proceedings of the 27th international conference on machine learning (ICML-10)},
  pages={807--814},
  year={2010}
}

@article{fama1973risk,
  title={Risk, return, and equilibrium: Empirical tests},
  author={Fama, Eugene F and MacBeth, James D},
  journal={Journal of political economy},
  volume={81},
  number={3},
  pages={607--636},
  year={1973},
  publisher={The University of Chicago Press}
}

@misc{barron2017continuously,
      title={Continuously Differentiable Exponential Linear Units}, 
      author={Jonathan T. Barron},
      year={2017},
      eprint={1704.07483},
      archivePrefix={arXiv},
      primaryClass={cs.LG}
}

@misc{clevert2016fast,
      title={Fast and Accurate Deep Network Learning by Exponential Linear Units (ELUs)}, 
      author={Djork-Arné Clevert and Thomas Unterthiner and Sepp Hochreiter},
      year={2016},
      eprint={1511.07289},
      archivePrefix={arXiv},
      primaryClass={cs.LG}
}

@inproceedings{he2015delving,
  title={Delving deep into rectifiers: Surpassing human-level performance on imagenet classification},
  author={He, Kaiming and Zhang, Xiangyu and Ren, Shaoqing and Sun, Jian},
  booktitle={Proceedings of the IEEE international conference on computer vision},
  pages={1026--1034},
  year={2015}
}

@article{ludvigson2007empirical,
  title={The empirical risk--return relation: A factor analysis approach},
  author={Ludvigson, Sydney C and Ng, Serena},
  journal={Journal of financial economics},
  volume={83},
  number={1},
  pages={171--222},
  year={2007},
  publisher={Elsevier}
}

@article{erichson2020sparse,
  title={Sparse principal component analysis via variable projection},
  author={Erichson, N Benjamin and Zheng, Peng and Manohar, Krithika and Brunton, Steven L and Kutz, J Nathan and Aravkin, Aleksandr Y},
  journal={SIAM Journal on Applied Mathematics},
  volume={80},
  number={2},
  pages={977--1002},
  year={2020},
  publisher={SIAM}
}

@article{fan2016projected,
  title={Projected principal component analysis in factor models},
  author={Fan, Jianqing and Liao, Yuan and Wang, Weichen},
  journal={Annals of statistics},
  volume={44},
  number={1},
  pages={219},
  year={2016}
}

@article{gao2019res2net,
  title={Res2net: A new multi-scale backbone architecture},
  author={Gao, Shang-Hua and Cheng, Ming-Ming and Zhao, Kai and Zhang, Xin-Yu and Yang, Ming-Hsuan and Torr, Philip},
  journal={IEEE transactions on pattern analysis and machine intelligence},
  volume={43},
  number={2},
  pages={652--662},
  year={2019},
  publisher={IEEE}
}

@inproceedings{devlin2019bert,
  title={Bert: Pre-training of deep bidirectional transformers for language understanding},
  author={Devlin, Jacob and Chang, Ming-Wei and Lee, Kenton and Toutanova, Kristina},
  booktitle={Proceedings of the 2019 conference of the North American chapter of the association for computational linguistics: human language technologies, volume 1 (long and short papers)},
  pages={4171--4186},
  year={2019}
}

@article{hinton2006reducing,
  title={Reducing the dimensionality of data with neural networks},
  author={Hinton, Geoffrey E and Salakhutdinov, Ruslan R},
  journal={science},
  volume={313},
  number={5786},
  pages={504--507},
  year={2006},
  publisher={American Association for the Advancement of Science}
}

@article{nelson1987parsimonious,
  title={Parsimonious modeling of yield curves},
  author={Nelson, Charles R and Siegel, Andrew F},
  journal={Journal of business},
  pages={473--489},
  year={1987},
  publisher={JSTOR}
}

@article{litterman1991common,
  title={Common factors affecting bond returns},
  author={Litterman, Robert},
  journal={Journal of fixed income},
  pages={54--61},
  year={1991}
}

@article{gibbons1989test,
  title={A test of the efficiency of a given portfolio},
  author={Gibbons, Michael R and Ross, Stephen A and Shanken, Jay},
  journal={Econometrica: Journal of the Econometric Society},
  pages={1121--1152},
  year={1989},
  publisher={JSTOR}
}

@article{cong2025growing,
  title={Growing the efficient frontier on panel trees},
  author={Cong, Lin William and Feng, Guanhao and He, Jingyu and He, Xin},
  journal={Journal of Financial Economics},
  volume={167},
  pages={104024},
  year={2025},
  publisher={Elsevier}
}

@article{driscoll1998consistent,
  title={Consistent covariance matrix estimation with spatially dependent panel data},
  author={Driscoll, John C and Kraay, Aart C},
  journal={Review of economics and statistics},
  volume={80},
  number={4},
  pages={549--560},
  year={1998},
  publisher={MIT Press 238 Main St., Suite 500, Cambridge, MA 02142-1046, USA journals~…}
}

@article{newey1986simple,
  title={A simple, positive semi-definite, heteroskedasticity and autocorrelationconsistent covariance matrix},
  author={Newey, Whitney K and West, Kenneth D},
  year={1986},
  publisher={National Bureau of Economic Research Cambridge, Mass., USA}
}

@article{hodrick1992dividend,
  title={Dividend yields and expected stock returns: Alternative procedures for inference and measurement},
  author={Hodrick, Robert J},
  journal={The Review of Financial Studies},
  volume={5},
  number={3},
  pages={357--386},
  year={1992},
  publisher={Oxford University Press}
}

@inproceedings{Cen2006ForecastedEP,
  title={Forecasted Earnings per Share and the Cross Section of Expected Stock Returns},
  author={Ling Cen},
  year={2006},
  url={https://api.semanticscholar.org/CorpusID:204538423}
}

@article{chinco2019sparse,
  title={Sparse signals in the cross-section of returns},
  author={Chinco, Alex and Clark-Joseph, Adam D and Ye, Mao},
  journal={The Journal of Finance},
  volume={74},
  number={1},
  pages={449--492},
  year={2019},
  publisher={Wiley Online Library}
}

@article{bryzgalova2025forest,
  title={Forest through the trees: Building cross-sections of stock returns},
  author={Bryzgalova, Svetlana and Pelger, Markus and Zhu, Jason},
  journal={The Journal of Finance},
  volume={80},
  number={5},
  pages={2447--2506},
  year={2025},
  publisher={Wiley Online Library}
}

@article{bessembinder2003trade,
  title={Trade execution costs and market quality after decimalization},
  author={Bessembinder, Hendrik},
  journal={Journal of Financial and Quantitative Analysis},
  volume={38},
  number={4},
  pages={747--777},
  year={2003},
  publisher={Cambridge University Press}
}

@article{frazzini2012trading,
  title={Trading costs of asset pricing anomalies},
  author={Frazzini, Andrea and Israel, Ronen and Moskowitz, Tobias J},
  journal={Fama-Miller Working Paper, Chicago Booth Research Paper},
  number={14-05},
  year={2012}
}

@article{barber2001can,
  title={Can investors profit from the prophets? Security analyst recommendations and stock returns},
  author={Barber, Brad and Lehavy, Reuven and McNichols, Maureen and Trueman, Brett},
  journal={The Journal of finance},
  volume={56},
  number={2},
  pages={531--563},
  year={2001},
  publisher={Wiley Online Library}
}

@article{palley2025effect,
  title={The Effect of Dispersion on the Informativeness of Consensus Analyst Target Prices},
  author={Palley, Asa B and Steffen, Thomas D and Zhang, X Frank},
  journal={Management Science},
  volume={71},
  number={3},
  pages={2264--2288},
  year={2025},
  publisher={INFORMS}
}

@article{van2023man,
  title={Man versus machine learning: The term structure of earnings expectations and conditional biases},
  author={Van Binsbergen, Jules H and Han, Xiao and Lopez-Lira, Alejandro},
  journal={The Review of financial studies},
  volume={36},
  number={6},
  pages={2361--2396},
  year={2023},
  publisher={Oxford University Press}
}

@article{cao2024man,
  title={From man vs. machine to man+ machine: The art and AI of stock analyses},
  author={Cao, Sean and Jiang, Wei and Wang, Junbo and Yang, Baozhong},
  journal={Journal of Financial Economics},
  volume={160},
  pages={103910},
  year={2024},
  publisher={Elsevier}
}
}

\end{refsection}

\end{document}